%% file: pk30.tex
\documentclass{aa}  

\usepackage{color}
\usepackage{aalongtable,lscape}
\usepackage{graphicx}
\usepackage{natbib}
\usepackage{amssymb}
\begin{document} 

\title{CARMENES input catalogue of M dwarfs}

\subtitle{IV. New rotation periods from photometric time series}

\titlerunning{New rotation periods of CARMENES M dwarfs}
    
\authorrunning{D\'iez Alonso et~al.}


\author{E.~D\'iez Alonso\inst{1,2,3}
   \and J.~A.~Caballero\inst{4}
   \and D.~Montes\inst{1}
   \and F.~J. de~Cos~Juez\inst{2}
   \and S.~Dreizler\inst{5}
   \and F.~Dubois\inst{6}
   \and S.~V.~Jeffers\inst{5}  
   \and S.~Lalitha\inst{5}
   \and R.~Naves\inst{7}
   \and A.~Reiners\inst{5}
   \and I.~Ribas\inst{8,9}
   \and S.~Vanaverbeke\inst{10,6}
   \and P.~J.~Amado\inst{11}
   \and V.~J.~S.~B\'ejar\inst{12,13}
   \and M.~Cort\'es-Contreras\inst{4}
   \and E.~Herrero\inst{8,9}
   \and D.~Hidalgo\inst{12,13,1}
   \and M.~K\"urster\inst{14}
   \and L.~Logie\inst{6}
   \and A.~Quirrenbach\inst{15}
   \and S.~Rau\inst{6}
   \and W.~Seifert\inst{15}
   \and P.~Sch\"ofer\inst{5}
   \and L.~Tal-Or\inst{5,16}
}

\institute{
        Departamento de Astrof\'isica y Ciencias de la Atm\'osfera, Facultad de Ciencias F\'isicas, Universidad Complutense de Madrid, E-280140 Madrid, Spain; \email{enridiez@ucm.es}
        \and 
        Departamento de Explotaci\'on y Prospecci\'on de Minas, Escuela de Minas, Energ\'ia y Materiales, Universidad de Oviedo, E-33003 Oviedo, Asturias, Spain
                \and
                Observatorio Astron\'omico Carda, Villaviciosa, Asturias, Spain (MPC Z76)
                \and
                Centro de Astrobiolog\'{\i}a (CSIC-INTA), Campus ESAC, Camino Bajo del Castillo s/n, E-28692 Villanueva de la Ca\~nada, Madrid, Spain
                \and
                Institut f\"ur Astrophysik, Georg-August-Universit\"at G\"ottingen, Friedrich-Hund-Platz 1, D-37077 G\"ottingen, Germany
                \and
                AstroLAB IRIS, Provinciaal Domein ``De Palingbeek'', Verbrandemolenstraat 5, B-8902 Zillebeke, Ieper, Belgium
                \and
                Observatorio Astron\'omico Naves, Cabrils, Barcelona, Spain (MPC 213) 
                \and
                Institut de Ci\`encies de l'Espai (CSIC-IEEC), Campus UAB, c/ de Can Magrans s/n, E-08193 Bellaterra, Barcelona, Spain 
        \and 
        Institut d'Estudis Espacials de Catalunya (IEEC), E-08034 Barcelona, Spain
                \and
                Vereniging Voor Sterrenkunde, Brugge, Belgium \& Centre for mathematical Plasma Astrophysics, Katholieke Universiteit Leuven, Celestijnenlaan 200B, bus 2400, B-3001 Leuven, Belgium
                \and
                Instituto de Astrof\'isica de Andaluc\'ia (CSIC), Glorieta de la Astronom\'ia s/n, E-18008 Granada, Spain 
                \and
                Instituto de Astrof\'{\i}sica de Canarias, c/ V\'{\i}a L\'actea s/n, E-38205 La Laguna, Tenerife, Spain 
        \and 
        Departamento de Astrof\'{\i}sica, Universidad de La Laguna, E-38206 La Laguna, Tenerife, Spain
                \and
                Max-Planck-Institut f\"ur Astronomie, K\"onigstuhl 17, E-69117 Heidelberg, Germany
        \and 
                Landessternwarte, Zentrum f\"ur Astronomie der Universit\"at Heidelberg, K\"onigstuhl 12, D-69117 Heidelberg, Germany 
        \and
        School of Geosciences, Raymond and Beverly Sackler Faculty of Exact Sciences, Tel Aviv University, Tel Aviv, 6997801, Israel
}
             
\date{Received 27 April 2018; accepted dd Mmm 2018}


 
  \abstract
   {}
   { The main goal of this work is to measure rotation periods of the M-type dwarf stars being observed by the CARMENES exoplanet survey to help distinguish radial-velocity signals produced by magnetic activity from those produced by exoplanets. 
   Rotation periods are also fundamental for a detailed study of the relation between activity and rotation in late-type stars.}
   {We look for significant periodic signals in 622 photometric time series of 337 bright, nearby M dwarfs obtained by long-time baseline, automated surveys (MEarth, ASAS, SuperWASP, NSVS, Catalina, ASAS-SN, K2, and HATNet) and for 20 stars which we obtained with four 0.2--0.8\,m telescopes at high geographical latitudes.}
   {We present 142 rotation periods (73 new) from 0.12\,d to 133\,d and ten long-term activity cycles (six new) from 3.0\,a to 11.5\,a.
   We compare our determinations with those in the existing literature; we investigate the distribution of $P_{\rm rot}$ in the CARMENES input catalogue, the amplitude of photometric variability, and their relation to $v \sin{i}$ and pEW(H$\alpha$); and we identify three very active stars with new rotation periods between 0.34\,d and 23.6\,d.}
   {}

   \keywords{stars: activity -- stars: late type -- stars: rotation -- techniques: photometry}

   \maketitle
%


\section{Introduction}
\label{sec.intro}

In current exoplanet search programmes, knowledge of the stellar rotation periods is essential in order to distinguish radial-velocity signals induced by real planets or by the rotation of the star itself \citep{1997ApJ...485..319S,2001A&A...379..279Q,2011IAUS..273..281B}.
This is even more important when the goal is to detect weak signals induced by Earth-like exoplanets around low-mass stars \citep{2007AsBio...7...85S,2009A&A...506..287L,2012Natur.491..207D,2016Natur.536..437A}.
For this purpose, star spots on the photosphere of stars can help us because they induce a photometric modulation from which we can infer not only the rotation period of the stars \citep{1947PASP...59..261K,1993A&A...272..176B,2002A&A...393..225M,2009A&ARv..17..251S}, but also long-term activity cycles \citep{1985ARA&A..23..379B,2005AN....326..283B}.

M dwarfs are strongly affected by star spots because of the presence of large active regions on their surfaces, which are due to the depth of the convective layers \citep{1998A&A...331..581D,2001ApJ...559..353M,2008ApJ...676.1262B,2008ApJ...684.1390R,2011MNRAS.412.1599B}
As a result, these late-type stars are the most likely to present this kind of modulation in photometric series \citep{2011ApJ...727...56I,2012AcA....62...67K,2015ApJ...812....3W,2016A&A...595A..12S}.
The low masses and small radii of M dwarfs also make them  ideal targets for surveys aimed at detecting small, low-mass, potentially habitable Earth-like planets \citep{1997Icar..129..450J, 2005AsBio...5..706S, 2007AsBio...7...30T, 2010ApJ...710..432R, 2013ApJ...763..149F}.
Therefore, the inclusion of M-dwarf targets in exoplanet surveys has increased steadily from the first dedicated radial-velocity searches \citep{2004ApJ...617..580B, 2005A&A...443L..15B, 2007ApJ...670..833J}, through transit searches from the ground and space \citep{2013ApJ...775...91B, 2013ApJ...765..131K, 2015ApJ...804...10C, 2015ApJ...807...45D, 2016Natur.533..221G, 2017Natur.542..456G, 2017Natur.544..333D}, to up-to-date searches with specially designed instruments and space missions such as CARMENES \citep{2014SPIE.9147E..1FQ}, HPF \citep{2014SPIE.9147E..1GM}, IRD \citep{2012SPIE.8446E..1TT}, SPIRou \citep{2014SPIE.9147E..15A}, {\em TESS} \citep{2015JATIS...1a4003R}, or GIARPS \citep{2016SPIE.9908E..1AC}.
As a result, there is a growing number of projects aimed at photometrically following up large samples of M dwarfs in the solar neighbourhood with the goals of determining their rotation periods and discriminating between signals induced by rotation from those induced by the presence of planets \citep{2011ApJ...727...56I, 2015MNRAS.452.2745S, 2016ApJ...821...93N}.
Some of the targeted M dwarfs have known exoplanets, or are suspected to harbour them, while others are just being monitored by radial-velocity surveys with high-resolution spectrographs.
Some exoplanets may transit their stars, although transiting exoplanets around bright M dwarfs are rare \citep{2007A&A...472L..13G,2009Natur.462..891C}.

This work is part of the CARMENES project\footnote{{\tt http://carmenes.caha.es}}.
It is also the fourth item in the series of papers devoted to the scientific preparation of the target sample being monitored during CARMENES guaranteed time observations \citep[see also][]{2015A&A...577A.128A, 2017A&A...597A..47C,2018A&A...614A..76J}.
Here we present the results of analysing long-term, wide-band photometry of 337 M dwarfs currently being monitored by CARMENES \citep{2018A&A...612A..49R}.
For many of them we had not been able to find rotation periods in the existing literature (see below).

To determine the rotation periods of our M dwarfs, we make extensive use of public time series of wide-area photometric surveys and databases such as the
All-Sky Automated Survey \citep[ASAS; ][]{1997AcA....47..467P}, 
Northern Sky Variability Survey \citep [NSVS;][]{2004AJ....127.2436W}, 
Wide Angle Search for Planets \citep[SuperWASP;][]{2006PASP..118.1407P}, 
Catalina Real-Time Transient Survey \citep[Catalina;][]{2009ApJ...696..870D}, 
and The MEarth Project \citep[MEarth;][]{2009Natur.462..891C, 2011ApJ...727...56I}.
Since the amplitude of the modulations induced by star spots, in the range of millimagnitudes, is also within reach of current amateur facilities \citep{2011A&A...526L..10H,2015MNRAS.450.3101B}, we also collaborate with amateur astronomers to obtain data for stars that have never been studied by systematic surveys or that need a greater number of observations. 

After collecting and cleaning the time series, we look for significant peaks in power spectra, determine probable rotation periods and long activity cycles, compare them with previous determinations and activity indicators when available, and make all our results available to the whole community in order to facilitate the disentanglement of planetary and activity signals in current and forthcoming radial-velocity surveys of M dwarfs.


\section{Data}
\label{section.data}

\subsection{Sample of observed M dwarfs}
\label{sec.sample}

During guaranteed time observations (GTOs), the double-channel CARMENES spectrograph has so far observed a sample of 336 bright, nearby M dwarfs with the goal of detecting low-mass planets in their habitable zone with the radial-velocity method \citep{2015csss...18..897Q, 2018A&A...612A..49R}:
 324 have been presented by  \cite{2018A&A...612A..49R}, {3} did not have enough CARMENES observations at the time  the spectral templates were being prepared for the study, and {9} are new spectroscopic binaries \citep{2018arXiv180806895B}.
Here we investigate the photometric variability of these 336 M dwarfs and of \object{G~34--23}\,AB (J01221+221AB), which \cite{2017A&A...597A..47C} found to be a close physical binary just before the GTOs started.
This results in a final sample size of 337 stars.

As part of the full characterisation of the GTO sample, for each target we have collected all relevant information:   astrometry, photometry, spectroscopy, multiplicity, stellar parameters, and activity, including X-ray count rates, fluxes, and hardness ratios, H$\alpha$ pseudo-equivalent widths, reported flaring activity, rotational velocities $v \sin{i}$, and rotational periods $P_{\rm rot}$ \citep{2016csss.confE.148C}.
In particular, for {69} stars of the CARMENES GTO sample we had already collected rotation periods from the existing literature 
\citep[e.g.][see below]{2007AcA....57..149K, 
2007A&A...467..785N,
2011AJ....141..166H,
2011ApJ...727...56I,
2012AcA....62...67K,
2015MNRAS.452.2745S,
2015ApJ...812....3W,
2016ApJ...821...93N}.

Of the 337 investigated stars, we were unable to collect or measure any photometric data useful for variability studies for only {3}.
In these {three} cases, the M dwarfs are physical companions at relatively small angular separations of bright primaries (J09144+526 = \object{HD~79211}, J11110+304 =  \object{HD~97101}\,B, and J14251+518 = \object{$\theta$~Boo}\,B).
For the other 334 M dwarfs, we looked for peaks in the periodograms of two large families of light curves that we obtained
($i$) from wide-area photometric surveys and public databases, and ($ii$) with 20--80\,cm telescopes at amateur and semi-professional observatories.
A more detailed photometric survey of particular GTO targets is being carried out within the CARMENES consortium with more powerful telescopes, such as the
Las Cumbres Observatory Global Telescope Network \citep{2013PASP..125.1031B} and
the Instituto de Astrof\'isica de Andaluc\'ia 1.5\,m and 0.9\,m telescopes at the Observatorio de Sierra Nevada.
Results of this photometric monitoring extension will be published elsewhere.

In the first four columns of Table~A.1 we give the Carmencita identifier \citep{2016csss.confE.148C}, discovery name, and 2MASS \citep{2006AJ....131.1163S} equatorial coordinates of the 337 M dwarfs investigated in this work.

\subsection{Photometric monitoring surveys}
\label{sec.surveys}

   \begin{table*}
      \caption[]{Number of investigated light curves and basic parameters of used public surveys and observatories.} %
         \label{table.surveys+observatories}
     $$ 
         \begin{tabular}{llll c}
            \hline
            \hline
            \noalign{\smallskip}
Survey          & Location                                      & Instrument configuration                           & Band                  & No. of \\
                        &                                                       &                                                                         &                               & light curves \\
            \noalign{\smallskip}
            \hline
            \noalign{\smallskip}
MEarth          & Mount Hopkins, USA                    & 8 $\times$ Apogee U42                                     & RG715, $I^a$  & {184} \\ 
ASAS            & Las Cumbres, Chile                    & Apogee AP10                                           & $V$                     & {174} \\ 
SuperWASP       & Roque de los Muchachos, Spain & 8 $\times$ Andor DW436                                & clear, broad$^{b}$      & {89} \\  
                        & Sutherland, South Africa              &                                                                       & broad$^{b}$             & \\ 
NSVS            & Los \'Alamos, USA                             & 4 $\times$ Apogee AP10                             & clear$^{c}$           & {86} \\ Catalina                & Mt. Lemmon/Mt. Bigelow, USA   & variable$^{d}$                                                & $V_{\rm CSS}^d$ & {37} \\ 
ASAS-SN         & worldwide$^{e}$                               & FLI ProLine PL230                                   & $V$                   & {14} \\
K2                      & (Earth-trailing heliocentric orbit)   & 0.95\,m {\em Kepler} + 42 $\times$ CCD  & clear                 & {13} \\
AstroLAB IRIS   & Zillebeke, Belgium                            & 0.68\,m Keller + SBIG STL-6303E                 & $B,~V$                        & {7} \\ 
Montcabrer      & Cabrils, Spain                                & 0.30\,m Meade LX200 + SBIG ST-8XME      & $R,~I$                        & {6} \\ 
HATNet          & Mt. Hopkins, USA                              & 5+2 $\times$ Apogee AP10$^{f}$                       & $R_C,~I_C$            & {5} \\
                        & Mauna Kea, USA                                &                                                                         &                               & \\
Montsec                 & San Esteban de la Sarga, Spain        & 0.80\,m Joan Or\'o + MEIA2$^{g}$                        & $R_C$                 & {4} \\ 
Carda           & Villaviciosa, Spain                           & 0.20\,m Celestron SC 8'' + SBIG ST-7E           & clear                 & {3} \\ 
            \noalign{\smallskip}
            \hline
            \noalign{\smallskip}
Total           &                                                       &                                                                       &                               & {622} \\
        \noalign{\smallskip}
            \hline
         \end{tabular}
     $$ 
\begin{list}{}{}
\item[$^{a}$] MEarth: broad 715\,nm long-pass filter in first (2008--2010) and third seasons (2011+), custom-made $I_{715-895}$ interference filter in second season (2010--2011).
\item[$^{b}$] SuperWASP: from 2006 onwards a broad-band filter was installed with a passband from 400 to 700\,nm.
\item[$^{c}$] NSVS: unfiltered optical response 450--1000\,nm, effective wavelength of $R$ band.
\item[$^{d}$] Catalina: see \citet{2015DPS....4730819C} for the latest camera configurations.
\item[$^{e}$] ASAS-SN: the network consists of 20 telescopes, distributed among five units in Hawai'i and Texas in the USA, two sites in Chile, and South Africa.
\item[$^{f}$] HATNet: see \citet{2018arXiv180100849B} for the latest camera configurations.
\item[$^{g}$] Montsec: the MEIA2 instrument at the Telescopi Joan Or\'o on the Observatori Astronomic del Montsec consists of a camera iKon-L with an Andor CCD42-40 chip and a Custom Scientific filter wheel. 
\end{list}
   \end{table*}

We searched for photometric data of our target stars, available through either VizieR \citep{2000A&AS..143...23O} or, more often, the respective public web pages of the {four} main wide-area surveys listed below and summarised in the top part of Table~\ref{table.surveys+observatories}.

\begin{itemize}

        \item{MEarth:} The MEarth Project\footnote{\tt http://www.cfa.harvard.edu/MEarth} \citep{2008PASP..120..317N,2011ApJ...727...56I}. 
This consists of two robotically-controlled 0.4\,m telescope arrays, MEarth-North at the the Fred Lawrence Whipple Observatory on Mount Hopkins, Arizona, and MEarth-South telescope at the Cerro Tololo Inter-American Observatory, Coquimbo.     
The project monitors the brightness of about 2000 nearby M dwarfs with the goal of finding transiting planets \citep{2013ApJ...775...91B,2017Natur.544..333D}, but it has also successfully measured rotation periods of M dwarfs \citep{2009IAUS..253...37I, 2011ApJ...727...56I,2016ApJ...821...93N}. 
On every clear night and for about six months, each star is observed with a cadence of 20\,min. 
In general we used data from three observing batches (2008--2010, 2010--2011, and 2011--2015) from the MEarth-North array ($\delta$ = +20 to +60) provided by the fifth MEarth data release DR5, but we also used a few DR6 light curves (2011--2016). 

        \item{ASAS:} All-Sky Automated Survey\footnote{\tt http://www.astrouw.edu.pl/asas} \citep{2002AcA....52..397P}. 
This is a Polish project devoted to constant photometric monitoring of the whole available sky (approximately 20 million stars brighter than $V \sim$ 14\,mag).
It consists of two observing stations in Las Cumbres Observatory, Chile (ASAS-3 from 1997 to 2010, ASAS-4 since 2010), and Haleakal\=a Observatory, Hawai'i (ASAS-3N since 2006).
Both are equipped with two wide-field instruments observing simultaneously in the $V$ and $I$ bands, and a set of smaller narrow-field telescopes and wide-field cameras.  
We used only the $V$-band ASAS-3 data for stars with $\delta <$ +28\,deg observed between 1997 and 2006, which are available through the ASAS All Star Catalogue.
In particular, we retrieved all light curves within a search radius of 15\,arcsec, discarded all data points with C or D quality flags, and computed an average magnitude per epoch of the five ASAS-3 apertures weighted by each aperture magnitude error.

        \item{SuperWASP:} Super-Wide Angle Search for Planets\footnote{\tt http://www.superwasp.org} \citep{2006PASP..118.1407P}. 
The UK-Spanish WASP consortium runs two identical robotic telescopes of eight lenses each, SuperWASP-North at the Observarorio del Roque de los Muchachos, La Palma, and SuperWASP-South at the South African Astronomical Observatory, Sutherland.
WASP has discovered over a hundred exoplanets through transit photometry \citep{2007MNRAS.375..951C,2011A&A...525A..54B}. 
For our work, we downloaded light curves of the first SuperWASP public data release \citep[DR1 --][]{2010A&A...520L..10B} from the Czech site\footnote{\tt http://wasp.cerit-sc.cz}.
The SuperWASP DR1 contains light curves for about 18 million sources in both hemispheres.
Although the two robotic telescopes are still operational, DR1 only provided data collected from 2004 to 2008.  
The average number of data points per light curve is approximately 6700. 
           
        \item{NSVS:} Northern Sky Variability Survey\footnote{\tt https://skydot.lanl.gov/nsvs/} \citep{2004AJ....127.2436W}. 
This was located at Los \'Alamos National Laboratory, New Mexico.
The NSVS catalogue contains data from approximately 14 million objects in the range $V$ = 8--15.5\,mag with a typical baseline of one year from April 1999 to March 2000, and 100--500 measurements for each source.
In a median field, bright unsaturated stars have photometric scatter of about 20\,mmag.
It covered the entire northern hemisphere, and part of the southern hemisphere down to $\delta$ = --28\,deg. 

\end{itemize}

In addition to these four main catalogues, which accounted for a total of {533} light curves  ({86\,\%}), we complemented  our dataset with light curves compiled from the Catalina Surveys CSDR2\footnote{\tt https://catalina.lpl.arizona.edu} \citep[Catalina;][]{2009ApJ...696..870D, 2014ApJS..213....9D}, K2 \citep{2014PASP..126..398H}, the All-Sky Automated Survey for Supernovae\footnote{\tt http://www.astronomy.ohio-state.edu/asassn/} \cite[ASAS-SN; ][]{2017PASP..129j4502K}, and the Hungarian-made Automated Telescope Network\footnote{\tt https://hatnet.org/} \citep[HATNet;][]{2004PASP..116..266B,2018arXiv180100849B}.
None of our targets was in the catalogues of  the {\em CoRoT} \citep{2009A&A...506..411A}, {\em Kepler} \citep{2010Sci...327..977B}, or the Transatlantic Exoplanet Survey \citep[TrES; ][]{2007ASPC..366...13A} surveys.
Finally, we did not use other wide surveys, such as 
Lincoln Near-Earth Asteroid Research \citep[LINEAR; ][]{2000Icar..148...21S};
XO \citep{2005PASP..117..783M}; 
Kilodegree Extremely Little Telescope \citep[KELT;][]{2007PASP..119..923P}, which will soon have an extensive data release (J.~Pepper, priv.~comm.); HATSouth \citep{2013PASP..125..154B}; or the Asteroid Terrestrial-impact Last Alert System \citep[ATLAS; ][]{2018arXiv180402132H}, with the first data release published after collecting all light curves included in this work.
All the public surveys used, amounting {602} light curves, are summarised in Table~\ref{table.surveys+observatories}.

\subsection{Our observations}

For {20} GTO M dwarfs without data in public surveys or published periods, or with unreliable or suspect periods in the existing literature (e.g. short $P_{\rm rot,lit}$ but low $v \sin{i}$ and faint H$\alpha$ emission, or vice versa), we made our own observations in collaboration with amateur and semi-professional astronomical observatories in Spain and Belgium:
Carda\footnote{\tt http://www.auladeastronomia.es/} (MPC~Z76) in Asturias, Montcabrer\footnote{\tt http://cometas.sytes.net} (MPC~213) and Montsec\footnote{\tt http://www.oadm.cat} in Barcelona (MPC~C65), and AstroLAB IRIS\footnote{\tt http://astrolab.be} near Ypres.
Most of our targets had northern declinations, which suitably fit the geographical latitudes of our observatories (between +41.5\,deg and +50.8\,deg).
                
The cadence of observations for each target was either continuous every night or just one observation per night during a long run, depending on rule-of-thumb estimations of their rotation periods based on literature values of rotational velocity and H$\alpha$ emission.
Exposure times varied widely, from a few seconds to several minutes.
We took special care to select enough reference stars with stable light curves and relatively red $J-K_s$ colours within the respective fields of view (f.o.v.; variable from 12.3$\times$12.3\,arcmin$^2$ to 43$\times$43\,arcmin$^2$). 

For the image reduction and light-curve generation, we used standard calibration procedures (bias and flat-field correction) and widely distributed  software such as {\tt MaxIm}\footnote{\tt http://diffractionlimited.com/product/maxim-dl/}, {\tt AstroImageJ}\footnote{\tt http://www.astro.louisville.edu/software/astroimagej/} \citep{2017AJ....153...77C}, {\tt FotoDif}\footnote{\tt http://www.astrosurf.com/orodeno/fotodif/}, and {\tt LesvePhotometry} \footnote{\tt http://www.dppobservatory.net/AstroPrograms/ Software4VSObservers.php} with parameters appropriate for each observatory, atmospheric condition, and star brightness (see again Table~\ref{table.surveys+observatories}).


\section{Analysis}

\begin{table*}
\caption{Cycle periods obtained for stars in our sample.}             
\label{table.ourPcycle}      
\centering          
\begin{tabular}{ll ccc l ccc c}
            \hline
            \hline
            \noalign{\smallskip}
Karmn           & Name          & $P_{\rm cycle}$       & FAP   & $A_\lambda$   & Survey  & $P_{\rm cycle,lit}$   & FAP$_{\rm lit}$       & $A_{\lambda,{\rm lit}}$  & Ref.$^{a}$    \\
                        &                       & [a]                           &       [\%]    & [mag]           &               & [a]                           &               [\%]            & [mag]                           &                       \\
            \noalign{\smallskip}
            \hline
            \noalign{\smallskip}
J06105-218      & HD 42581 A    & 8.3$\pm$3.1           &$<$ 10$^{-4}$&0.026& ASAS &8.4$\pm$0.3&$<$ 0.1&0.0128&SM16 \\
J07361-031      & BD--02 2198   & 11.5$\pm$1.9  & $<$ 10$^{-4}$&0.035& ASAS &...&...&...&...\\
J08161+013      & GJ 2066       & 4.1$\pm$0.7           &$<$ 10$^{-4}$&0.012& ASAS &...&...&...&...\\
J10122-037      &AN Sex         & 3.2$\pm$0.4&$<$ 10$^{-4}$&0.020& ASAS &...&...&...&...\\
J11477+008      & FI Vir                & 4.5$\pm$2.0&0.15&0.012& ASAS &4.1$\pm$0.3&1.7&0.0071&SM16 \\
J15218+209      & OT Ser                &6.5$\pm$0.8&$<$ 10$^{-4}$&0.055& ASAS &...&...&...&...\\
J16303-126      & V2306 Oph     &3.9$\pm$1.0&$<$ 10$^{-4}$&0.014& ASAS &4.4$\pm$0.2&$<$ 0.1&0.0083&SM16 \\
J19169+051N     & V1428 Aql     &3.3$\pm$0.4&$<$ 10$^{-4}$&0.013& ASAS &9.3$\pm$1.9&$<$ 0.1&0.0077&SM16 \\
J19346+045      & BD+04 4157    &3.0$\pm$0.8&$<$ 10$^{-4}$&0.013& ASAS &...&...&...&...\\
J22532-142      & IL Aqr                &4.5$\pm$0.7&$<$ 10$^{-4}$&0.013& ASAS &...&...&...&...\\
\noalign{\smallskip}
\hline
\end{tabular}
\begin{list}{}{}
\item[$^{a}$] {\em Reference --}
SM16: \cite{2016A&A...595A..12S}.
\end{list}
\end{table*}

\begin{figure}
  \centering
    \includegraphics[width=0.49\textwidth]{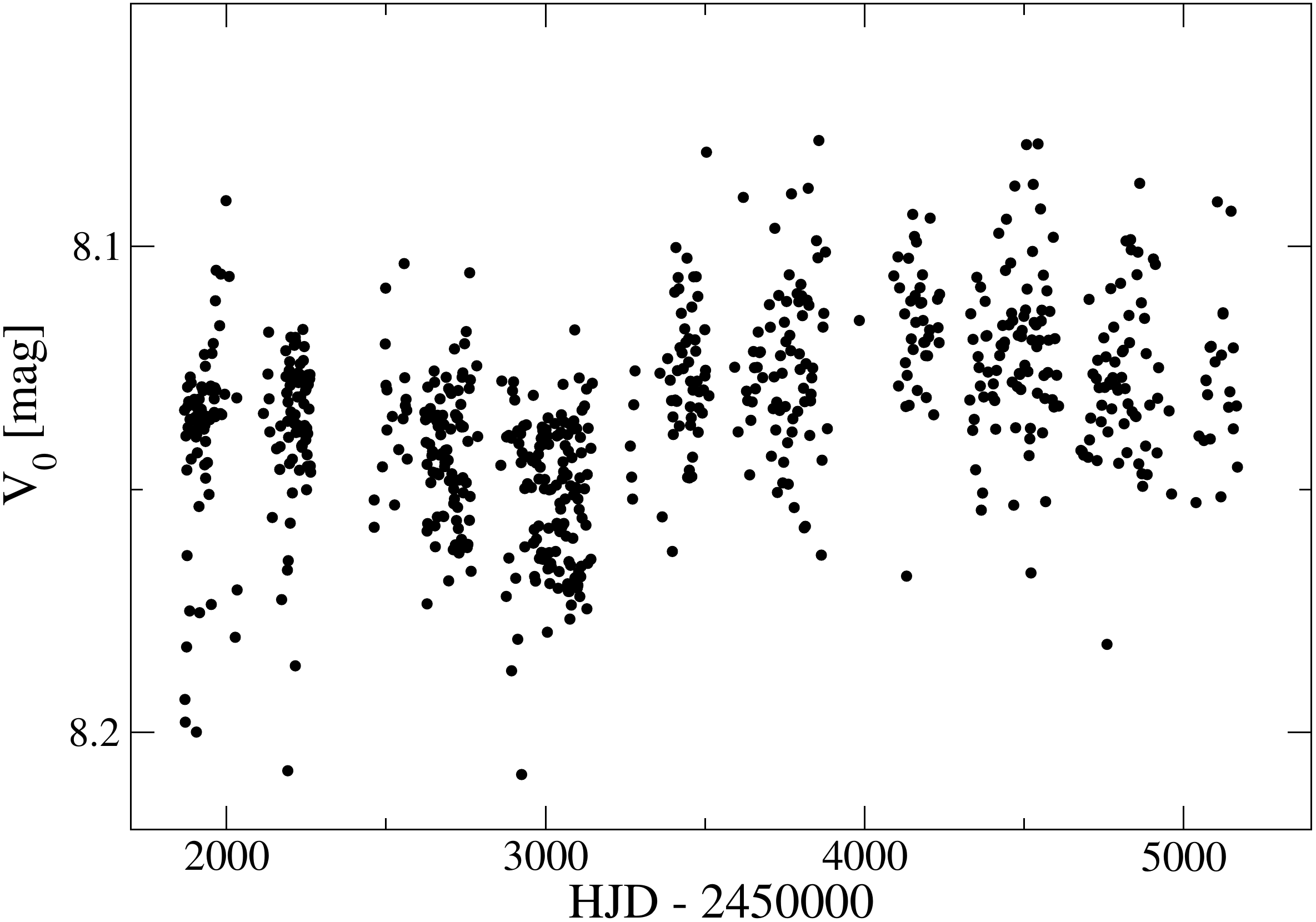}
  \caption{ASAS $V$-band photometric data for the M0.5\,V star J06105-218 = \object{HD~42581A}. 
  It may display an activity cycle modulation with $P_{\rm cycle} \sim$ 8.3\,a superimposed to a rotation period $P_{\rm rot}$ = 27.3\,d (the modulation in the light curve of J15218+209 is similar).}
  \label{fig.pcycleexample}
\end{figure}

We searched for significant signals in the periodograms of the {622} light curves compiled or obtained by us as described in Section~\ref{section.data}.
Prior to this, we cleaned our light curves by discarding outlying data points caused by the combination of target faintness and sub-optimal weather.
Some of our M dwarfs could also undergo flaring activity (most flares in our light curves are difficult to identify because of the typical low cadence).
We used the same cleaning procedure as \citet{2015MNRAS.452.2745S,2016A&A...595A..12S}, rejecting iteratively all data points that deviated more than 2.5$\sigma$ from the mean magnitude (which might eventually bias the amplitude of the variations).

The last six columns of Table~\ref{table.allstars} show the corresponding survey, number of data points $N_{\rm obs}$ before and after the cleaning, time interval length $\Delta t$ between first and last visit, mean $\overline{m}$ and standard deviation $\sigma_m$ of the individual magnitudes, and mean error $\overline{\delta m}$.
See Section~\ref{section.aperiodicvariables} for a discussion on the variation of $\sigma_m$ as a function of $\overline{m}$, and the possibility of finding aperiodic or irregular variable stars in our data.

Next, we used the Lomb--Scargle (LS) periodogram \citep{1982ApJ...263..835S} with the {\tt peranso} analysis software \citep{2016AN....337..239P}, which implements multiple light-curve and period analysis functions. 
We evaluated the significance of the signals found in the periodograms with the \citet{2004MNRAS.354.1165C} modification of the \citet{1986ApJ...302..757H} formula. 
In this way, the   false alarm probability (FAP) becomes
\begin{equation}
FAP = 1 - [1 - p(z>z_0)]^M,
\end{equation}
with $p(z>z_0)$ being the probability that $z$, the target spectral density, is greater than $z_0$, the measured spectral density; and $M$ the number of independent frequencies.
In our case, $p(z>z_0) = e^{-z_0}$, where $z_0$ is the peak of the hypothetical signal;  $M = \Delta t ~ \Delta f$, where $\Delta t$ is the time baseline, $\Delta f = f_2 - f_1$; and $f_2$ and $f_1$ are the maximum and minimum search frequencies, respectively. The
{\tt peranso} software also measures amplitudes of light curves.

We searched for significant frequencies with FAP $<$ 2\,\% and within the standard frequency interval.
On the one hand, we set the highest frequency at the Nyquist frequency to about {half} of the minimum time interval between consecutive visits (from $f \sim$ 100\,d$^{-1}$ 
for some of our intensive campaigns to $f \sim$ 1\,d$^{-1}$ for most of the public-survey light curves). 
On the other hand, we set the lowest frequency at half the length of the monitoring (typically $f \sim$ 0.005\,d$^{-1}$).
Only in the case of ASAS, the public survey with the longest time baseline (up to ten years), did we search for low frequencies down to 0.0005\,d$^{-1}$.  
For stars with significant signals shorter than 2\,d we used a significant frequency oversampling ($\sim 10 \times$).
When a periodogram displayed several significant peaks, we paid special attention to identifying aliases and picked up the strongest signal with astrophysical meaning (e.g. with $P_{\rm rot}$ consistent with existing additional information on the star, especially its rotational velocity, $v \sin{i}$; see Section~\ref{section:activity}). Figures~\ref{fig.mearth} to \ref{fig.montsec} illustrate our analysis with one representative example of a raw stellar light curve and corresponding LS periodogram and phase-folded light curve for each dataset with identified period.

There is a justified concern in the literature about the use of the LS periodogram for period determination \citep[e.g.][]{2013MNRAS.432.1203M}.
As a result, for the light curves of stars classified as new spectroscopic binaries by \cite{2018arXiv180806895B}, with K2 data, or with periods shorter than 1\,d, or different by more than 10\,\% from those in the literature (see Section~\ref{sec.Prot}) we also applied the generalised LS periodogram method \citep[GLS;][]{2009A&A...496..577Z}.

For the spectroscopic binaries and K2 stars, we additionally applied the Gaussian process regression \citep[GP;][]{RW05} with the {\tt celerite} package \citep{2017AJ....154..220F}. 
To build a celerity model for our case, we defined a covariance/kernel for the GP model, which consisted of a stochastically-driven damped oscillator and a jitter term \citep{2018MNRAS.474.2094A}. 
We then wrapped the  kernel defined in this way in the GP and computed the likelihood.
In all investigated cases {except two}, the GLS and GP periods were identical within their uncertainties to the LS periods computed with {\tt peranso}.
{The only two significant differences were J18356+329, for which GLS did not recover $P_{\rm rot}$ = 0.118\,d, the shortest period in our sample and identical to that reported by \citet{2008ApJ...684..644H}, and J16254+543, for which GLS found $P_{\rm rot}$ = 76.8\,d, a value similar to that in the literature \citep{2015MNRAS.452.2745S}, but for which the LS algorithm found 100$\pm$5\,d.}

In Table~\ref{table.ourProt} we show the periods, amplitudes, FAPs, and corresponding surveys of {142} M dwarfs.
When available, we show the GLS periods.
We tabulate the uncertainty in $P_{\rm rot}$ from the full width at half maximum of the corresponding peak in the periodogram \citep{1991MNRAS.253..198S}.
For GLS periods, we tabulate formal uncertainties, which are significantly smaller than the real ones.

When several datasets are available, and even though the periodogram peaks are detected in both, we list the $P_{\rm rot}$ of the dataset with the lowest FAP.  
For the sake of completeness, Table~\ref{table.ourProt} includes {five} stars with FAP = 2--10\,\% for which we recover periods similar to those previously published \citep{2004ApJ...617..508T,2015MNRAS.452.2745S,2016A&A...595A..12S}, but which did not pass our initial FAP criterion. 
One  such period is for J11477+008 = \object{FI~Vir}, for which \citet{2016A&A...595A..12S} reported a rotation period of 165\,d consistent with ours.
Its K2 light curve, which spans only 80\,d, shows a clear modulation that matches such a long period, the longest one reported by us.
Something similar occurs with J13458-179 = \object{LP~798--034}, whose K2 light curve shows a 20\,mmag peak-to-peak modulation of about three months.
We did not find any periods in the existing literature or our ASAS data for this star, which is not listed in Table~\ref{table.ourProt}.

We also looked for long-period cycles in ASAS light curves of stars with identified $P_{\rm rot}$ and time baseline $\Delta t \approx$ 9\,a. 
Ten stars have significant signals at $P_{\rm cycle} \ge$ 3.0\,a, and are listed in Table~\ref{table.ourPcycle}. 
Two of these (J06105-218 and J15218+209) have cycle periods longer than half $\Delta t$, and one has a cycle period even longer than the full $\Delta t$, but its modulation is very clear (see Fig.~\ref{fig.pcycleexample}).
Curiously, the star with the longest rotation period and highest FAP, J11477+008, also displays  a long-term activity cycle.
This flaring M4 dwarf is also the star with the smallest ratio $P_{\rm cycle} / P_{\rm rot} \sim$~10.

Of the ten values of $P_{\rm cycle}$ shown in Table~\ref{table.ourPcycle}, {six} are new, {three} are identical within their uncertainties to previous determinations by \cite{2016A&A...595A..12S}, and {one} is revised from 9.3\,a to about 3.3\,a. 
There are  two additional stars, not listed in Table~\ref{table.ourPcycle}, with significant signals (FAP $<$ 2\,\%)  at 2--3\,a, but without identified or reported rotation periods (J06371+175 and J22559+178).
It may be that they truly display a long-term modulation overlaid on a low-amplitude rotation period not detected yet, as they could be pole-on.


\section{Results and discussion}

\subsection{Rotation periods}
\label{sec.Prot}

\begin{table}
\caption{Stars with published rotation periods not recovered in this work.}             
\label{table.protnotrecovered}      
\centering          
\begin{tabular}{llcl}
            \hline
            \hline
            \noalign{\smallskip}
    Karmn               &  Name                 & $P_{\rm rot,lit}$     & Ref.$^{a}$      \\ 
                        &                               & [d]                            &                       \\ 
            \noalign{\smallskip}
            \hline
            \noalign{\smallskip}
   J03133+047   & CD Cet                        & 126.2                         & New16           \\ 
   J05019-069   & LP 656--038           & 88.5                          & Kir12           \\ 
   J07274+052   & Luyten's star                 & 115.6$\pm$19.4        & SM15            \\ 
   J09003+218   & LP 368--128           & 0.439                         & New16           \\ 
   J09360-216   & GJ 357                        & 74.3$\pm$1.7          & SM15            \\ 
   J11033+359   & Lalande 21185         & 48.0                          & KS07            \\ 
   J13299+102   & BD+11 2576            & 28 $\pm$ 2.9          & SM15          \\ 
   J11421+267   & Ross 905              & 39.9                          & SM15            \\ 
   J13457+148   & HD 119850             & 52.3$\pm$1.7          & SM15          \\ 
   J14010-026   & HD 122303             & 43.4                          & SM16            \\ 
   J15194-077   & HO Lib                        & 132$\pm$6.3           & SM15            \\ 
    J17578+046  & Barnard's star        & 130                           & KS07            \\ 
   J18051-030   & HD 165222             & 127.8$\pm$3.2         & SM15          \\ 
   J22096-046   & BD--05 5715           & 39.2$\pm$6.3          & SM15          \\ 
\noalign{\smallskip}
\hline
\end{tabular}
\begin{list}{}{}
\item[$^{a}$] {\em References:}
KS07: \cite{2007AcA....57..149K};
Har11: \cite{2011AJ....141..166H};
Kir12: \cite{2012AcA....62...67K};
SM15: \cite{2015MNRAS.452.2745S};
New16: \cite{2016ApJ...821...93N};
SM16: \cite{2016A&A...595A..12S}.
\end{list}
\end{table}

Of the {142} stars with periods in Table~\ref{table.ourProt}, {69 (49\,\%)} had rotation periods previously reported in the existing literature and tabulated in Carmencita \citep{2016csss.confE.148C}; see Table~\ref{table.ourProt} for full references. 
Therefore, we present {73} new photometric periods of M dwarfs in this study.
This number is comparable to the number of new rotation periods of M dwarfs reported by \cite{2007A&A...467..785N}, \cite {2011ApJ...727...56I}, and \cite{2016A&A...595A..12S}, but is far lower than other more comprehensive searches by \cite{2011AJ....141..166H}, \cite{2012AcA....62...67K}, and \cite{2017ApJ...834...85N}.
However, there is a fundamental difference in the samples; ours exclusively contains bright, nearby M dwarfs that are targets of dedicated exoplanet surveys.
In particular, the mean $J$-band magnitude and heliocentric distance of our CARMENES GTO sample are only 7.7\,mag and 11.6\,pc \citep[cf.][]{2016csss.confE.148C}, much brighter and closer than any M dwarf sample photometrically investigated previously.

We were not able to recover the rotation periods of {14} stars reported by \cite{2007AcA....57..149K}, \cite{2012AcA....62...67K}, \cite{2015MNRAS.452.2745S}, and \cite{2016ApJ...821...93N}, and listed in Table~\ref{table.protnotrecovered}. 
In most of these cases, unrecovered published $P_{\rm rot,lit}$ values were long, of low significance, and came from low-cadence, relatively noisy, ASAS data.

\begin{figure}
  \centering
    \includegraphics[width=0.49\textwidth]{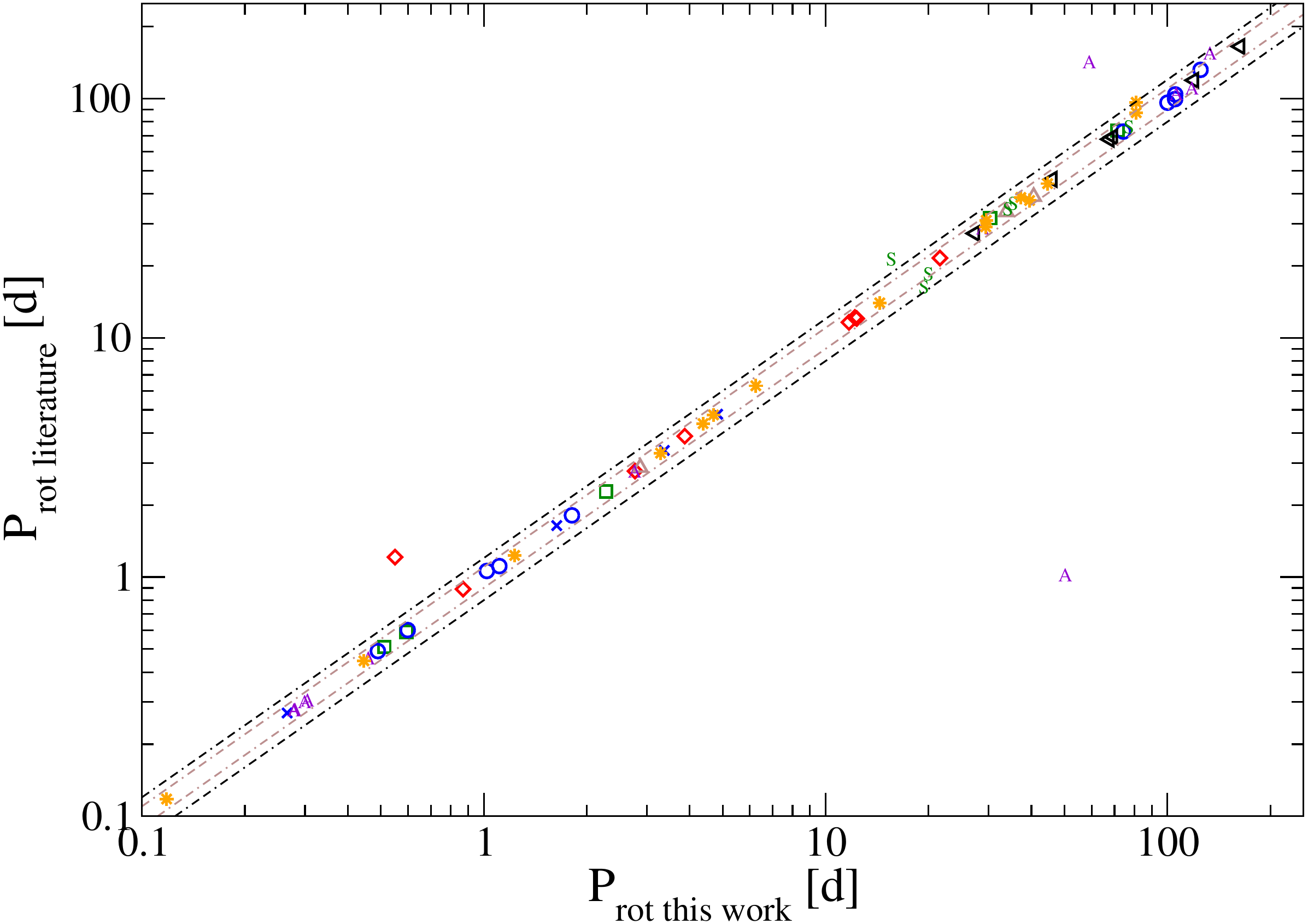}     
\caption{ Periods from the literature $P_{\rm rot,lit}$ as a function of our period, $P_{\rm rot, this work}$, for stars with previously published rotation periods. 
Dashed lines show 10\,\% and 20\,\% deviations from the 1:1 relation.
Different symbols and colours stand for the literature source: 
blue crosses: \cite{2007A&A...467..785N};
green squares: \cite{2011AJ....141..166H}; %
blue circles: \cite{2011ApJ...727...56I}; %
red diamonds: \cite{2012AcA....62...67K}; %
brown up-triangles: \cite{2007AcA....57..149K}; %
black left-triangles: \cite{2016A&A...595A..12S}; %
violet A's: \cite{2016ApJ...821...93N}; %
green S's: \cite{2018A&A...612A..89M}; %
orange asterisks: \cite{1998IBVS.4652....1G}, \cite{2000AJ....120.3265F}, \cite{2004ApJ...617..508T}, \cite{2005ApJ...634..625R}, \cite{2008ApJ...684..644H}, \cite{2013AcA....63...53K}, \cite{2015MNRAS.452.2745S}, \cite{2015ApJ...812....3W}, \cite{2016Natur.534..658D}, \cite{2017A&A...608A..35C}, \cite{2017ApJ...841..124V}, \cite{2018arXiv180100412L}.
}
\label{fig.Protlit}
\end{figure}

In Fig.~\ref{fig.Protlit} we compare the rotation periods that we found and those reported in the existing literature for the {69} stars in common. 
Except for {four} discordant stars, the agreement is excellent, with maximum deviations rarely exceeding 10\,\%, and with virtually all values identical within their uncertainties.
There are {four} outliers with deviations between $P_{\rm rot,lit}$ and $P_{\rm rot, this work}$ larger than 20\,\%:

\begin{itemize}

\item J00051+457 = \object{GJ~2}.
A rotation period of 21.2\,d was previously measured by \cite{2015MNRAS.452.2745S} from a series of high-resolution spectroscopy of activity indicators, while we measure 15.37\,d on an ASAS light curve;

\item J16570--043 = \object{LP 686--027}. 
This is an active, `RV-loud star' (pEW(H$\alpha$) = --4.2$\pm$0.1\,{\AA} and $v \sin{i} \approx$ 10.1\,km\,s$^{-1}$ \citep{2018A&A...614A..76J,2018A&A...612A..49R,2018A&A...614A.122T}.
Our analysis of the ASAS data reproduced the 1.21\,d period of \cite{2012AcA....62...67K}, but with an FAP $\gg$ 5\,\%.
However, our analysis of NSVS data, with $\sim$4 more epochs and a baseline $\sim$9 times longer than ASAS, resulted in a shorter $P_{\rm rot} \sim$ 0.55\,d with a lower FAP = 1.4\,\%; 

\item J18363+136 = \object{Ross~149}.
We found two significant signals in our ASAS data of roughly the same power at 1.02 \,d and 50.2\,d, which are aliases of one another.
The first period is identical to $P_{\rm rot,lit}$ = 1.017\,d discovered by \citet{2016ApJ...821...93N} in MEarth data. 
However, we tabulate the 50.2\,d period based on the very slow rotation velocity of the star ($v \sin{i} <$ 2\,km\,s$^{-1}$; \citealt{2018A&A...612A..49R});

\item J20260+585 = \object{Wolf~1069}.
\citet{2016ApJ...821...93N} reported a possible or uncertain period of 142.09\,d.
With MEarth data, we found instead a signal at 59.0\,d.
However, neither of the two signals is visible in the SuperWASP dataset.

\end{itemize}

\subsection{Period distribution}

\begin{figure}
  \centering
    \includegraphics[width=0.49\textwidth]{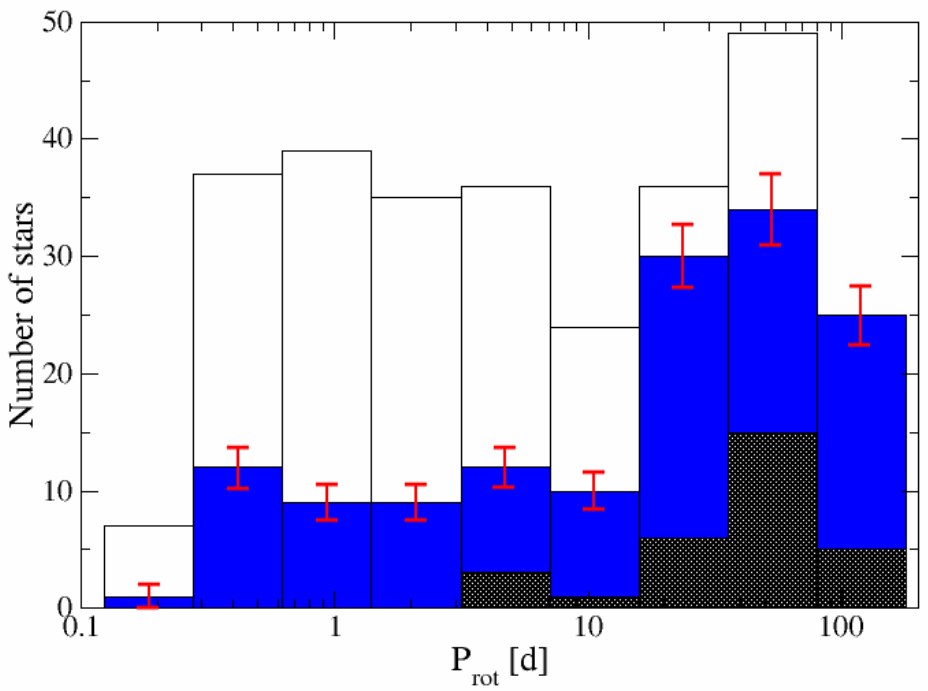}
  \caption{Distribution of rotation periods in our work (shaded, with Poissonian uncertainties) and in Carmencita (white). Rotation periods found with ASAS are shaded darker.
  The size of the bins follows the definitions given by \cite{FD81}.
  Compare with Fig.~3 in \citet{2016ApJ...821...93N}.}
  \label{fig.nodoublepeak}
\end{figure}

Using MEarth data, \citet{2016ApJ...821...93N} reported that the distribution of M-dwarf rotation periods displays a lack of stars with intermediate rotation periods around 30\,d.
This lack at intermediate periods could be understood as a bimodal $P_{\rm rot}$ distribution, with peaks at 0.5--2\,d and 50--100\,d, and a valley in between.
While there is a significant overabundance of periods larger than 30\,d in our sample, as illustrated in Fig.~\ref{fig.nodoublepeak}, the distribution of periods shorter than 30\,d (and longer than 0.5\,d) is flat within Poissonian statistics. 
Therefore, we see only one wide peak in the period distribution at $P_{\rm rot}$ = 20--120\,d or, conversely, a lack of stars with intermediate and short rotation periods $P_{\rm rot}$ = 0.6--20\,d.
However, rather than contradicting the results presented by \cite{2016ApJ...821...93N}, we attribute the lack of short periods to the different temporal sampling of the  surveys used (see below).
The authors also stated that the rotators with the highest quality grade were biased towards such short periods, while the sparse temporal sampling of most surveys used here (excluding SuperWASP and MEarth) prevented us from detecting periods of a few days.  
If the available periods for the $\sim$2200 Carmencita stars are taken into account (cf. Fig.~\ref{fig.nodoublepeak}), the peak at $P_{\rm rot}$ = 20--120\,d becomes much less apparent, and the period distribution is rather flat from 0.6\,d to 8\,d, approximately.
There might be a dip at $P_{\rm rot}$ = 8--20\,d.

Regarding survey sampling, we analysed the distribution of time elapsed between consecutive exposures of our light curves.
The SuperWASP distribution has two peaks, one very narrow at slightly less than one minute (the CCD read-out time) and  another  wider peak, centred at 10--20\,min, a bit more optimistic than the 40\,min stated by \cite{2006PASP..118.1407P}.
The NSVS distribution also has  two peaks, one at around 1 min and the other at 1\,d, much wider and sparser than SuperWASP and with a baseline of only one year \citep{2004AJ....127.2436W}.
Together with SuperWASP, the MEarth dataset is, with a field cadence of between 15 and 40\,min ($\sim$20\,min according to \citealt{2011ApJ...727...56I}) and `sequentially for the entire time [that] they are above airmass 2', the most appropriate one for sampling the $P_{\rm rot}$ = 0.6--20\,d interval. 
The ASAS sampling is much poorer than the other three main surveys, with a peak of the distribution centred at around one week.
The nominal sampling value of 2\,d \citep{2002AcA....52..397P} is attained only for some bright stars and during a fraction of the observing time.
However, the sampling strategy and the very long duration of the ASAS survey, of several years, makes it excellent for searching for periods longer than 20\,d.

This  partly explains the apparent lack of rotation periods in the range between 0.6 and 20\,d of GTO stars with respect to Carmencita stars in Fig.~\ref{fig.nodoublepeak}. Our GTO stars are brighter than the rest of the Carmencita stars \citep{2015A&A...577A.128A} so the observed lack of shorter periods in our sample is probably caused by the inclusion of bright ASAS targets, combined with the observational cadence of ASAS, which is optimised towards detecting long periods.  We thus have a combination of a Malmquist bias and a sampling effect.

A larger (volume- or magnitude-limited) sample of M dwarfs with a homogeneous, denser, longer monitoring could be needed to settle the question if there is a real lack of rotation periods between 0.6\,d or 8\,d and 20\,d at low stellar masses.
We may have to wait for the 8.4\,m Large Synoptic Survey Telescope, which will monitor the entire available sky every three nights on average.

\subsection{Activity}
\label{section:activity}

\subsubsection{RV-loud stars }

A comprehensive study linking rotational velocity, H$\alpha$ emission, and rotation periods of M dwarfs was already presented by \cite{2018A&A...614A..76J} in a previous publication of this series.
Using different tracers of magnetic activity, it can be seen that the rotational velocity increases with M-dwarf activity until saturation in rapid rotators \citep[][and references therein]{1984ApJ...280..202S,1996AJ....112.2799H,1998A&A...331..581D,2007AcA....57..149K,2003ApJ...583..451M,2009ApJ...692..538R,2010AJ....139..504B,2017ApJ...834...85N}.

\begin{figure}
  \centering
    \includegraphics[width=0.49\textwidth]{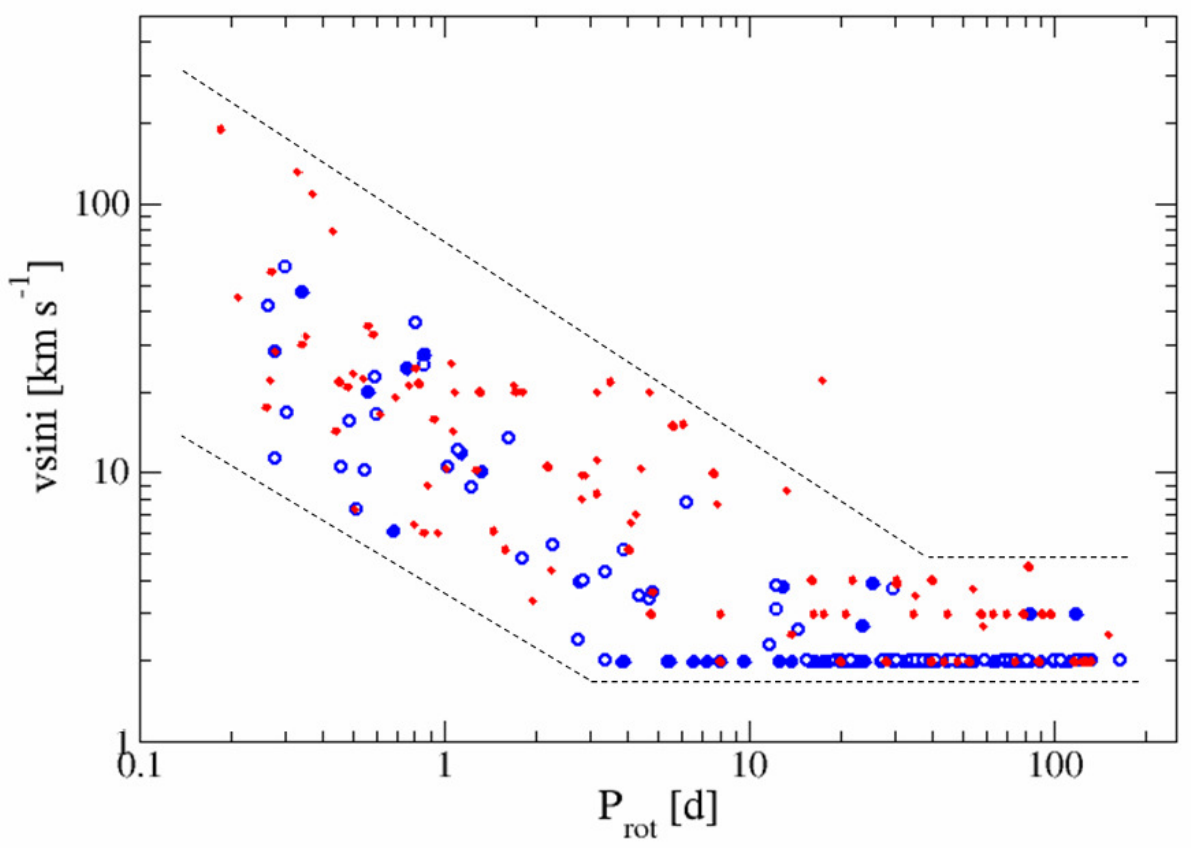}  
    \includegraphics[width=0.49\textwidth]{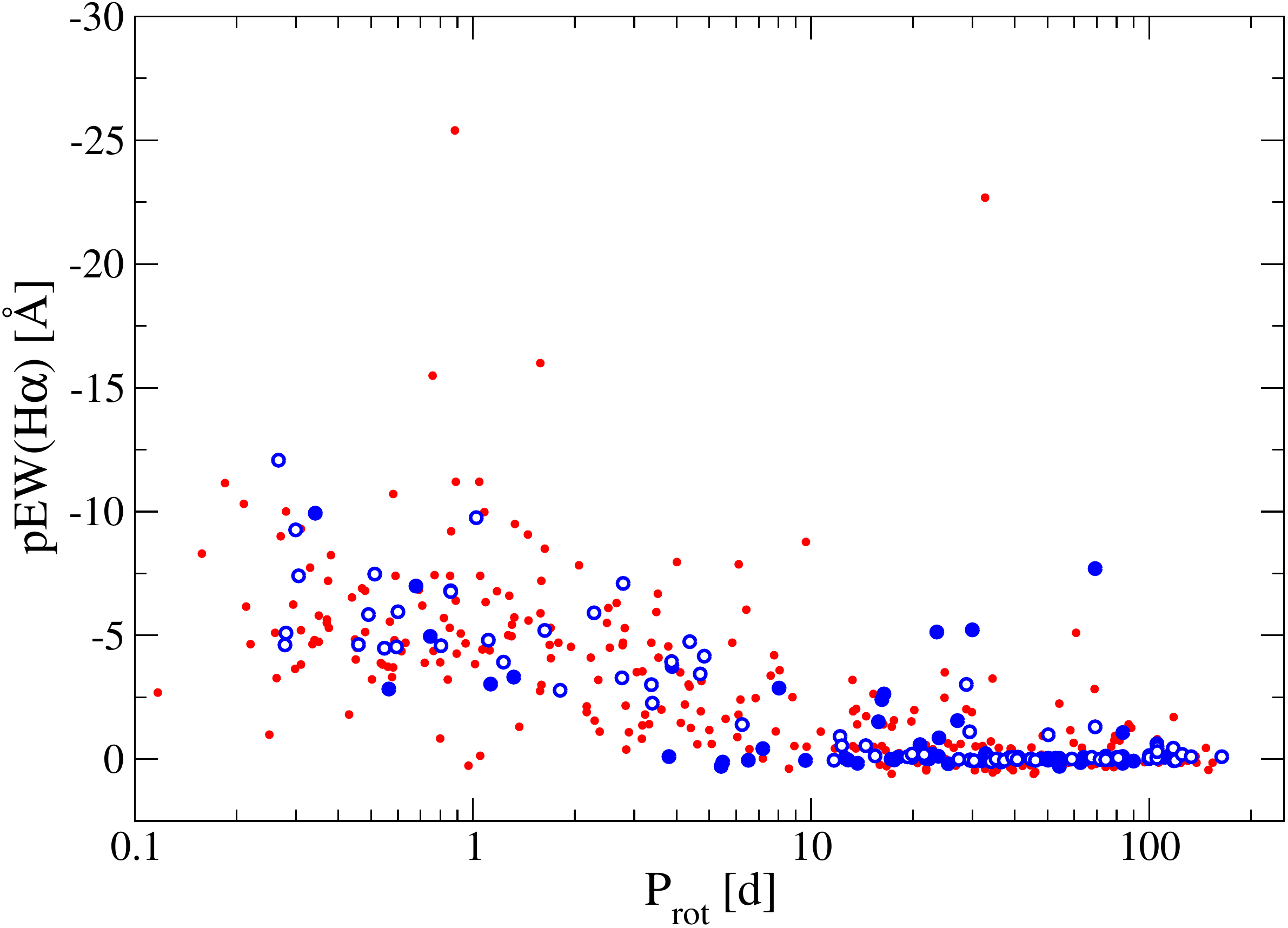}  
  \caption{$v \sin{i}$ vs. $P_{\rm rot}$ ({\em top panel}) and pEW(H$\alpha$) vs. $P_{\rm rot}$ ({\em bottom panel}) for stars with new (blue filled circles) and re-computed rotation periods in this study (blue open circles) and stars in Carmencita with previously published rotation periods (red dots).
In the top panel, the dotted lines indicate the lower and upper envelopes of the $v \sin{i}$-$P_{\rm rot}$ relation, and most of the values at $v \sin{i}$ = 2 or 3\,km\,s$^{-1}$ are upper limits.
Compare with Figs.~6 and~8 in \cite{2018A&A...614A..76J}.
 Here we present {154} new or improved $v \sin{i}$-$P_{\rm rot}$ and pEW(H$\alpha$)-$P_{\rm rot}$ pairs, with periods from this work and velocities and H$\alpha$ equivalent widths from \cite{2018A&A...612A..49R}.
Outlier stars in both panels are discussed in the text.}
  \label{fig.Protactivity}
\end{figure}

As a complement to these studies, we present the following brief discussion on the rotation periods of the most active stars in our sample, most of which were tabulated by \cite{2018A&A...614A.122T}.
They presented a list of 31 RV-loud stars: CARMENES GTO M-dwarf targets that displayed large-amplitude radial-velocity variations due to activity.
Most of these stars, close to activity saturation, were also part of the investigation on wing asymmetries of H$\alpha$, Na~{\sc i}~D, and He~{\sc i} lines in CARMENES spectra by \cite{2018A&A...615A..14F}.
We found rotation periods from photometric time series for all of them except for nine stars\footnote{
J01352--072 = \object{Barta~161~12},
J09449--123 = \object{G~161--071},
J12156+526 = \object{StKM 2--809},
J14173+454 = \object{RX~J1417.3+4525},
J15499+796 = \object{LP~022--420},
J16555--083 = \object{vB~8},
J18189+661 = \object{LP~071--165},
J19255+096 = \object{LSPM~J1925+0938},
and J20093--012 = \object{2MASS~J20091824--0113377}.},
which are either extremely active (e.g. Barta~161~12, LP~022--420), too faint and late for the surveys used (e.g. vB~8, 2MASS~J20091824--0113377), or both (e.g. LSPM~J1925+0938).
{The most active of these nine stars without an identified period could actually be variable, but irregular (see below).}

As expected from their typically large $v \sin{i}$ values, most of the other 22 RV-loud stars  have short or very short rotation periods: 20 stars have $P_{\rm rot} <$ 10\,d, and {11} have $P_{\rm rot} <$ 1\,d. 
The list of  RV-loud stars with identified periods contains some very well-known variable stars, such as
\object{V388~Cas}, \object{V2689~Ori}, \object{YZ~CMi}, \object{DX~Cnc}, \object{GL~Vir}, \object{OT~Ser}, \object{V1216~Sgr}, \object{V374~Peg}, \object{EV~Lac}, and \object{GT~Peg}. 
However, there are {three} such RV-loud stars for which there was no published value of rotation period.
Of these, {two} have identified periods shorter than one day: 
J02088+494 = \object{G~173--039} ($P_{\rm rot}$ = 0.748\,d) and
J04472+206 = \object{RX~J0447.2+2038} ($P_{\rm rot}$ = 0.342\,d).
They both have high rotational velocities of $v \sin{i} \approx$ 24--48\,km\,s$^{-1}$ \citep{2018A&A...609L...5R,2018A&A...614A.122T}.
The {third} star without a previously reported rotation period is J19169+051S = \object{V1298~Aql} (vB~10, $P_{\rm rot}$ = 23.6\,d), which has a slow rotational velocity of only 2.7\,km\,s$^{-1}$, but it does have an M8.0\,V spectral type \citep{2015A&A...577A.128A,2018arXiv180801183K} and has long been known to display variability of up to 0.2\,mag peak-to-peak in $V$ band \citep{1956PASP...68..531H,1978PASP...90..718L}.

The 31 RV-loud stars in  \cite{2018A&A...614A.122T} are not the only CARMENES GTO M-dwarf targets that display large-amplitude radial-velocity variations due to activity, as all stars with ten or fewer observations during the first 20 months of operation were discarded from the study.
In total, {46} stars in our sample have rotation periods shorter than 10\,d, and {19} shorter than 1\,d.
Of the latter, only {seven} were not tabulated as RV-loud stars by \cite{2018A&A...614A.122T}: four M4.0--5.5 dwarfs with previously reported rotation period and $v \sin{i} >$ 16\,km\,s$^{-1}$, 
the strong X-ray emitter \object{RBS~365} (J02519+224, M4.0\,V, $v \sin{i}$ = 27.4$\pm$0.6\,km\,s$^{-1}$), the $\beta$~Pictoris moving group candidate \object{1RXS~J050156.7+010845} (J05019+011, M4.0\,V, $v \sin{i}$ = 6.1$\pm$0.9\,km\,s$^{-1}$;  see \citealt{2015A&A...583A..85A} and references therein), 
and the unconfirmed close astrometric binary G~34--23\,AB (J01221+221\,AB; not in the GTO sample, see Section~\ref{section.data}). 
Because of the magnitude difference between the two components in the pair ($\Delta I$ = 0.86$\pm$0.12\,mag according to \citealt{2017A&A...597A..47C}) the most likely variable is the primary, G~34--23\,A. 
New analyses of RV scatter and activity of CARMENES GTO stars, including the other {six} single stars just described and the {three} RV-loud stars with new periods described above, most of them with $P_{\rm rot} <$ 1\,d, are ongoing and will be presented elsewhere.

\subsubsection{Linking $P_{\rm rot}$, $v \sin{i}$, H$\alpha$, and $A$}

We go deeper into the $P_{\rm rot}$-$v \sin{i}$ discussion with the top panel in Fig.~\ref{fig.Protactivity}.
To  build this plot we took {305} rotation periods and rotation velocities from the existing literature and compiled in the Carmencita catalogue (see references in Section~\ref{sec.sample}).
As expected and widely discussed by  \cite{2015ApJ...812....3W} and \cite{2018A&A...614A..76J}, among others,  the shorter the rotation period $P_{\rm rot}$ of an M dwarf, the higher the rotation velocity $v \sin{i}$.
The vertical width is due to the indetermination in inclination angle $i$ and the spread in spectral types, and thus stellar radii $R$ according to the formula

\begin{equation}
v \sin{i} = \frac{2 \pi R  \sin{i}}{P_{\rm rot}}.
\end{equation}

\begin{figure}
        \centering
        \includegraphics[width=0.49\textwidth]{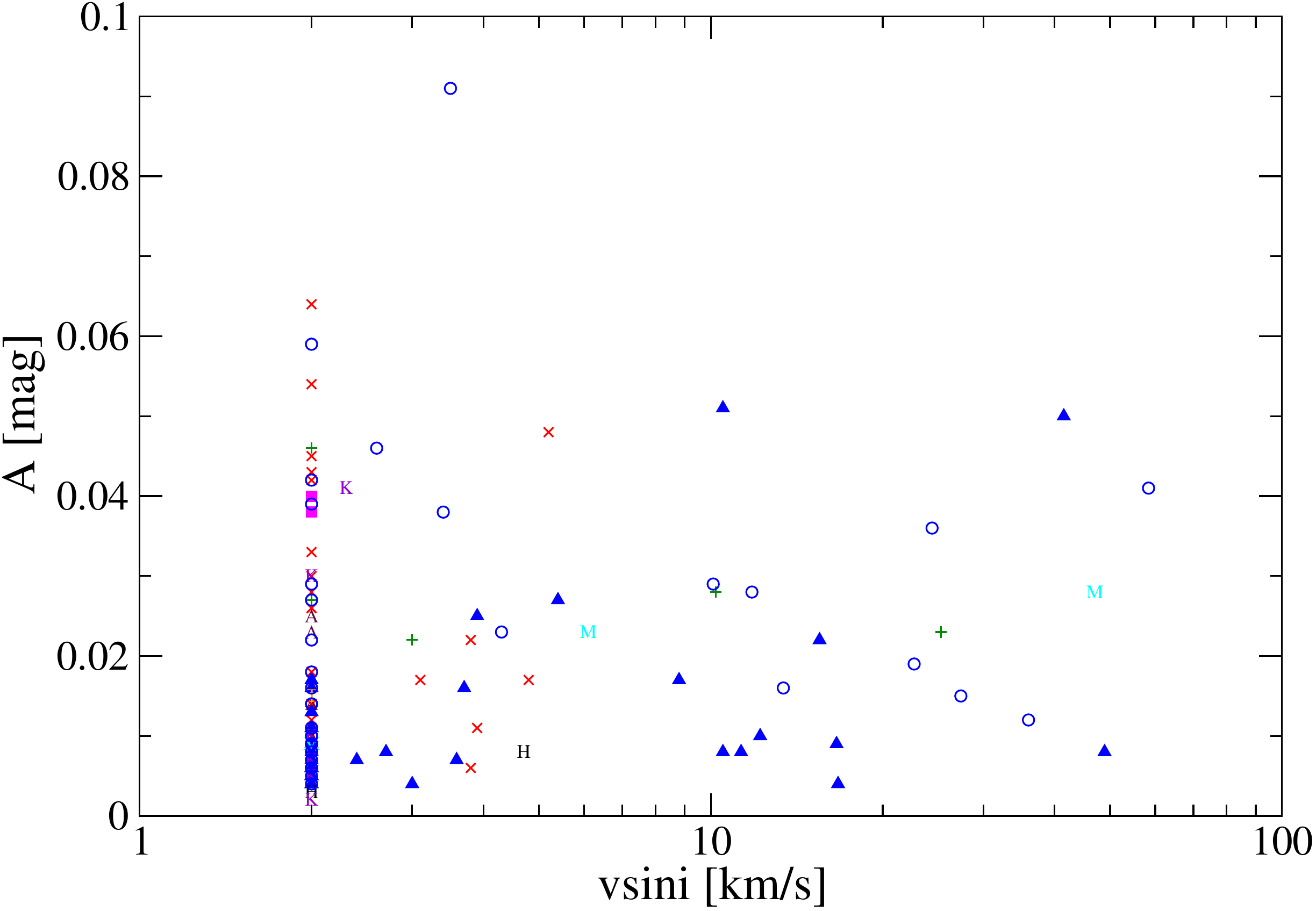}  
        \includegraphics[width=0.49\textwidth]{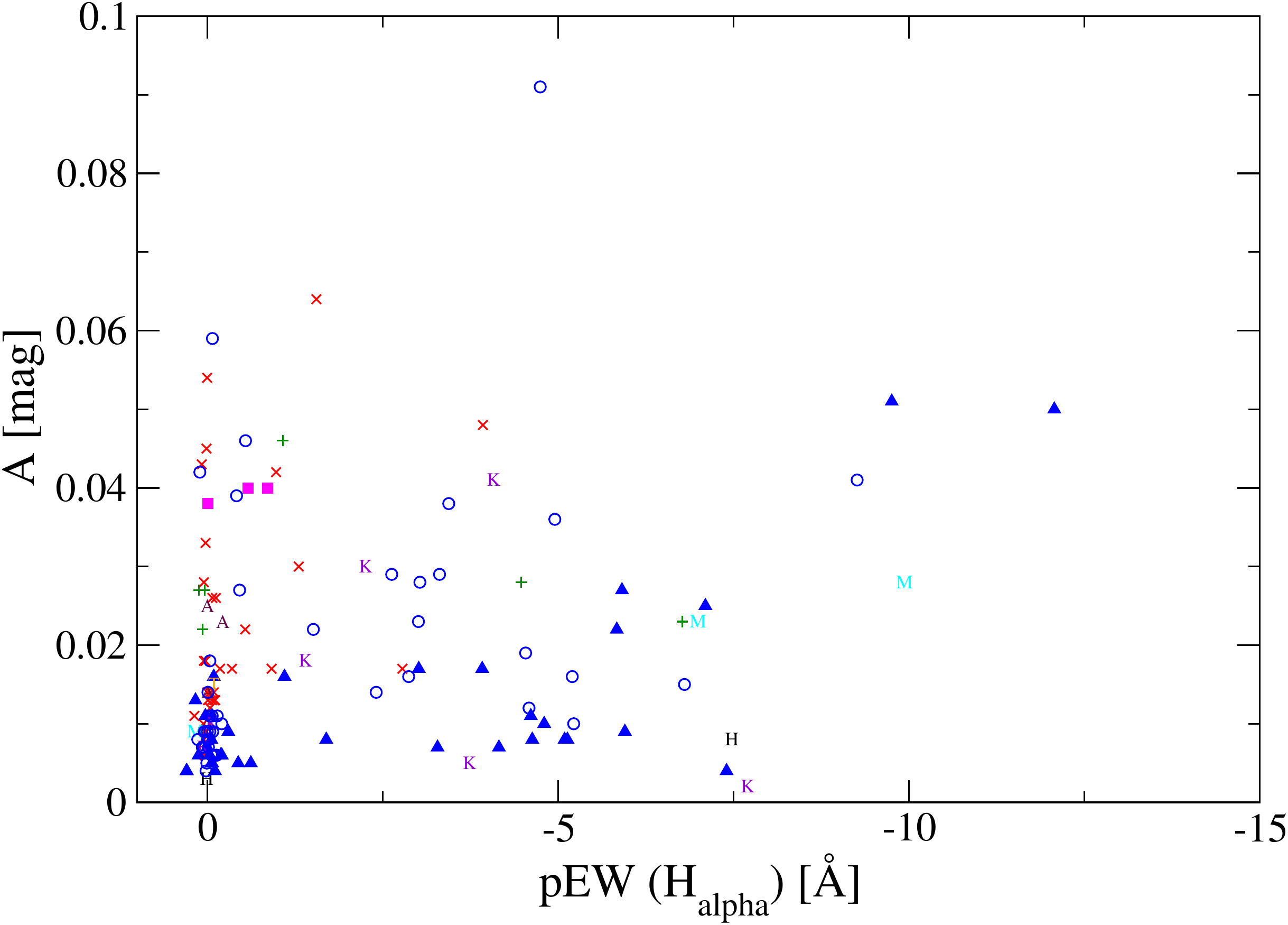}    
        \caption{Our amplitude of photometric variability as a function of $v \sin{i}$ (top panel) and pEW(H$\alpha$) (bottom panel).
Red crosses: ASAS data, 
blue triangles: MEarth, 
blue open circles: SuperWASP,
green pluses: NSVS, 
magenta squares: AstroLAB IRIS, 
maroon C's: ASAS-SN, 
black H's: HATNet, 
violet K's: K2, 
cyan M's: Montcabrer, 
orange T's: Montsec.}
\label{fig.Protamplitudea}
\end{figure}

\begin{figure}
        \centering
        \includegraphics[width=0.49\textwidth]{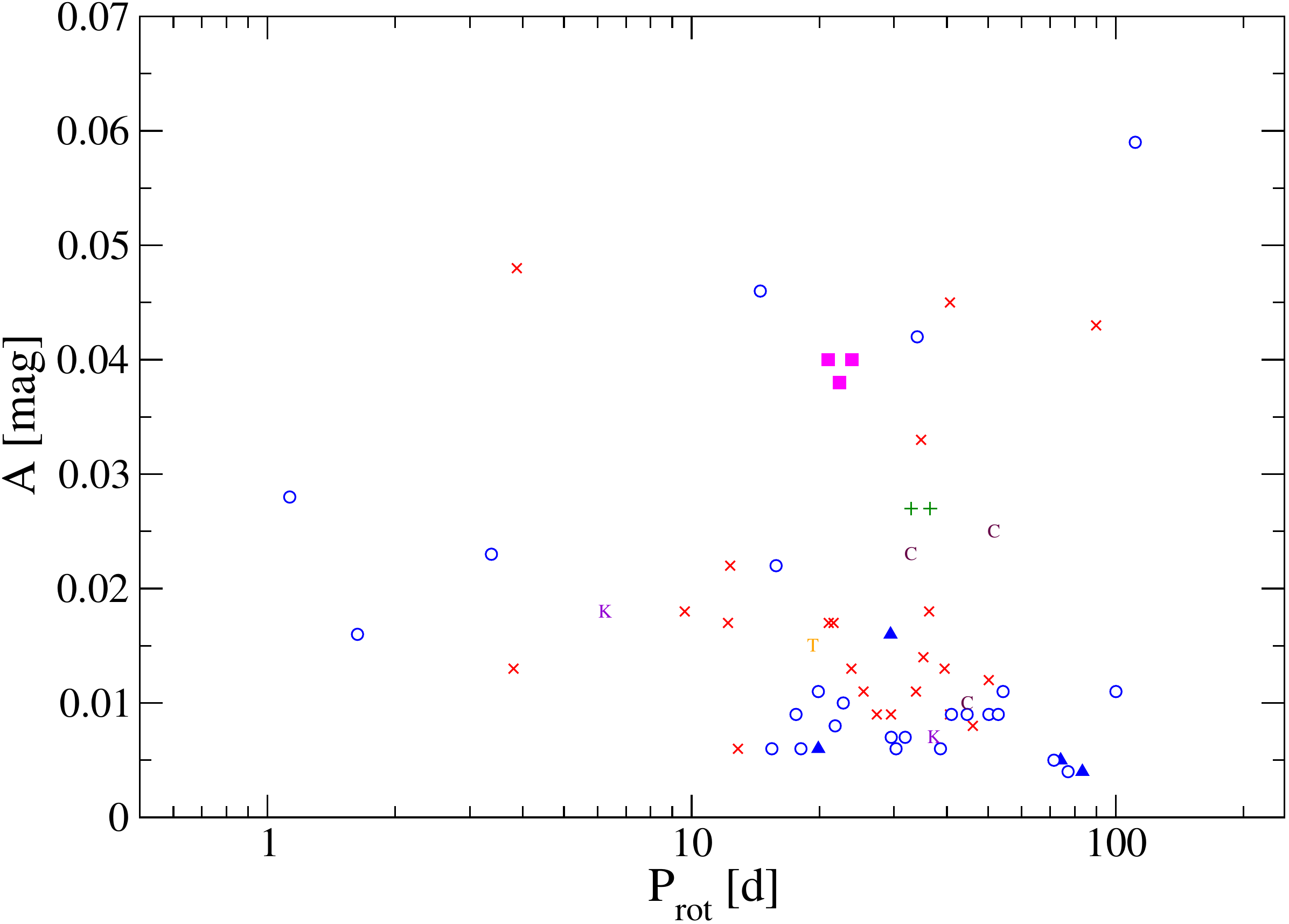}  
        \includegraphics[width=0.49\textwidth]{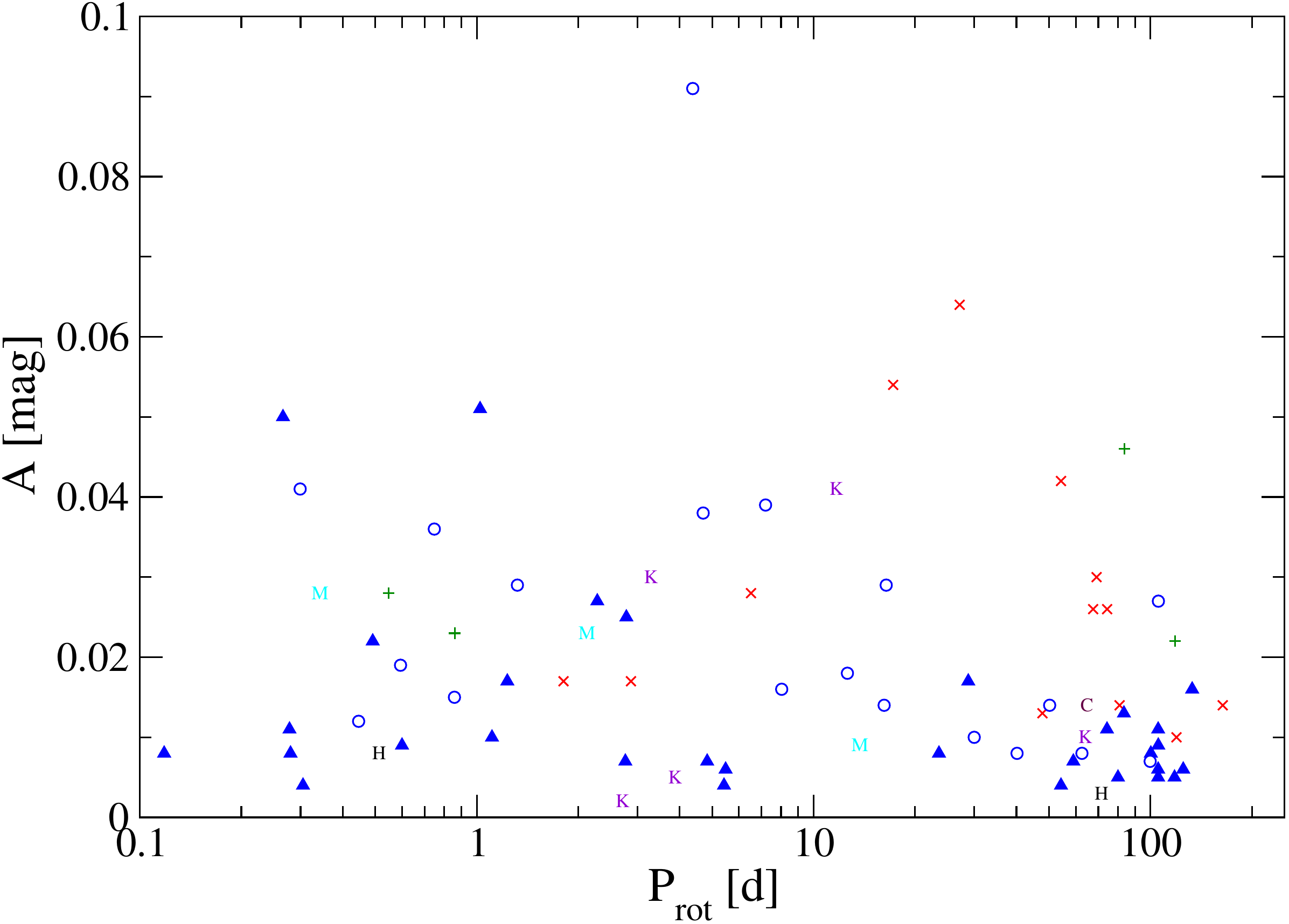}    
        \caption{Our amplitude of photometric variability as a function of rotation period for M0.0 -- M3.0\,V stars (top panel) and M3.5 -- M8.5\,V stars (bottom panel). Same symbol legend as in Fig.~\ref{fig.Protamplitudea}.}
\label{fig.Protamplitudeb}
\end{figure}

We can outline conservative lower and upper envelopes of the $P_{\rm rot}$-$v \sin{i}$ relation.
The horizontal regime at long periods and low rotational velocities is due to the instrumental limit in determining $v \sin{i}$ from high-resolution spectra, not to any real astrophysical effect on the stars.

\footnote{An ultra-high-resolution spectrograph of $\mathcal{R} \gtrsim$ 200,000 at an 8\,m-class telescope would be needed to measure rotational velocities of a large sample of M dwarfs with an accuracy of 1\,km\,s$^{-1}$ \citep{2018A&A...612A..49R}.}

In our plot there is only one outlier, \object{G~131--047}\,AB (J00169+200\,AB), which has $P_{\rm rot}$ = 17.3\,d from \cite{2015ApJ...812....3W} and $v \sin{i}$ = 22\,km\,s$^{-1}$ from \cite{2002AJ....124.2868M}.
The star, a non-GTO star from Carmencita, is another unconfirmed close astrometric binary with components separated by only 1.08\,arcsec \citep{2017A&A...597A..47C}, so it might also be a double-line spectroscopic binary.
A double narrow cross-correlation function could appear to \cite{2002AJ....124.2868M} as a single wide function, which would explain the high rotational velocity for its rotation period.  
In any case, our plot will be a reference for future studies of the $P_{\rm rot}$-$v \sin{i}$ relation, as we used the most precise values of rotational velocities \citep{2018A&A...612A..49R} and of rotation periods (this study), together with an exhaustive compilation of data for nearby bright M dwarfs.
 
In the bottom panel of Fig.~\ref{fig.Protactivity} we also link our rotation periods and those compiled in Carmencita with the corresponding pseudo-equivalent widths of the H$\alpha$ line, pEW(H$\alpha$), possibly the most widely used activity indicator in low-mass stars.
From the plot, again, the shorter the rotation period, the stronger the H$\alpha$ emission \cite[cf.][]{2018A&A...614A..76J}.
And again, there are outliers.
The most remarkable ones, with pEW(H$\alpha$) $<$ --20\,{\AA}, are J07523+162 = \object{LP~423-031} and J08404+184 = \object{AZ~Cnc}.
They just displayed well-documented flaring activity during their spectroscopic observations \citep{1966VA......8...89H,1978IBVS.1454....1J,1985IBVS.2796....1D,2009ApJ...699..649S,2010ApJ...710..924R,2015A&A...577A.128A}.
Other stars at the outer boundary of the general distribution have either moderate flaring activity (slightly larger $|$pEW(H$\alpha$)$|$ for their $P_{\rm rot}$) or early M spectral types (slightly smaller $|$pEW(H$\alpha$)$|$ for their $P_{\rm rot}$).

We searched for correlations between amplitude $A$ and $P_{\rm rot}$,  $v \sin{i}$, and pEW(H$\alpha$), as illustrated in Figs.~\ref{fig.Protamplitudea} and ~\ref{fig.Protamplitudeb}.
However, contrary to our expectations, we did not find that the most active stars (fastest rotators, strongest H$\alpha$ emitters) always have the largest photometric amplitudes.
For example, the star with the largest amplitude measured in our sample, $A_{\rm SuperWASP}$ = 0.091\,mag, is the most strongly active star J22468+443 = \object{EV~Lac} \citep{1984ApJ...282..214P,2000A&A...353..987F,2000A&A...364..641Z,2005ApJ...621..398O, 2010ApJ...721..785O}, while the star with the third largest amplitude, $A_{\rm SuperWASP}$ = 0.059\,mag, is J14257+236W = \object{BD+24~2733}A, a relatively inactive M0.0\,V star with $P_{\rm rot} \approx$ 111\,d, $v \sin{i} <$ 2\,km\,s$^{-1}$, and absent H$\alpha$ emission \citep{2017A&A...598A..27M,2018A&A...612A..49R,2018A&A...614A..76J}.

Previous works have investigated the relation between $A$ and $P_{\rm rot}$. \cite{2011AJ....141..166H} found no correlation for K and early to mid M dwarf stars if $P_{\rm rot}<$ 30 d, and an anti-correlation if  $P_{\rm rot}>$ 30 d. They found no correlation for mid to late M dwarfs. \cite{2016ApJ...821...93N} analysed the MEarth sample in two groups, $0.25 M_{\odot} < M < 0.5 M_{\odot}$ and $0.08 M_{\odot} < M < 0.25 M_{\odot}$, and found an anti-correlation $A - P_{\rm rot}$ in the first group and no correlation in the second group. We repeated the same exercise separating our sample in early to mid (M0.0 -- 3.0\,V; $N$=63) and mid to late (M3.5 -- 8.5\,V; $N$=78) stars (Fig ~\ref{fig.Protamplitudeb}). We performed a Spearman rank correlation analysis and obtained a relation coefficient of --0.24 with $p$-value 0.06 for early to mid stars, and a coefficient of   --0.21  with $p$-value 0.06 for mid to late stars. Therefore, in both cases we found a suggestive (but non-significant) anti-correlation between $A$ and $P_{\rm rot}$. We note the overabundace of M0.0 -- 3.0\,V stars with periods $P_{\rm rot}=$10 -- 100\,d with respect to M3.5 -- 8.5\,V stars).

\subsection{Aperiodic variables}
\label{section.aperiodicvariables}

\begin{figure}
        \centering
        \includegraphics[width=0.49\textwidth]{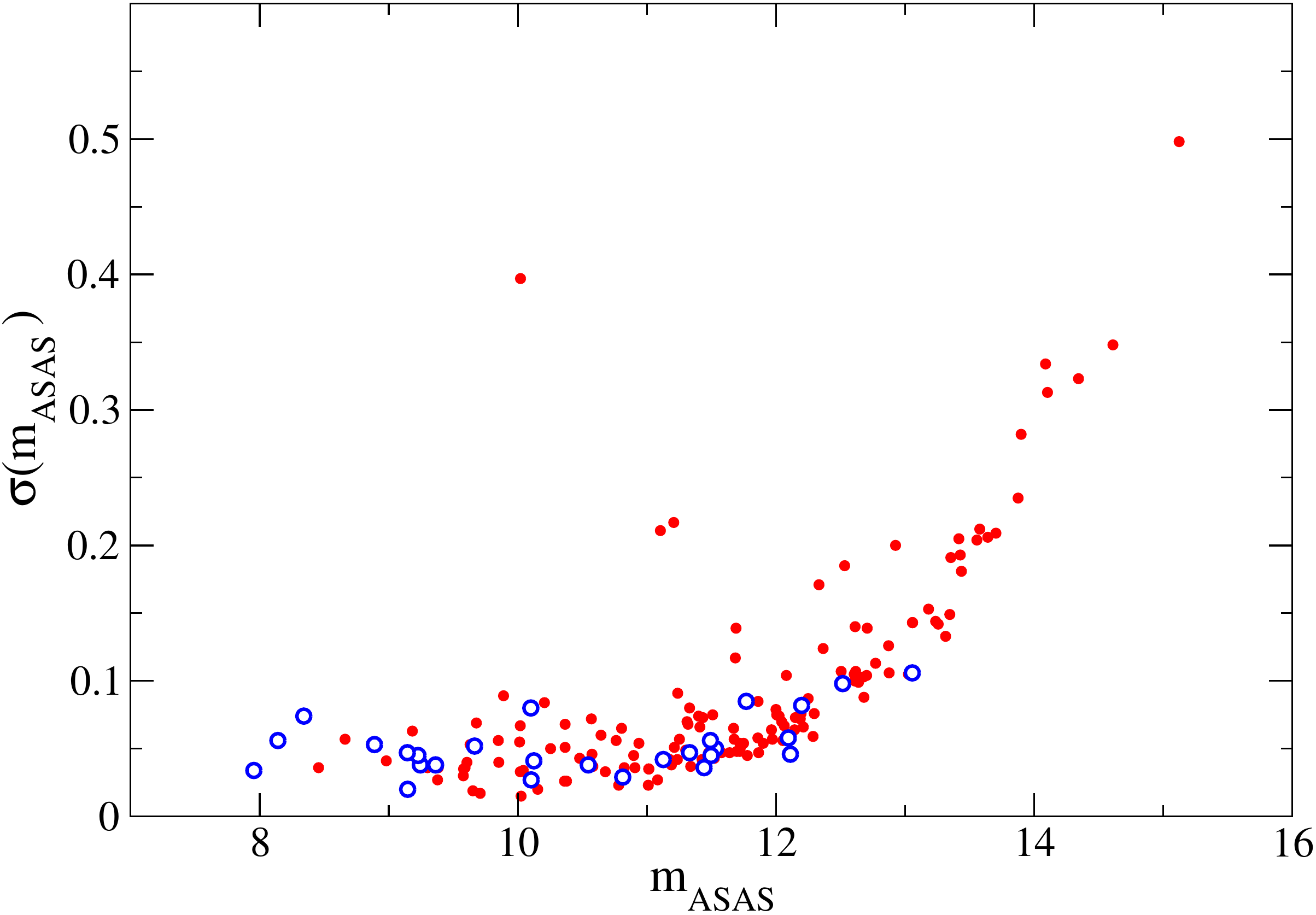}
        \includegraphics[width=0.49\textwidth]{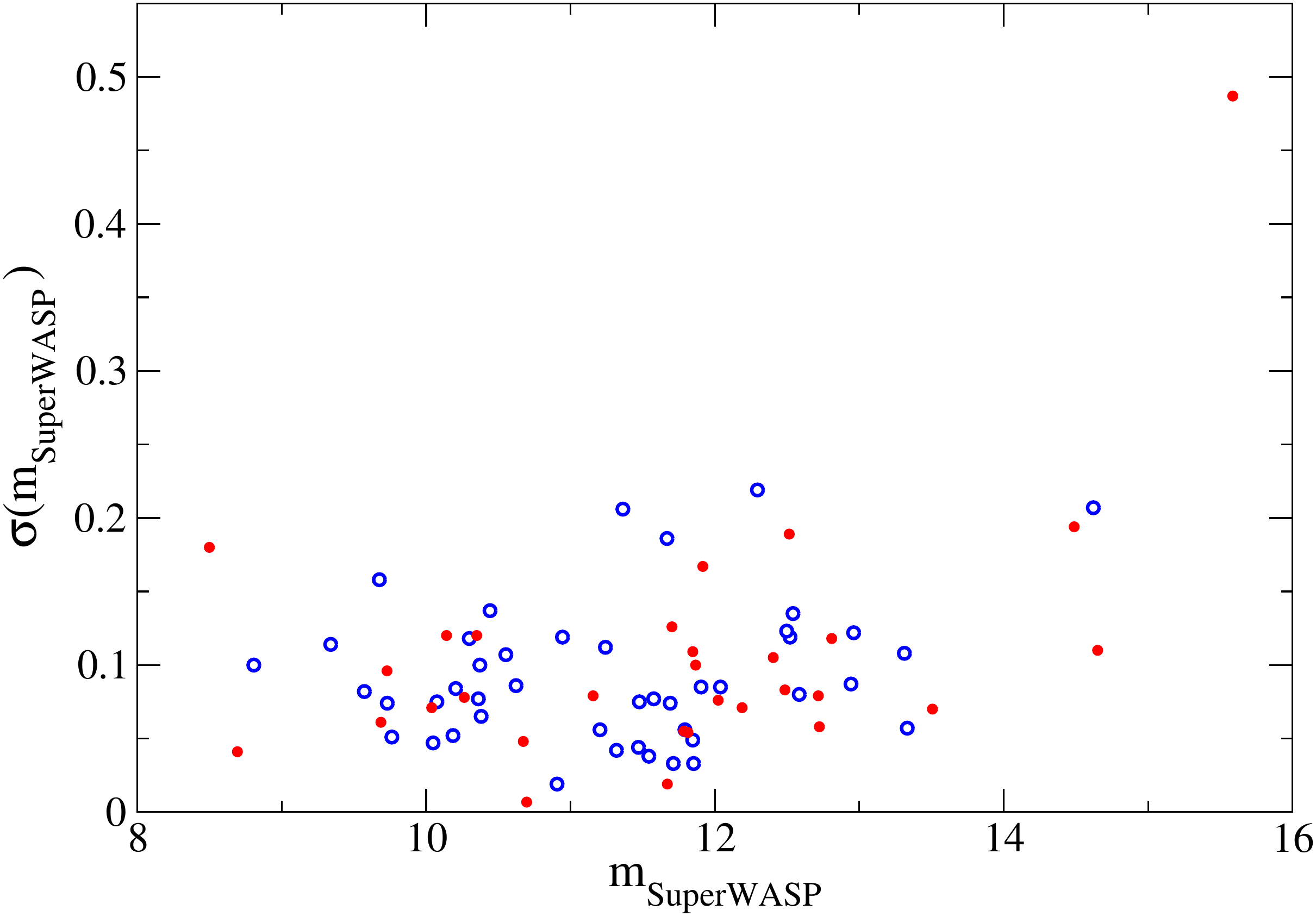}
        \includegraphics[width=0.49\textwidth]{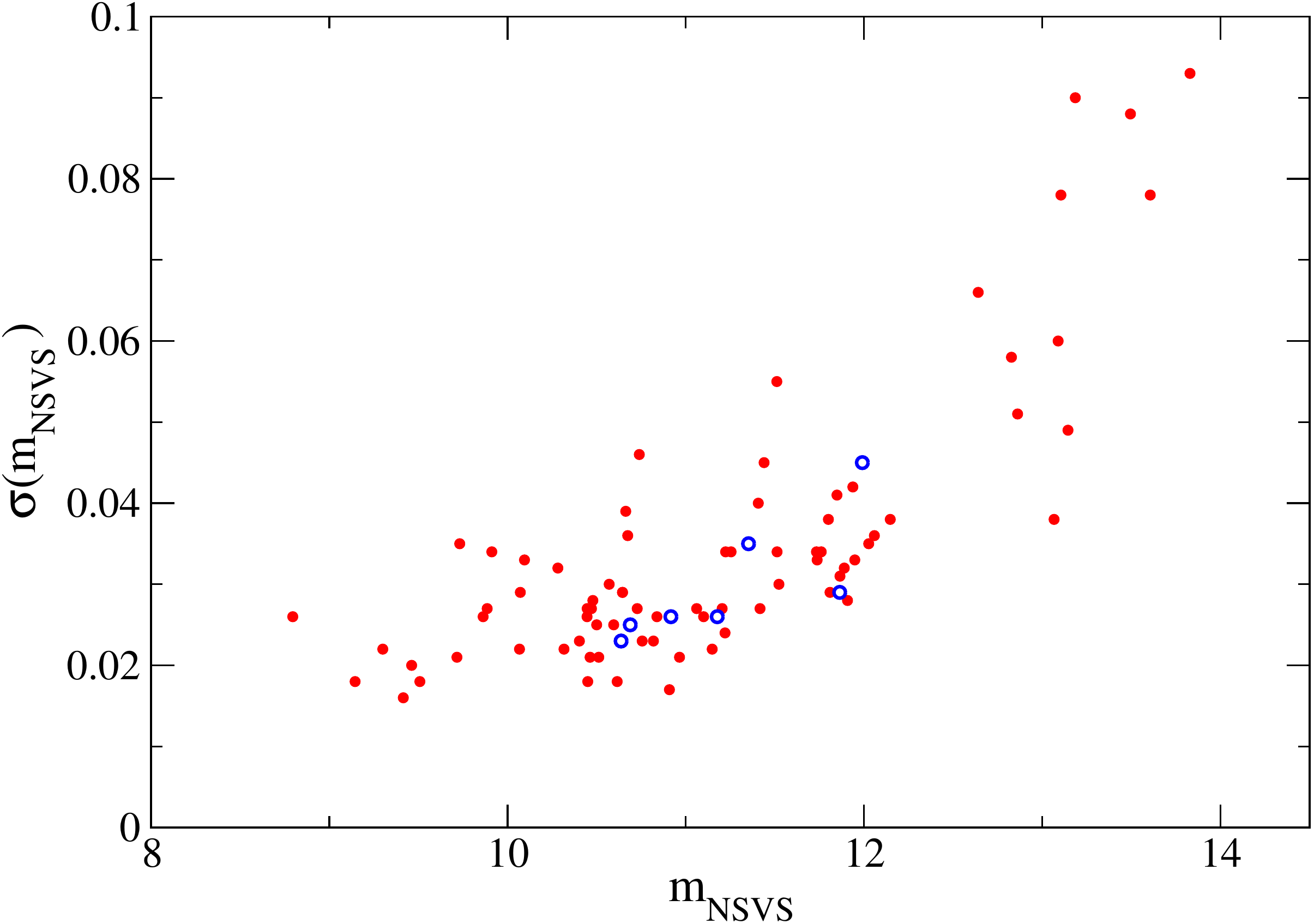}
         \caption{Standard deviation $\sigma(m)$ vs. mean $\overline{m}$ for ASAS, SuperWASP, and NSVS magnitudes. 
Blue open circles and red filled circles represent stars with and without $P_{\rm rot}$ computed in this work, respectively. Most of the structure, specifically in the top and bottom panels, is not of astrophysical origin but related to the different sources of noise.} 
\label{fig.Sigmamagvsmagmedia}
\end{figure}

We also looked for aperiodic or irregular variable stars that display a large scatter in their light curves, but for which we failed to find any significant periodicity.
For this reason, we used the common approach of plotting the standard deviation $\sigma(m)$ of each individual light curve as a function of the mean magnitude $\overline{m}$ (see e.g. \citealt{2010AN....331..257C} for a practical application). 
We performed this exercise for only three of the main surveys because average magnitudes of MEarth light curves were set to zero, different pass bands and instrumental set-ups must not be mixed up, and the complementary surveys and our own observations do not have enough data points for a robust analysis.

The $\sigma(m)$  versus $\overline{m}$ plots for ASAS, SuperWASP, and NSVS are shown in Fig.~\ref{fig.Sigmamagvsmagmedia}.
All NSVS stars follow the expected trend for their mean magnitude (see \citealt{2007MNRAS.375.1449I} for a detailed description of virtually all parameters affecting $\sigma(m)$ in a photometric survey).
Of the SuperWASP stars, only J09003+218 = \object{LP~368--128}, a faint M6.5\,V \citep{2012ApJS..201...19D,2015A&A...577A.128A}, has a much larger scatter than its peers. 
\cite{2016ApJ...821...93N} found a period of 0.439\,d that we did not reproduce in our datasets (Table~\ref{table.protnotrecovered}).
However, we attribute its large $\sigma(m)$, of about 0.5\,mag, to its magnitude, which is over 1\,mag fainter than the second-faintest star in the SuperWASP sample.
Finally, there are three outliers among the ASAS stars: J12248-182 = \object{Ross~695}, J04520+064 = \object{Wolf~1539}, J09307+003 = \object{GJ~1125}.
We ascribe the origin of their large $\sigma(m)$ to their short angular separation to bright background stars that contaminate the ASAS light curves.
In the three cases, they are located at variable angular separations $\rho \approx$ 25--50\,arcsec to stars 1.2--2.5\,mag brighter in the $V$ band than our M dwarfs.
As a result, the variability observed is not intrinsic to the stars themselves.

A few aperiodic or irregular variable stars also appeared in our monitoring with amateur and semi-professional telescopes.
For example, with Carda we measured intra-night trends of over 0.030\,mag in the known variable star J05337+019 = \object{V371~Ori}, and a 15\,min, 0.030\,mag-amplitude flare in the poorly investigated, X-ray emitting star J06574+740 = \object{2MASS J06572616+7405265}.

\subsection{Long-period cycles}

\cite{1996ApJ...460..848B} suggested the observable $P_{\rm cycle}/P_{\rm rot}$ to study the relation between cycle and rotation periods. They proposed $P_{\rm cycle}/P_{\rm rot}$ $\sim D^{i}$, being $D$ the dynamo number and $i$ the slope of the relation. Slopes i $\sim$ 1 would imply no correlation, while values different from unity would imply a correlation between the length of the cycle and rotation period. Previous works have explored this relation \citep{2009A&A...501..703O,2012A&A...541A...9G,2016A&A...595A..12S} and find a weak correlation for F, G, and K stars.
Here we repeated the exercise restricting the analysis to M dwarf stars. For this, we carefully selected 47 single, inactive M dwarfs with previous published $P_{\rm cycle}$ and $P_{\rm rot}$ \citep{2012ARep...56..716S,2013ApJ...764....3R,2016A&A...595A..12S,2017MNRAS.464.3281W,2018arXiv180402925K,2018A&A...612A..89M} and the stars of our sample for which we have found $P_{\rm cycle}$ and $P_{\rm rot}$ (see Table ~\ref{table.ourPcycle}). In our sample of field M dwarfs we did not include fast-rotating, probably very young M dwarfs tabulated by  \cite{2013AN....334..972V,2014MNRAS.441.2744V} and \cite{2016A&A...591A..43D,2017A&A...606A..58D}.

\begin{figure}
        \centering
        \includegraphics[width=0.49\textwidth]{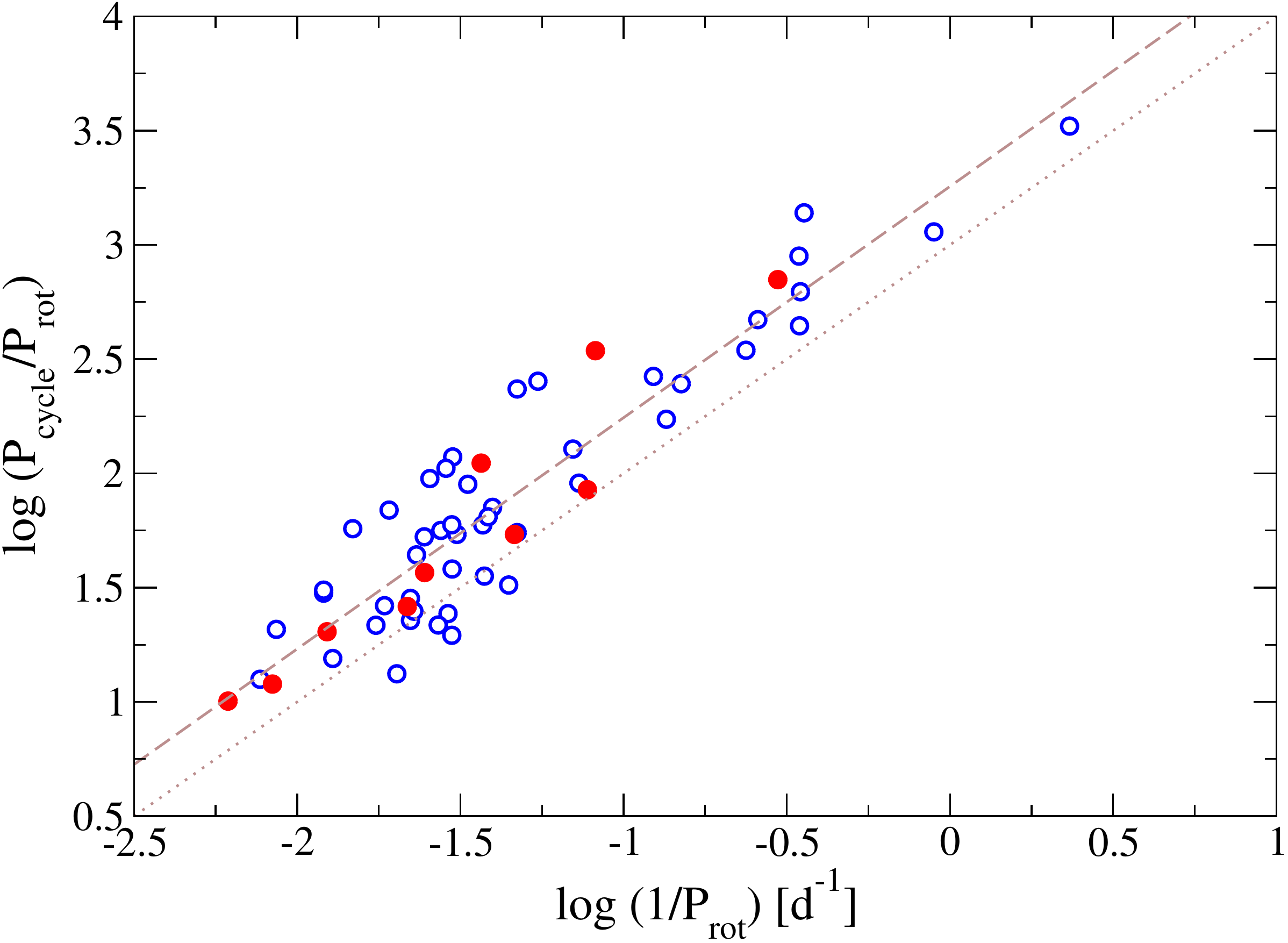}
        
         \caption {$P_{\rm cycle}/P_{\rm rot}$ vs. $1/P_{\rm rot}$ in log-log scale for M dwarf stars with previous published $P_{\rm cycle}$ and $P_{\rm rot}$ (blue open circles) and stars with new $P_{\rm cycle}$ from this work (red filled circles). Dashed ($i$ = 1.01) and dotted ($i$ = 1.00, with arbitrary offset) lines mark our sample fit and non-correlation, respectively.}
          
\label{fig.Cycles}
\end{figure}

Figure~\ref{fig.Cycles} shows the plot $P_{\rm cycle}/P_{\rm rot}$ versus $1/P_{\rm rot}$ in log-log scale. We found a slope $i = 1.01 \pm 0.06$, in agreement with the results of \cite{2012ARep...56..716S} and \cite{2016A&A...595A..12S}, who also did not find a correlation between $P_{\rm cycle}$ and $P_{\rm rot}$ in M dwarfs.

%
%


\section{Conclusions}

As a complement to the CARMENES survey for exoplanets, we searched for rotation periods from photometric series of 337 M dwarfs.
We collected public data from long-term monitoring surveys (MEarth, ASAS, SuperWASP, NSVS, Catalina, ASAS-SN, K2, and HATNet). 
For {20} stars without data in these public surveys or for those with poor data, we carried out photometric monitoring in collaboration with amateur observatories. 
In total, we investigated {622} light curves of 334 M dwarfs.
We analysed each light curve by computing LS periodograms and identifying significant signals.
In some cases, we also applied GLS and Gaussian processes analyses. 
We found {142} signals that we interpret as rotation periods.
Of these, {73} are new and {69} match or even improve previous determinations found in the existing literature. 
We also found long activity cycles for {ten} stars of the sample, {six} of which we report here for the first time. We explored the relation between $P_{\rm cycle}$ and $P_{\rm rot}$ for a sample of {47} M dwarfs with previous reported cycle and rotation periods, and the {ten} M dwarfs with cycle and rotation periods presented in this work, and did not find any correlation between $P_{\rm cycle}$ and $P_{\rm rot}$.  

Although the main aim of this work was to catalogue rotation periods of M dwarfs being searched for exoplanets with the radial-velocity method, and therefore to be able to discriminate between stellar activity and true exoplanet signals, we also presented results on
($i$) the absence of a lack of stars with intermediate periods at about 30\,d with respect to stars with shorter periods in the rotation period distribution;
($ii$) the link between rotation period and activity, especially through rotational velocity and H$\alpha$ pseudo-equivalent width;
($iii$) the identification of three very active, possibly young stars with new rotation periods between 0.34\,d and 23.6\,d; 
and
($iv$) the lack of apparent correlation between amplitude of photometric variability and $P_{\rm rot}$, $v \sin{i}$, and pEW(H$\alpha$).

The CARMENES Consortium will continue to improve or find new rotation periods, using them to discard or confirm radial-velocity exoplanets, and will make them public for use by other groups worldwide.


\begin{acknowledgements}

  We thank the anonymous referee for the careful review, F.~Garc\'ia~de~la~Cuesta for observing J05019+011 from Observatorio Astron\'omico La Vara, Luarca, Spain.
  CARMENES is an instrument for the Centro Astron\'omico Hispano-Alem\'an de
  Calar Alto (CAHA, Almer\'{\i}a, Spain). 
  CARMENES is funded by the German Max-Planck-Gesellschaft (MPG), 
  the Spanish Consejo Superior de Investigaciones Cient\'{\i}ficas (CSIC),
  the European Union through FEDER/ERF FICTS-2011-02 funds, 
  and the members of the CARMENES Consortium 
  (Max-Planck-Institut f\"ur Astronomie,
  Instituto de Astrof\'{\i}sica de Andaluc\'{\i}a,
  Landessternwarte K\"onigstuhl,
  Institut de Ci\`encies de l'Espai,
  Insitut f\"ur Astrophysik G\"ottingen,
  Universidad Complutense de Madrid,
  Th\"uringer Landessternwarte Tautenburg,
  Instituto de Astrof\'{\i}sica de Canarias,
  Hamburger Sternwarte,
  Centro de Astrobiolog\'{\i}a and
  Centro Astron\'omico Hispano-Alem\'an), 
  with additional contributions by the Spanish Ministerio de Ciencia, Innovacion y Universidades (under grants {AYA2016-79425-C3-1/2/3-P, AYA2017-89121-P, AYA2018-84089}), 
  the German Science Foundation (DFG), 
  the Klaus Tschira Stiftung, 
  the states of Baden-W\"urttemberg and Niedersachsen, 
  the Junta de Andaluc\'{\i}a, 
and by the Principado de Asturias (under grant FC-15-GRUPIN14-017).
  This research made use of 
  the SIMBAD and VizieR, operated at Centre de Donn\'ees astronomiques de Strasbourg, France, 
  and  NASA's Astrophysics Data System.

\end{acknowledgements}



\appendix

\section{Long tables}

\include{pkT30}


\section{Example light curves and periodograms}

\begin{figure}
\centering
    \includegraphics[width=0.49\textwidth]{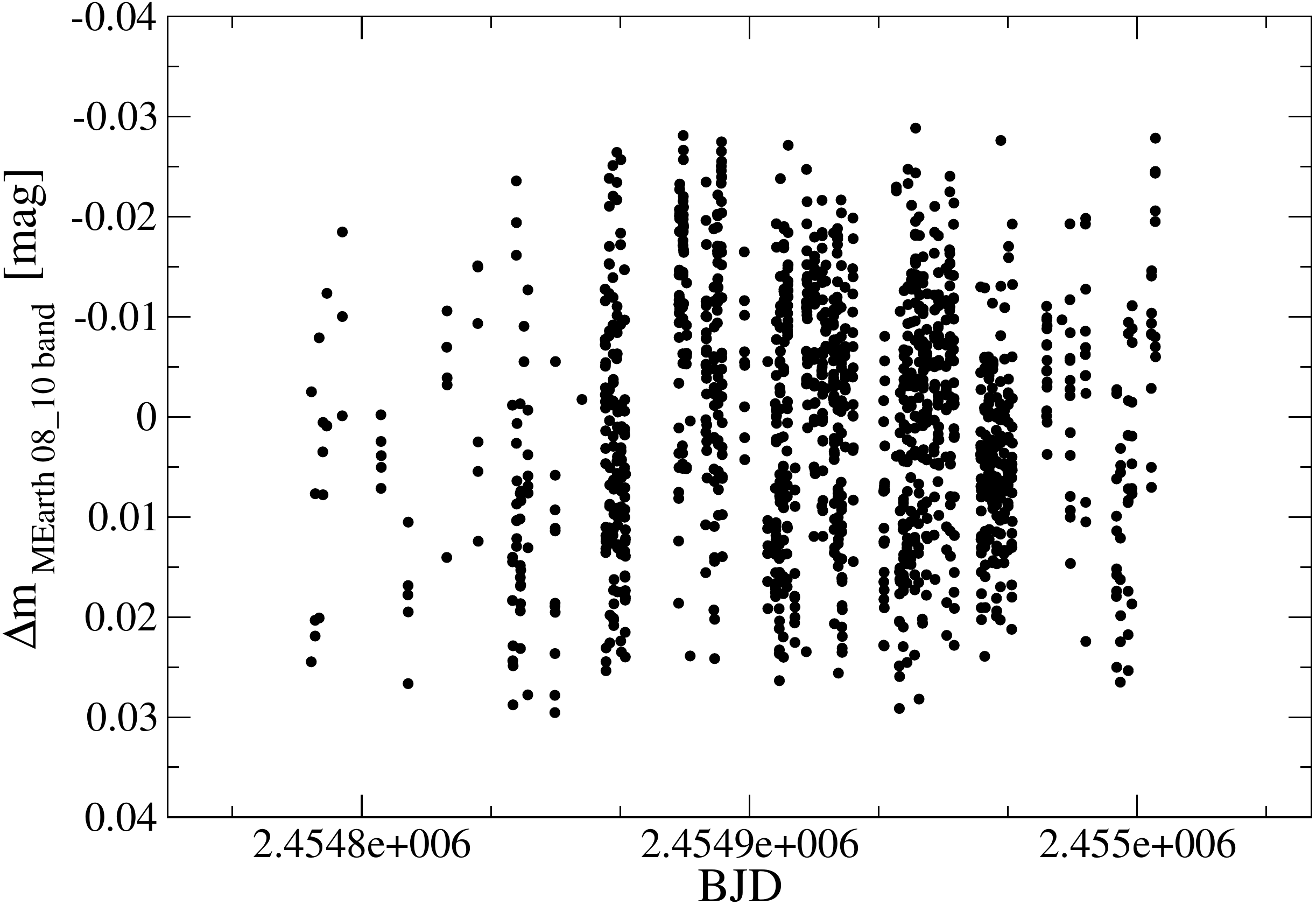}
    \includegraphics[width=0.49\textwidth]{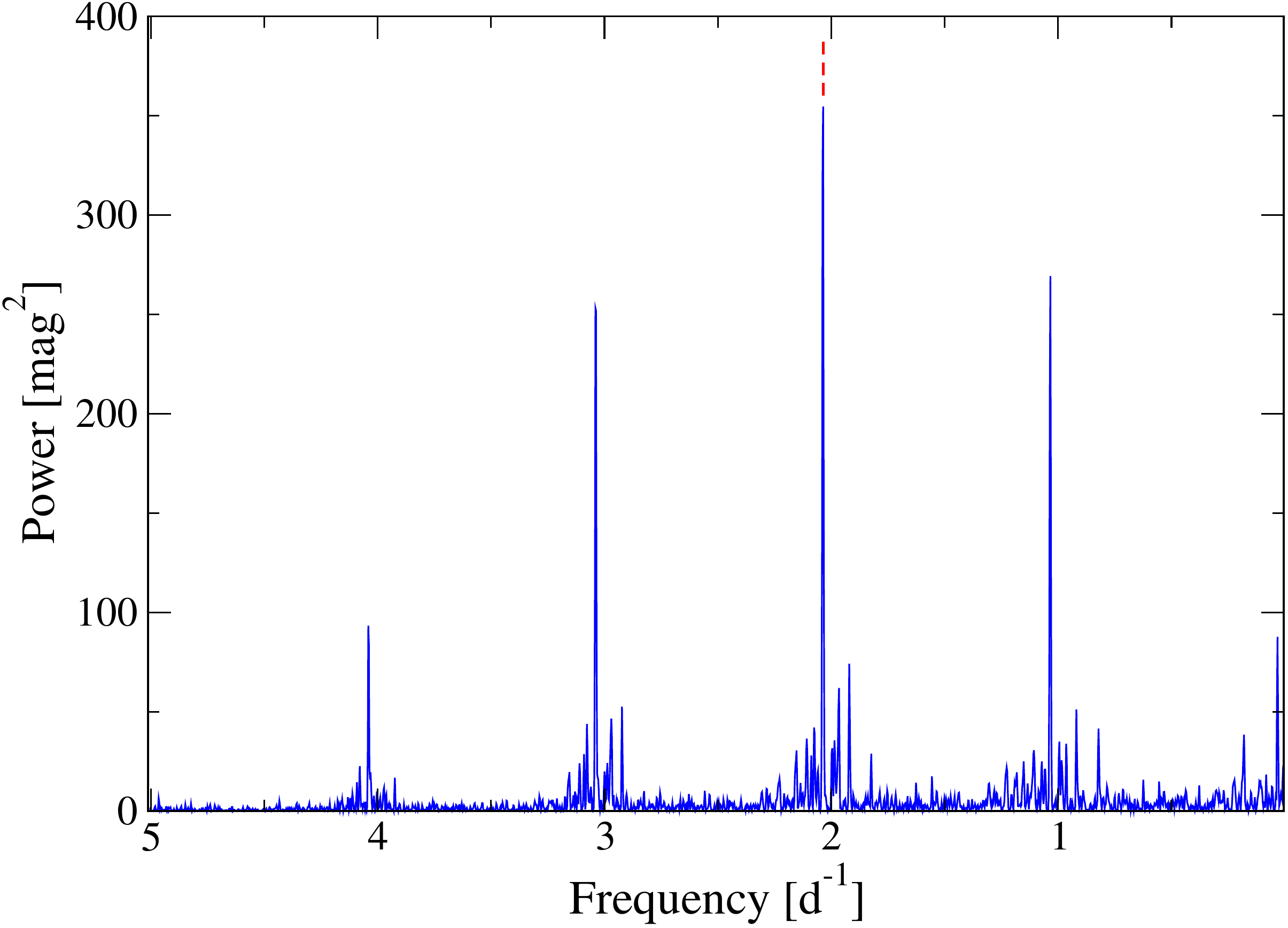}
    \includegraphics[width=0.49\textwidth]{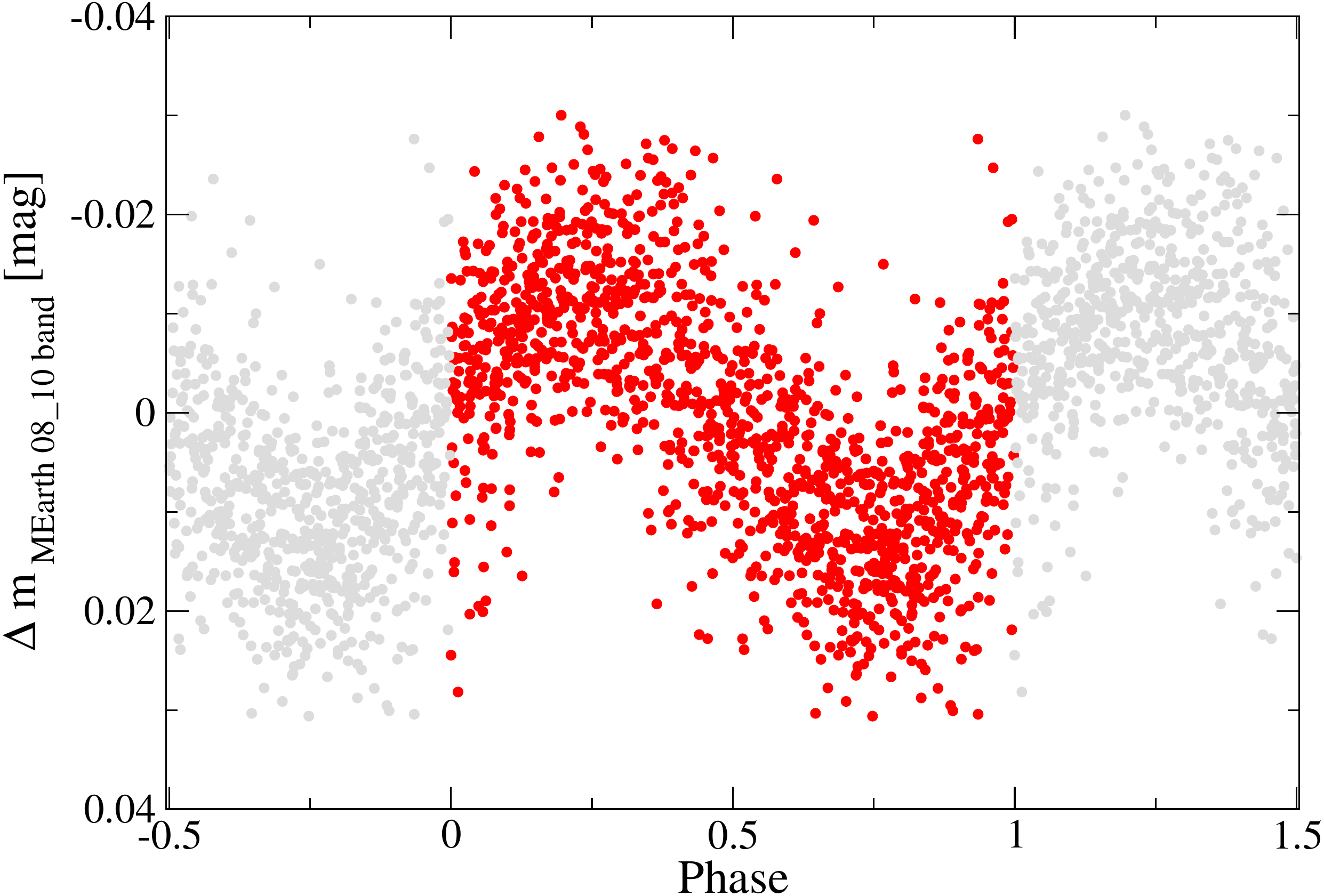}
  \caption{MEarth RG715-band photometric data ({\em top left}), Lomb--Scargle periodogram ({\em middle}), and phase-folded  rotation curve for $P$=0.491\,d ({\em bottom}) for the M5.0\,V star J12189+111 = \object{GL~Vir}.}
  \label{fig.mearth}
\end{figure}

\begin{figure}
\centering
    \includegraphics[width=0.49\textwidth]{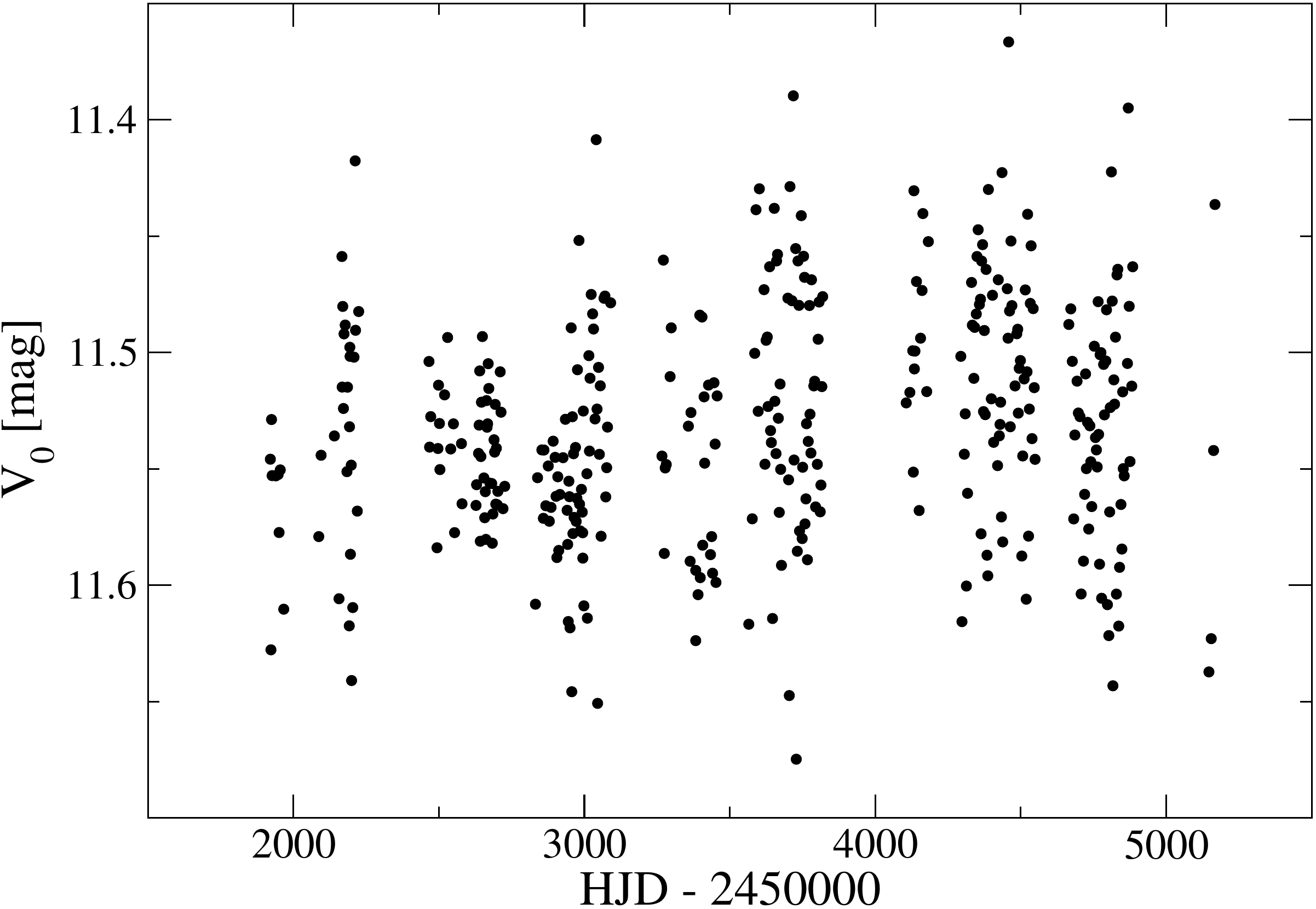}
    \includegraphics[width=0.49\textwidth]{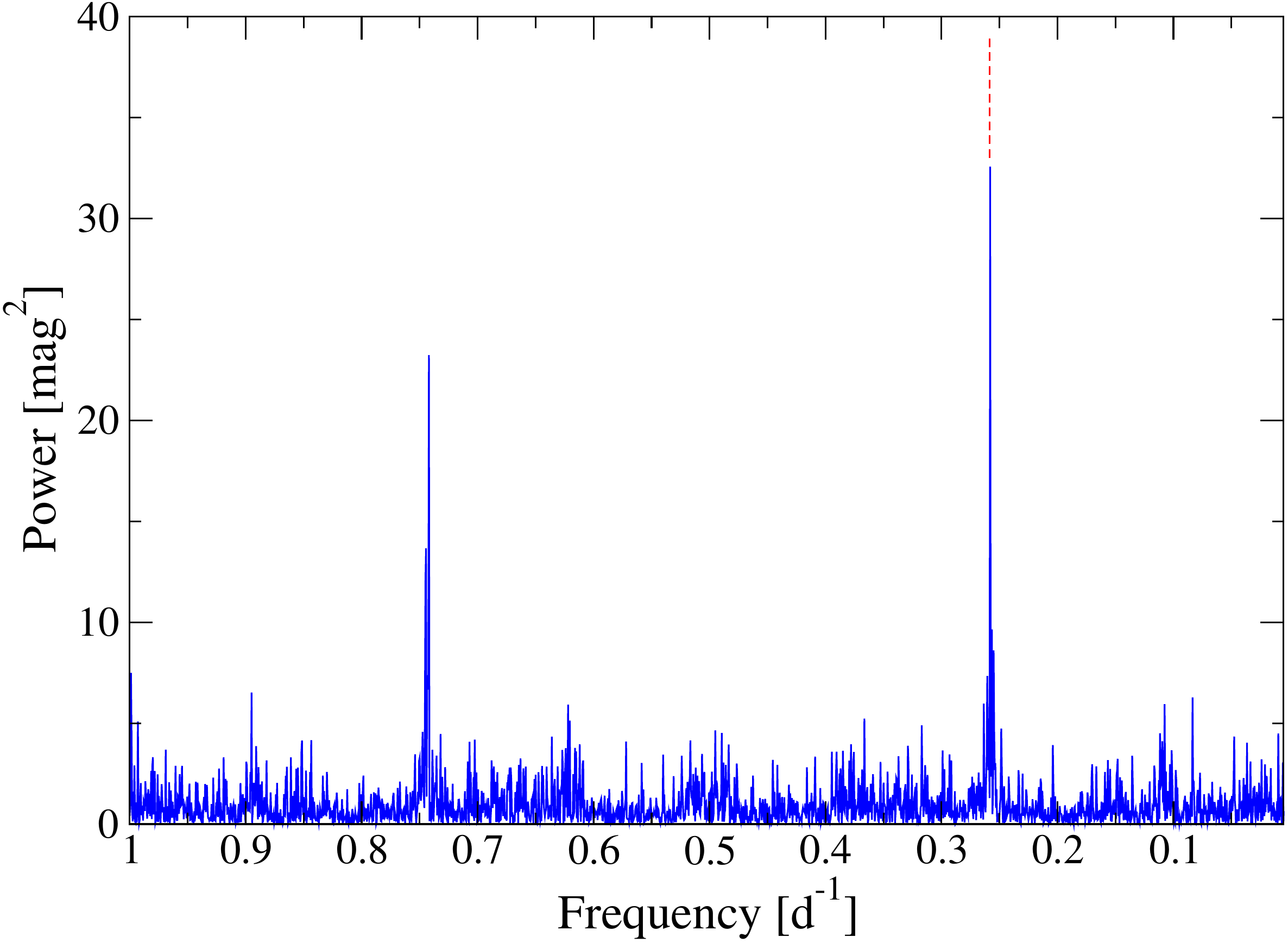}
    \includegraphics[width=0.49\textwidth]{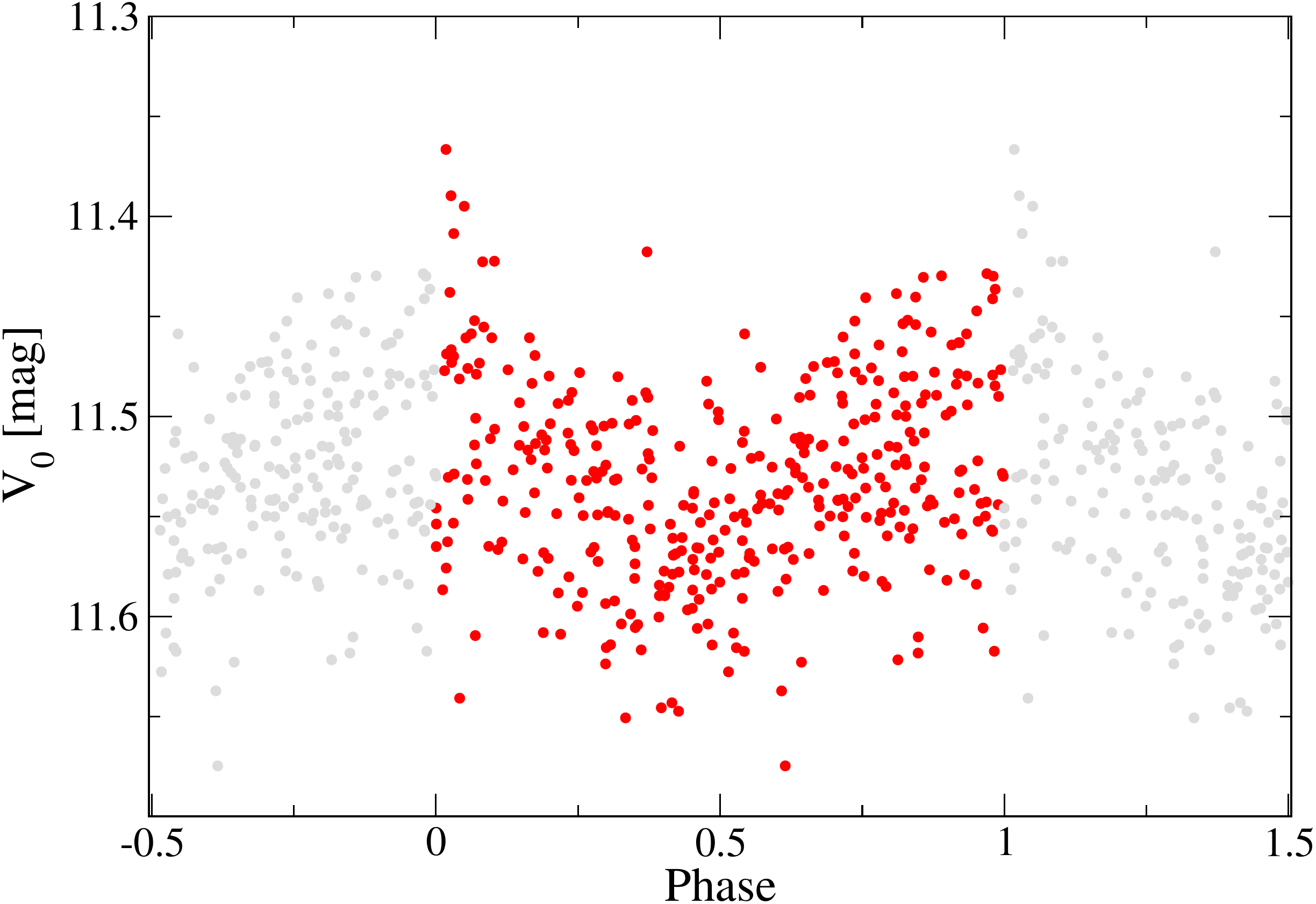}
  \caption{ASAS-3 $V$-band photometric data ({\em top}), Lomb--Scargle periodogram ({\em middle}), and phase-folded rotation curve for $P$=3.87\,d ({\em bottom}) for the M3.0\,V star J03473-019 = \object{G~080--021}.}
  \label{fig.asas}
\end{figure}

\begin{figure}
  \centering
    \includegraphics[width=0.49\textwidth]{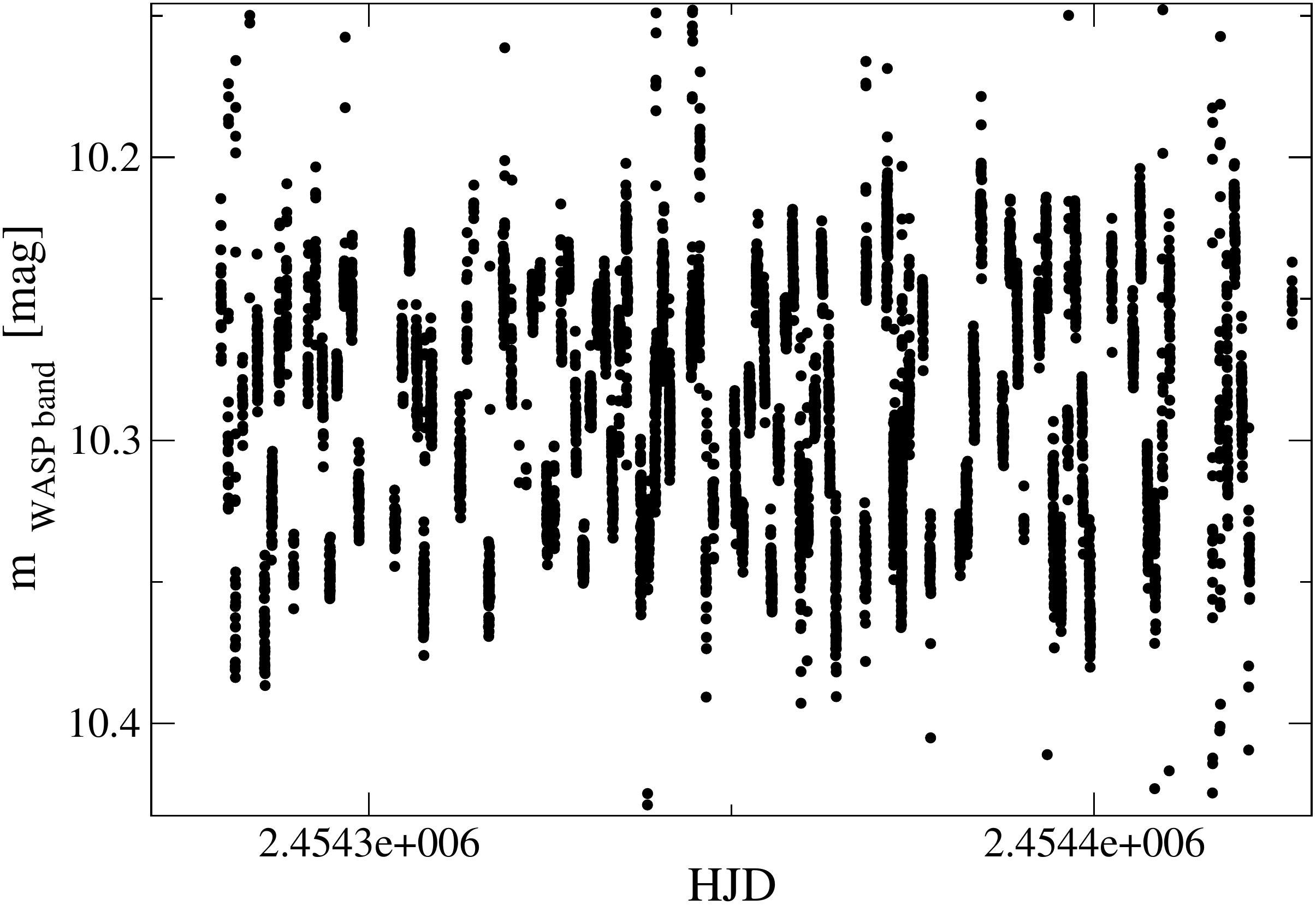}
    \includegraphics[width=0.49\textwidth]{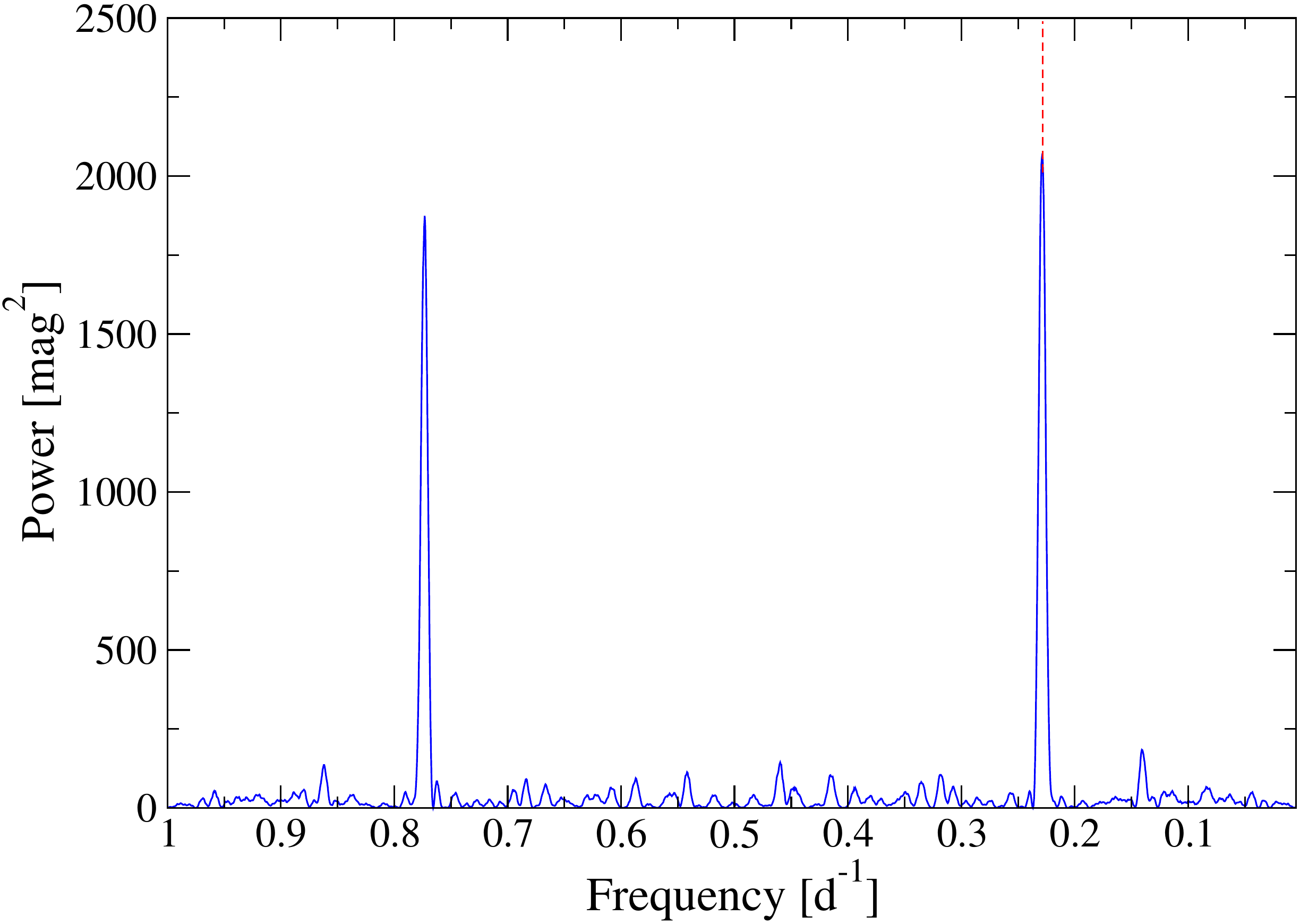}
    \includegraphics[width=0.49\textwidth]{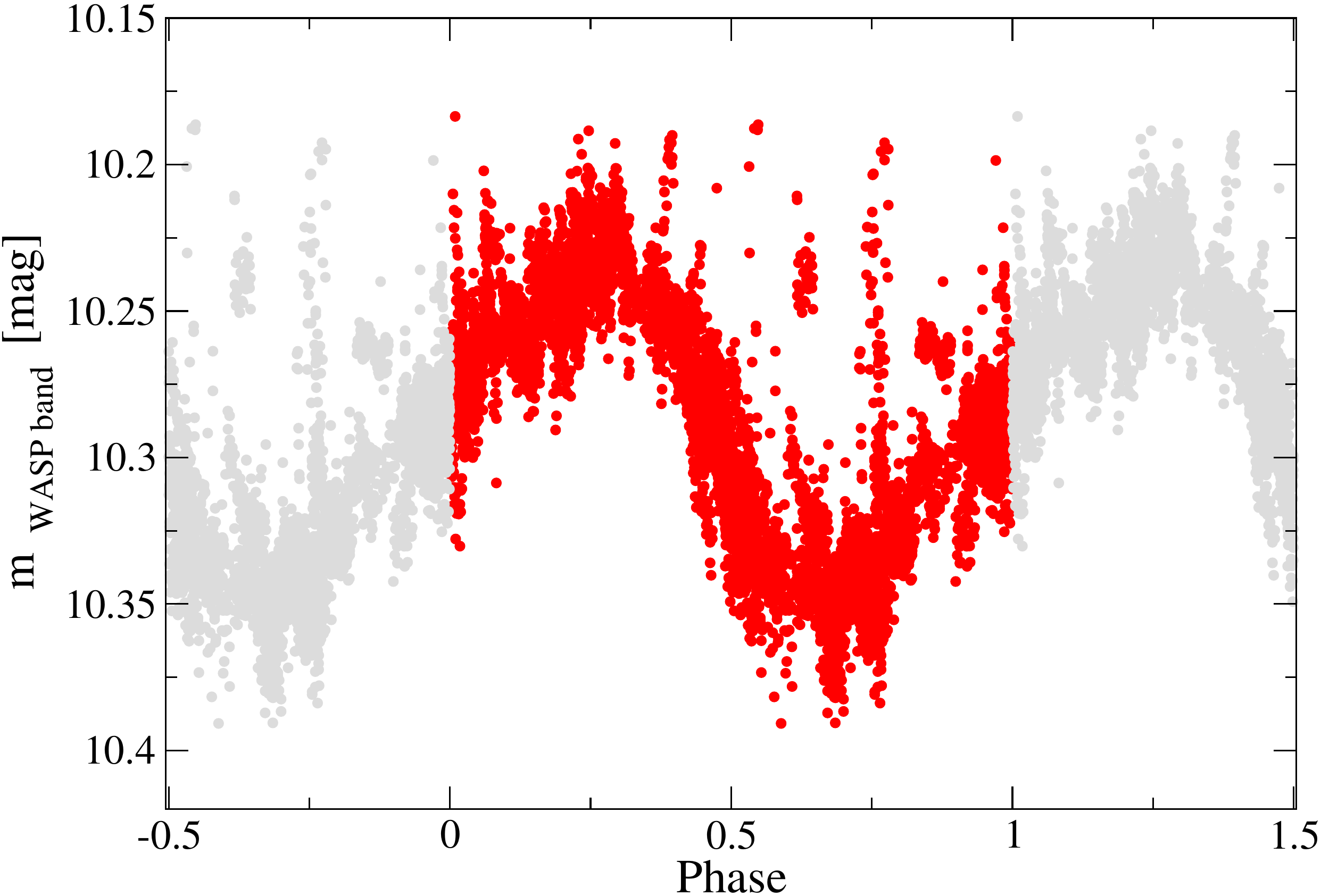}
  \caption{SuperWASP broad-band photometric data ({\em top}), Lomb--Scargle periodogram ({\em middle}), and phase-folded rotation curve for $P$=4.379\,d ({\em bottom}) for the M3.5\,V star J22468+443 = \object{EV~Lac}.}
  \label{fig.superwasp}
\end{figure}

\begin{figure}
  \centering
    \includegraphics[width=0.49\textwidth]{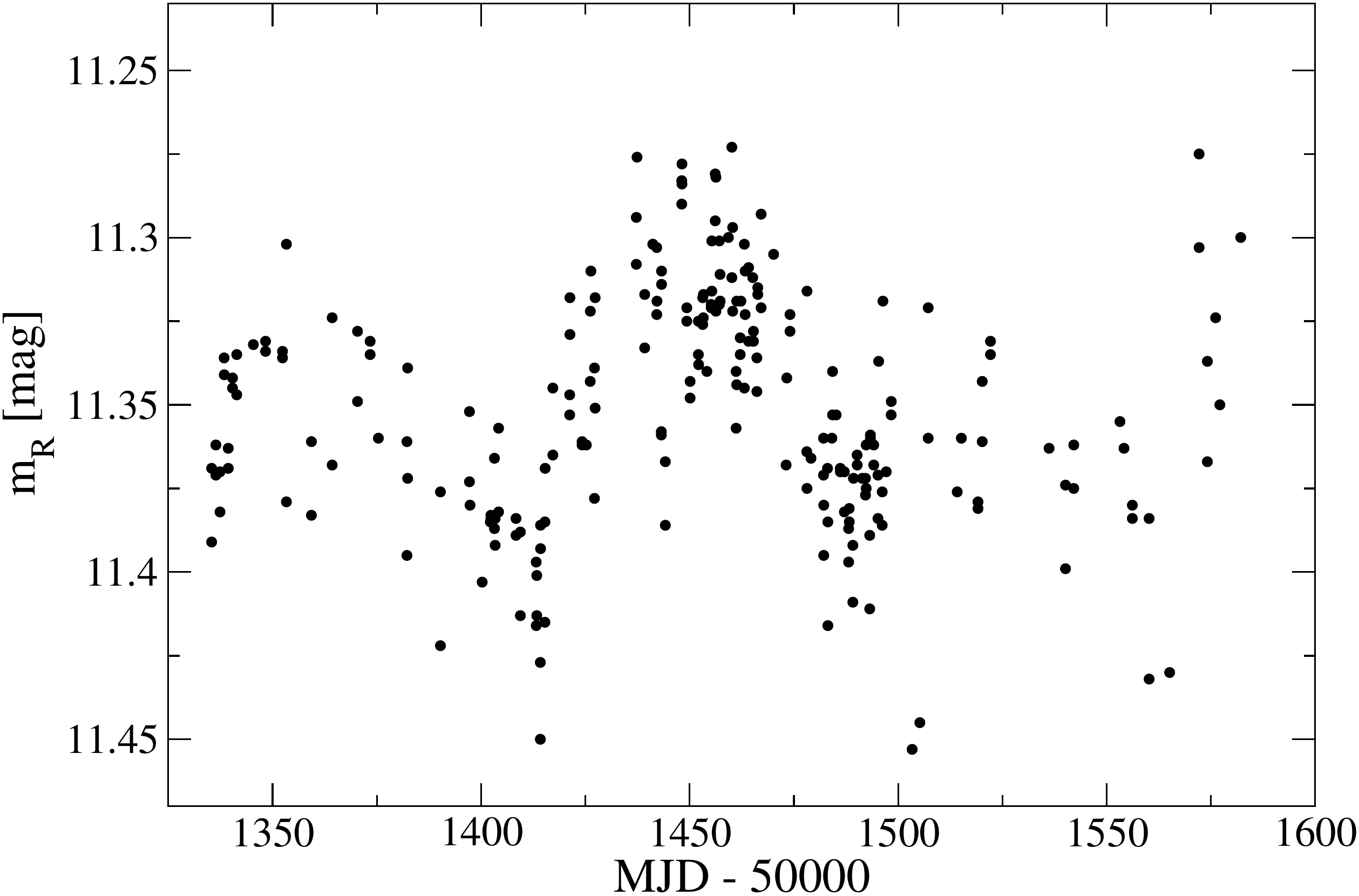}
    \includegraphics[width=0.49\textwidth]{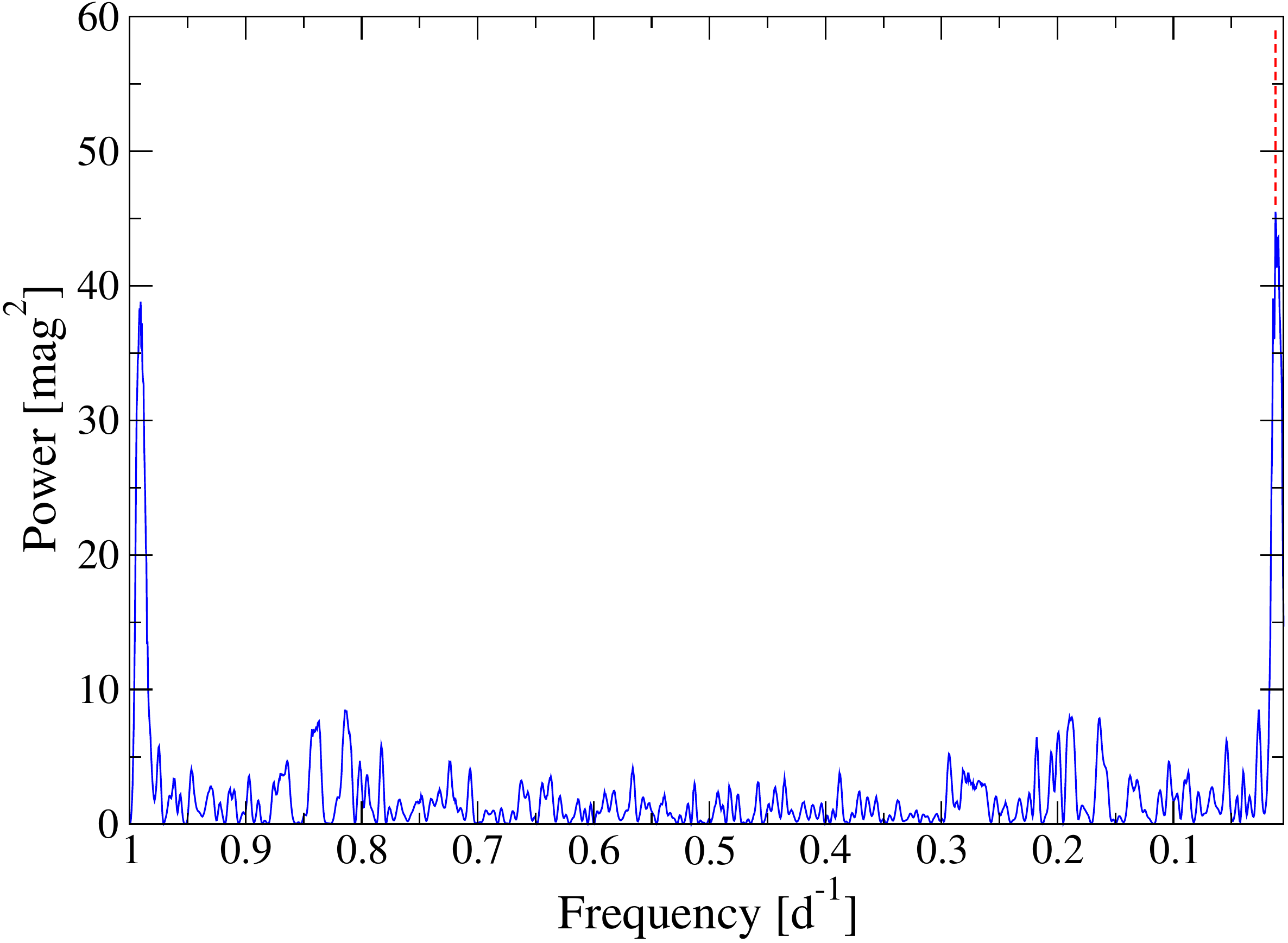}
    \includegraphics[width=0.49\textwidth]{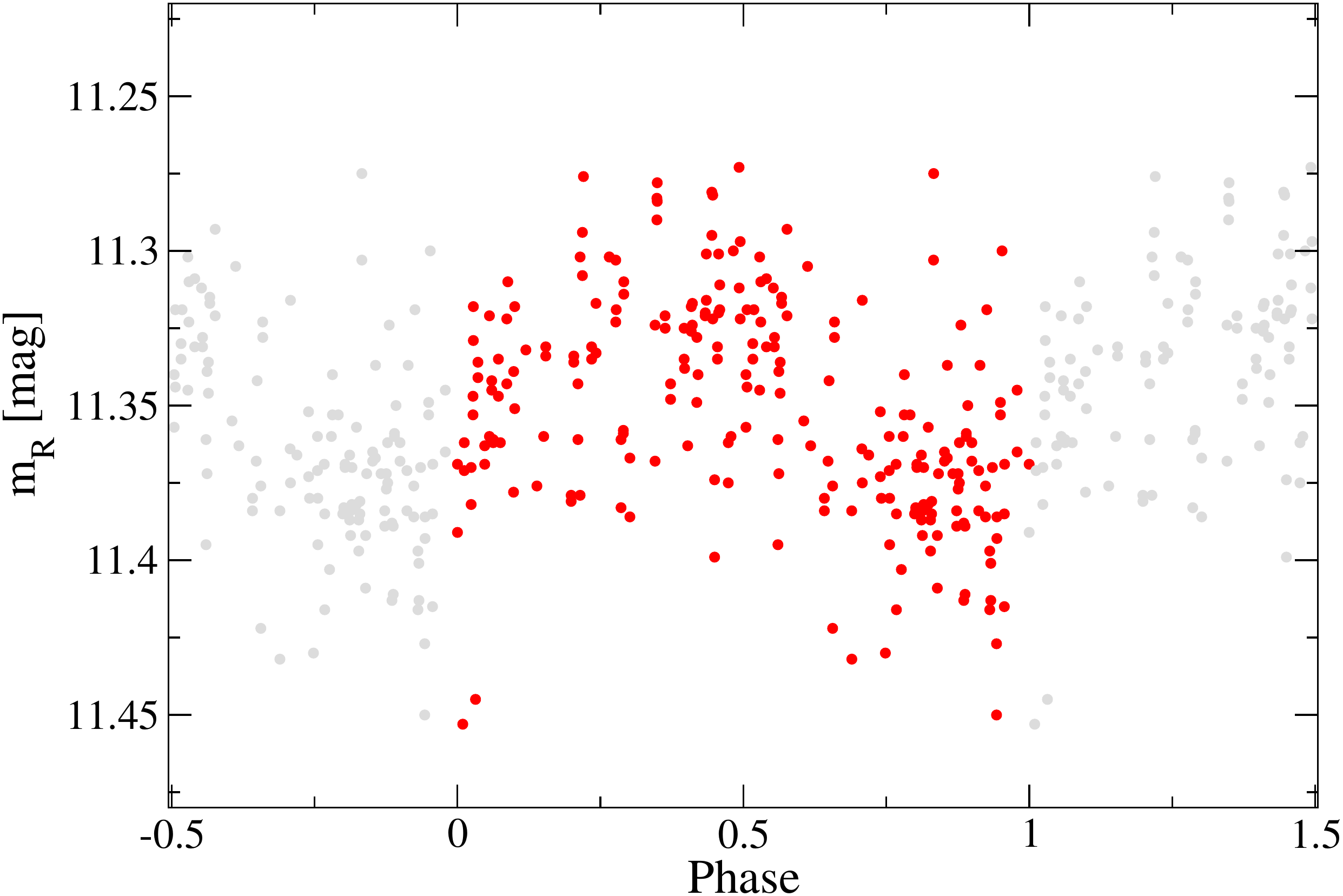}
  \caption{NSVS clear-band photometric data ({\em top}), Lomb--Scargle periodogram ({\em middle}), and phase-folded rotation curve for $P$=87\,d ({\em bottom}) for the M4.0\,V star J23431+365 = \object{GJ~1289}.}
  \label{fig.nsvs}
\end{figure}

\begin{figure}
  \centering
    \includegraphics[width=0.49\textwidth]{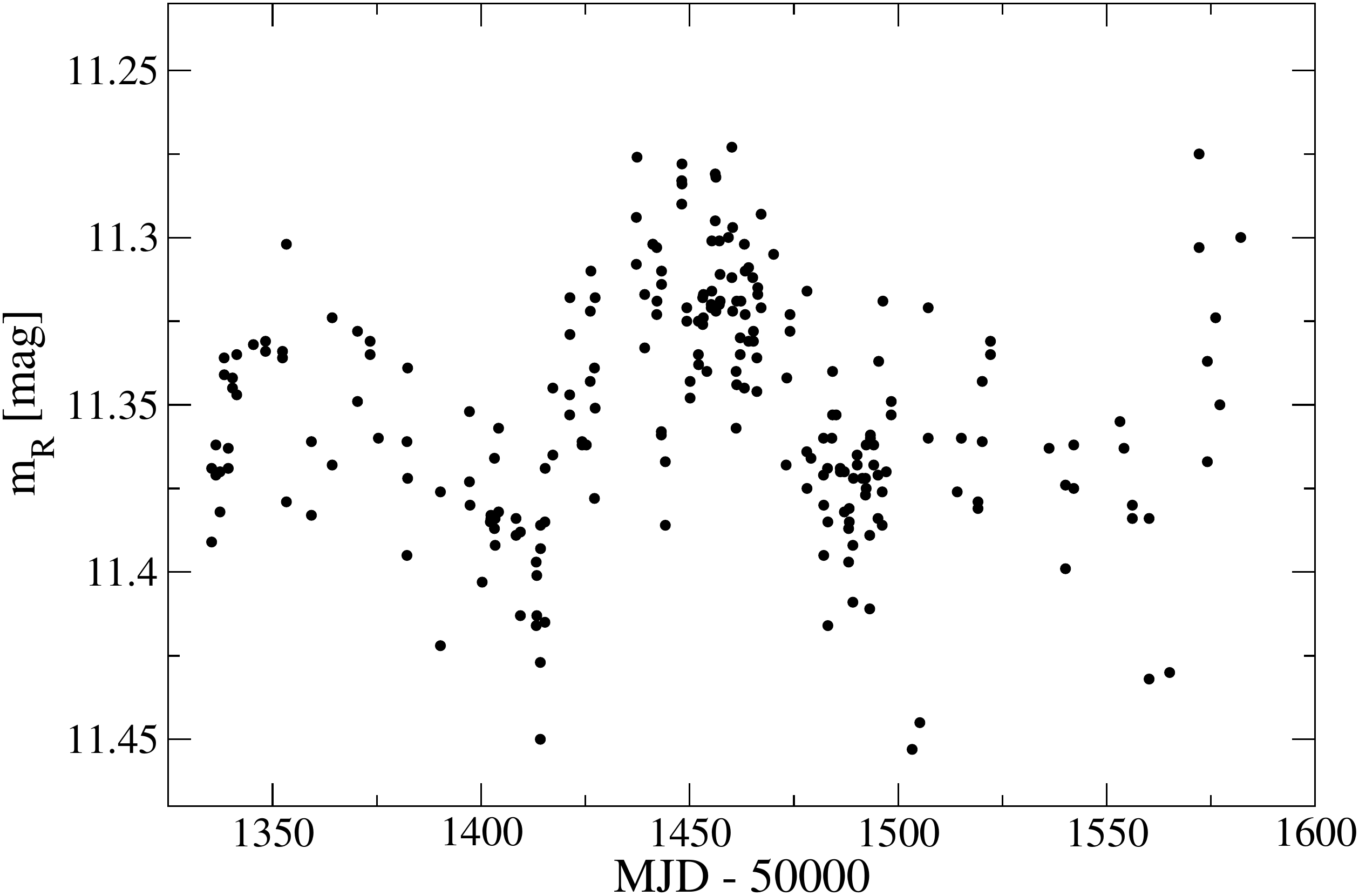}
    \includegraphics[width=0.49\textwidth]{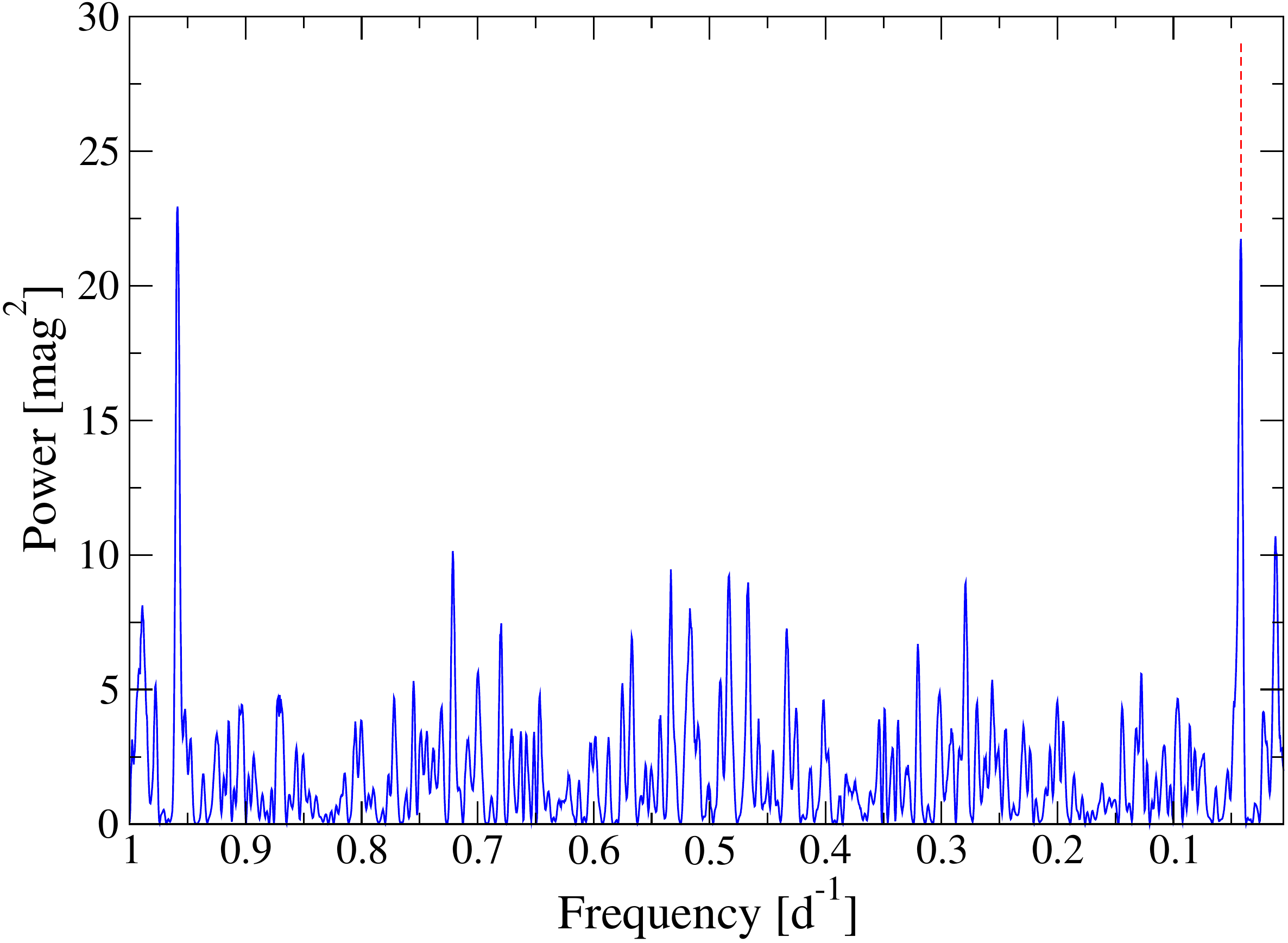}
    \includegraphics[width=0.49\textwidth]{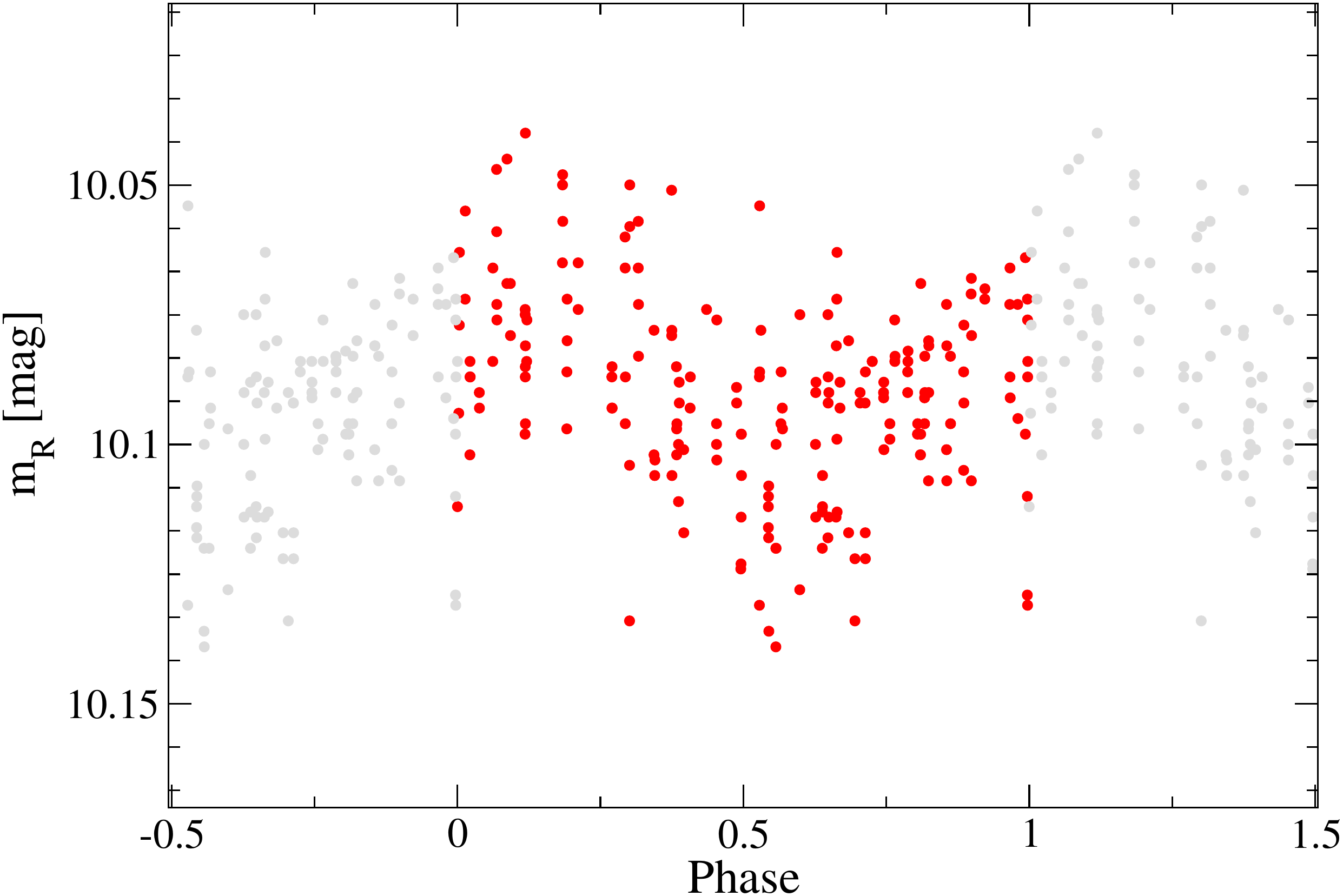}
  \caption{AstroLAB IRIS $R$-band photometric data ({\em top}), Lomb--Scargle periodogram ({\em middle}), and phase-folded rotation curve for $P$=23.9\,d ({\em bottom}) for the M3.0\,V star J09428+700 = \object{GJ~362}.}
  \label{fig.astrolab}
\end{figure}

\begin{figure}
  \centering
    \includegraphics[width=0.49\textwidth]{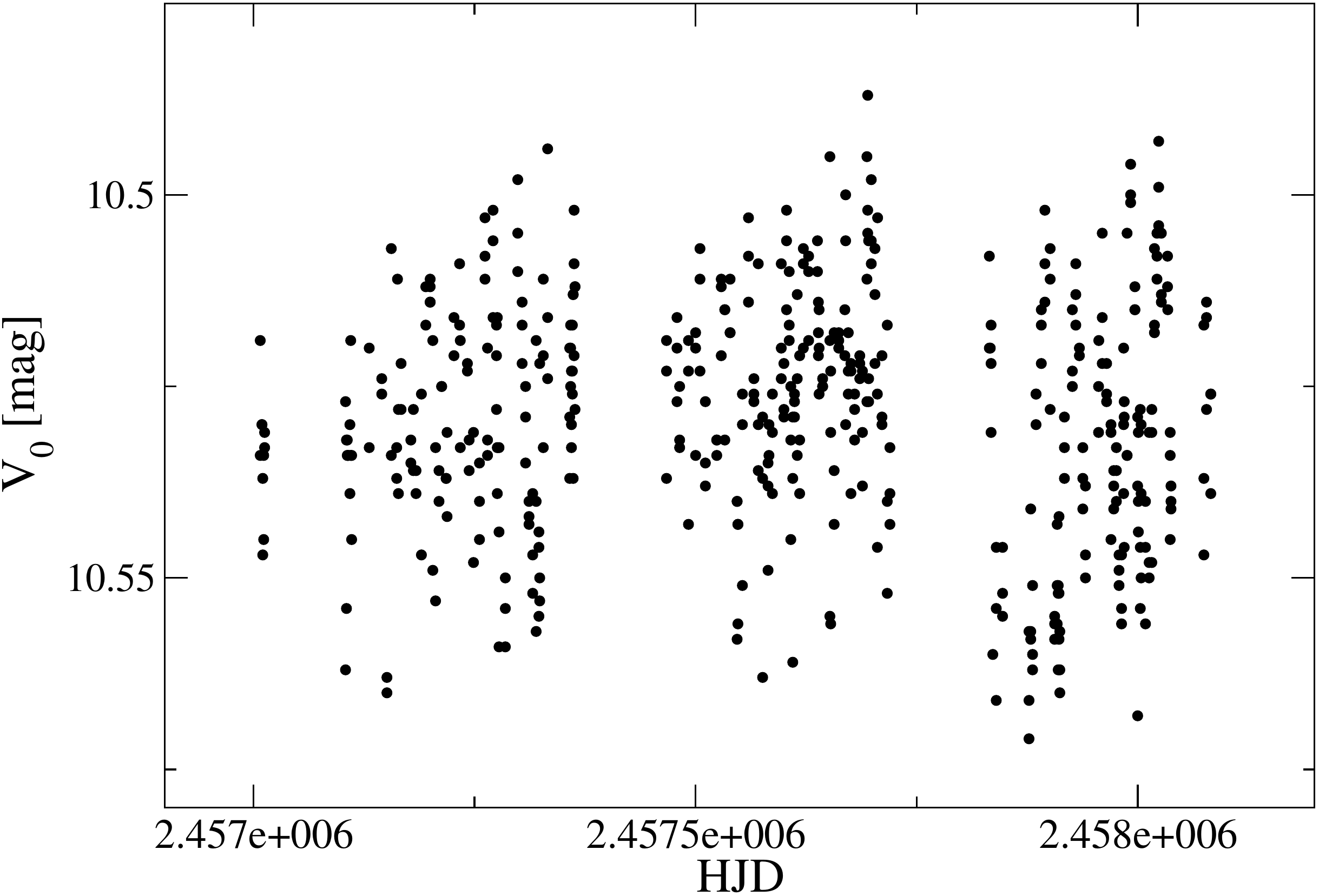}
    \includegraphics[width=0.49\textwidth]{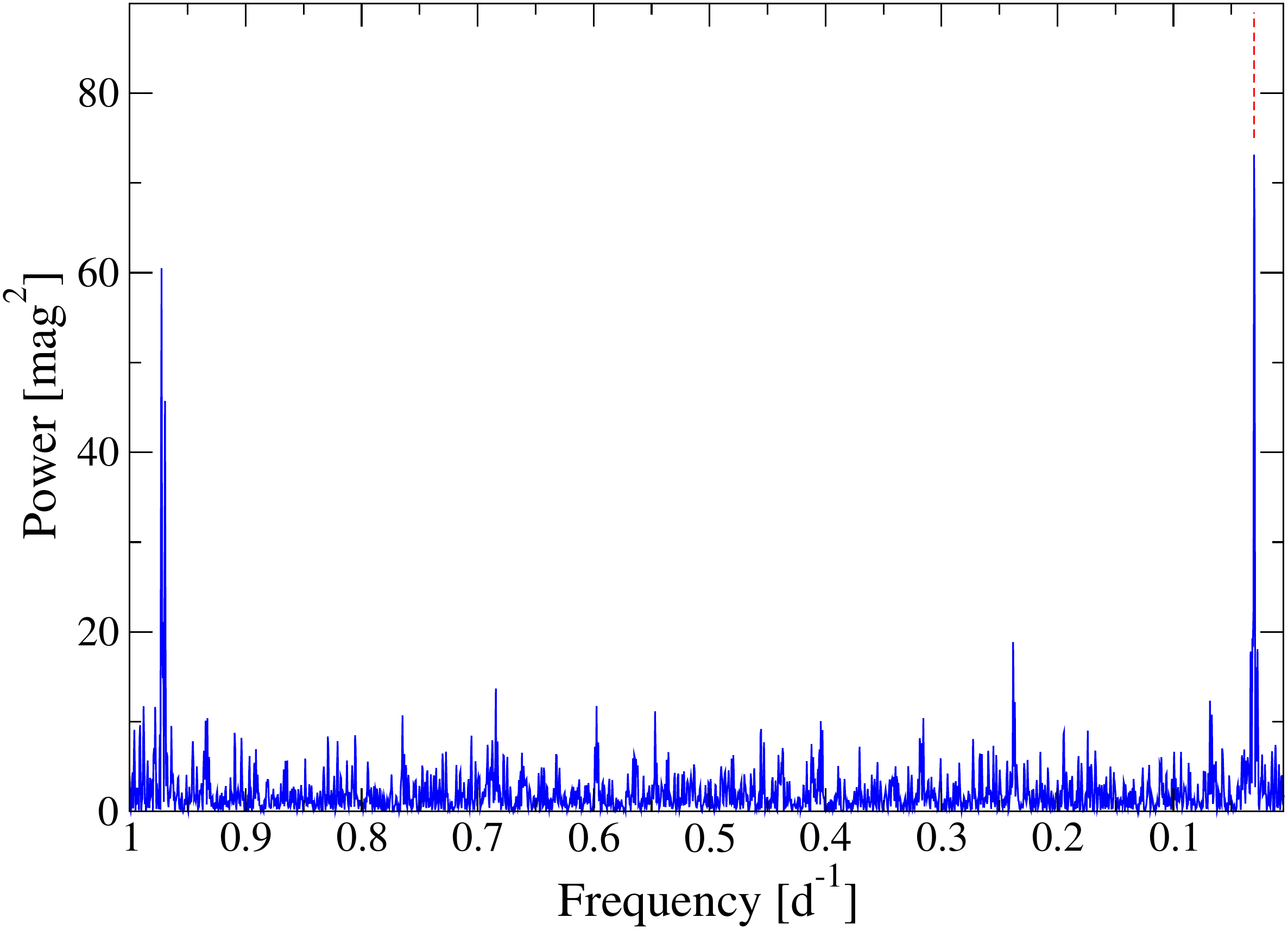}
    \includegraphics[width=0.49\textwidth]{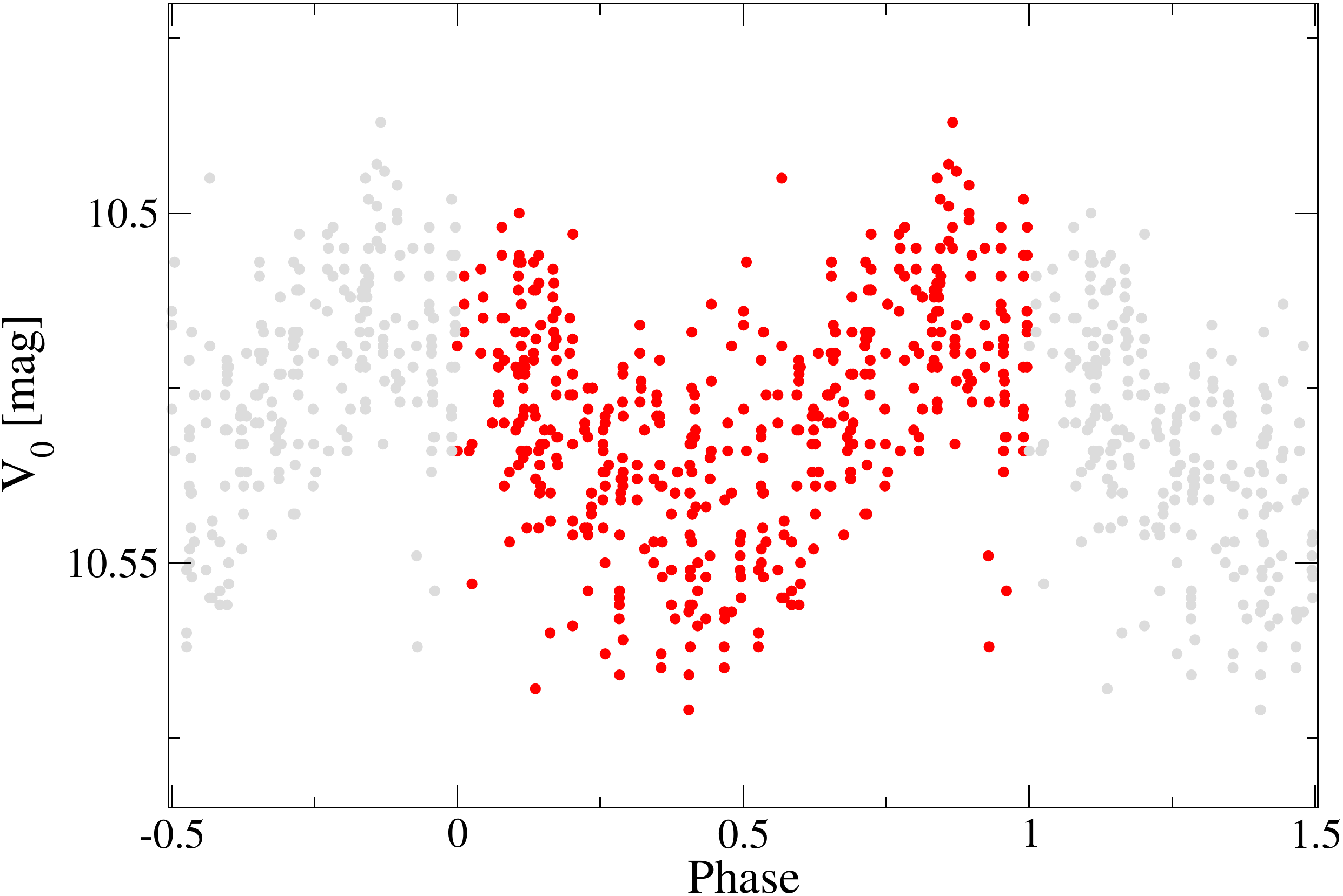}
  \caption{ASAS-SN $V$-band photometric data ({\em top}), Lomb--Scargle periodogram ({\em middle}), and phase-folded rotation curve for $P$=32.8\,d ({\em bottom}) for the M2.5\,V star J20305+654 = \object{GJ 793}.}
  \label{fig.asas-sn}
\end{figure}

\begin{figure}
  \centering
    \includegraphics[width=0.49\textwidth]{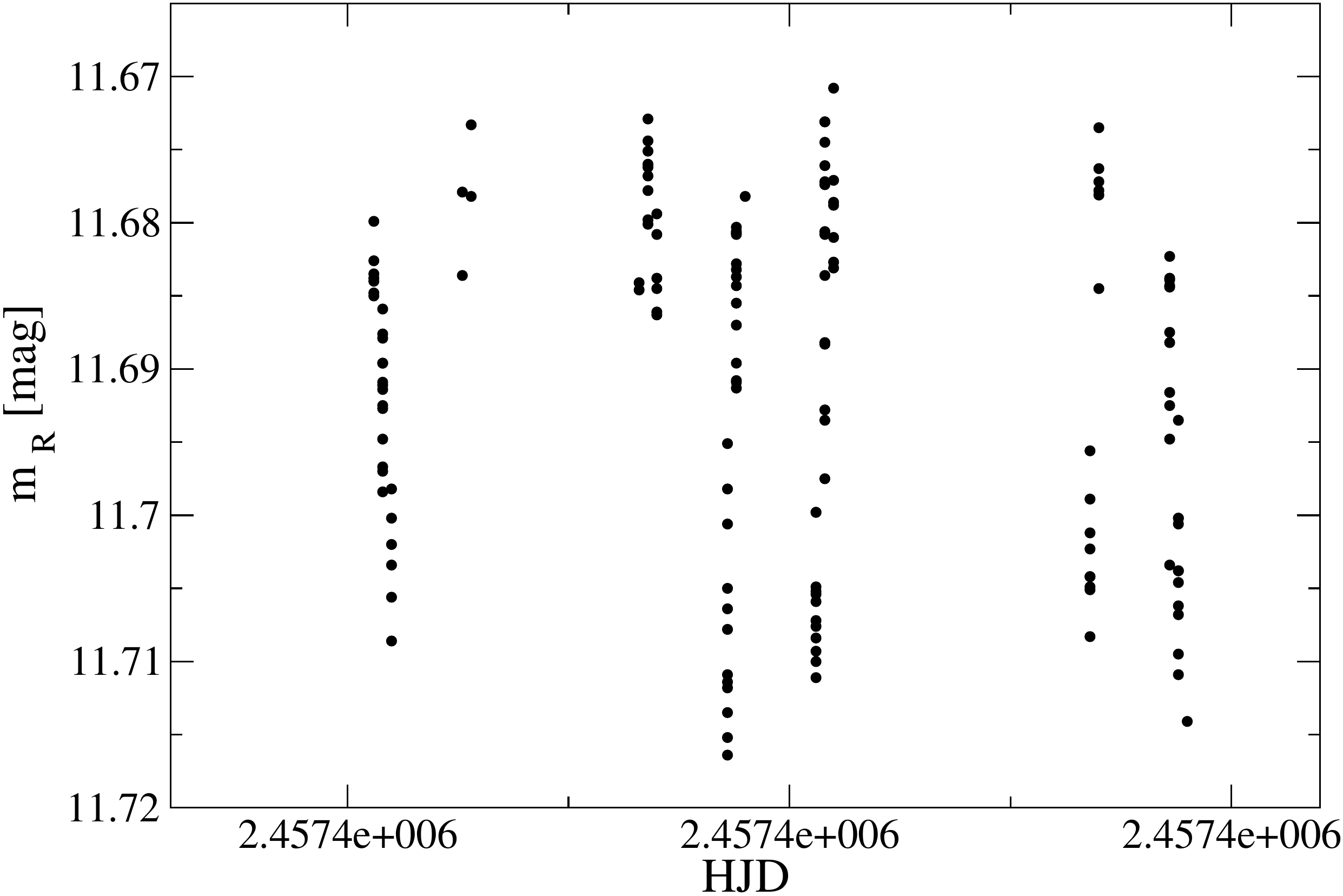}
    \includegraphics[width=0.49\textwidth]{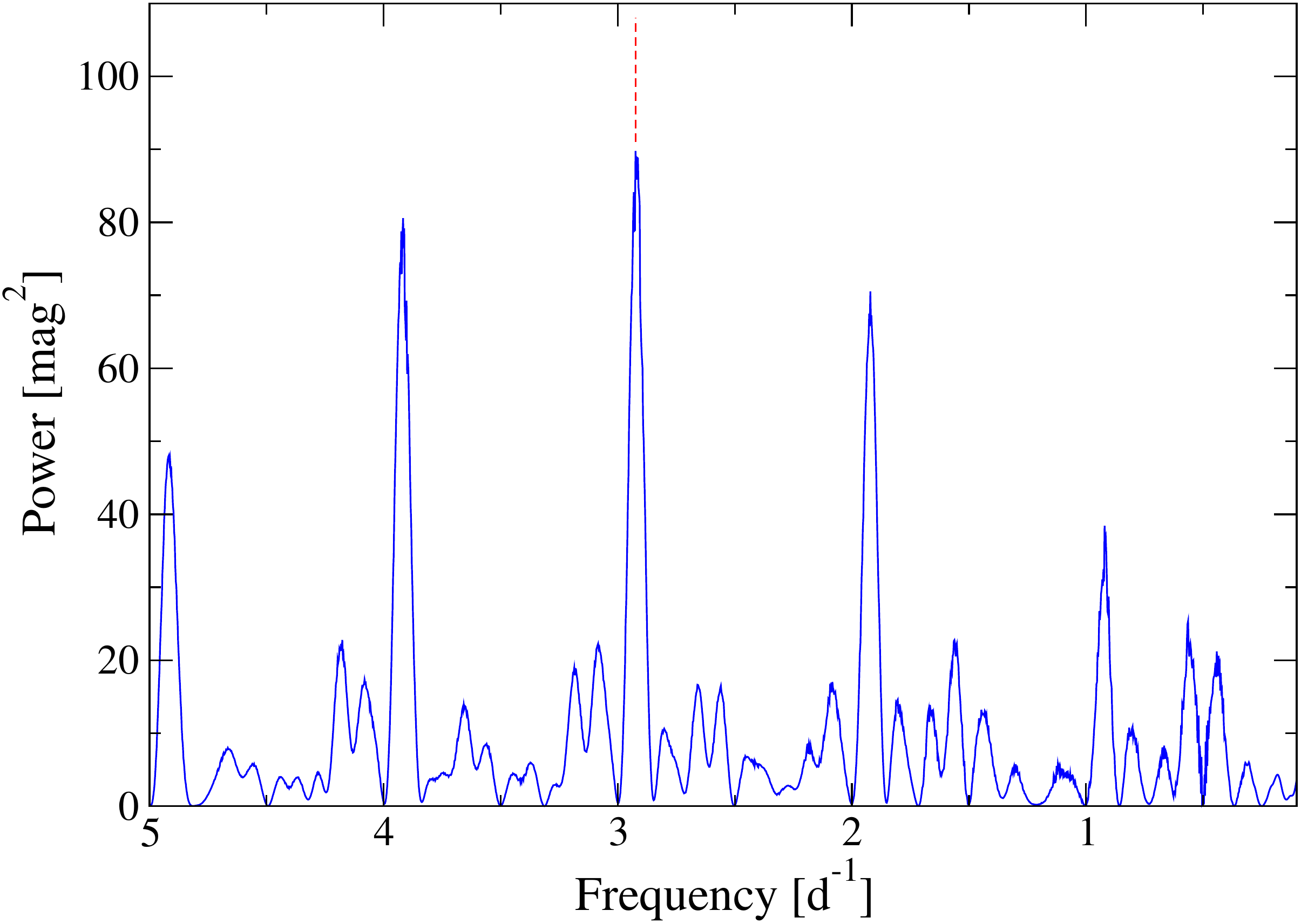}
    \includegraphics[width=0.49\textwidth]{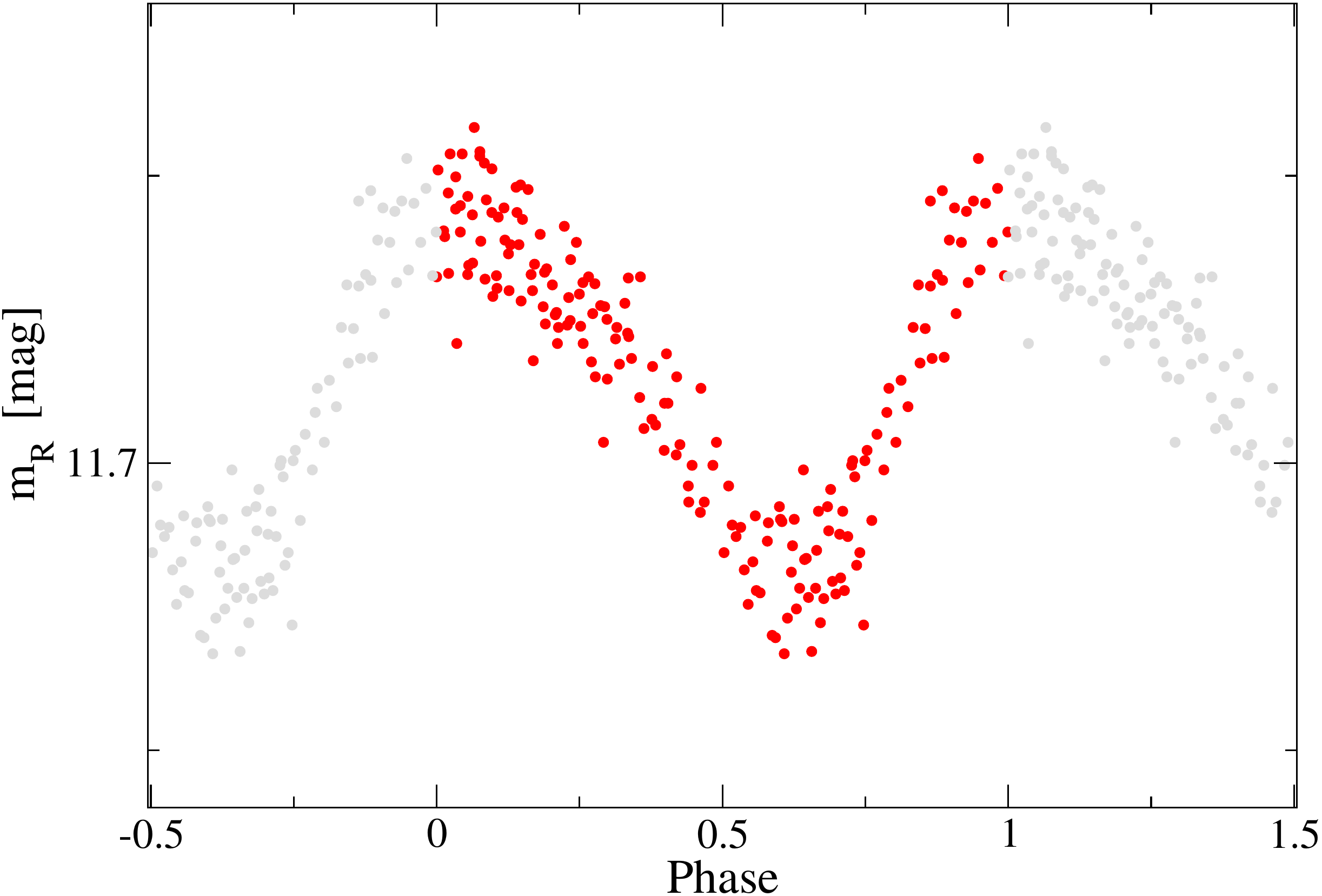}
  \caption{Montcabrer $R$-band photometric data ({\em top}), Lomb--Scargle periodogram ({\em middle}), and phase-folded rotation curve for $P$=0.342\,d ({\em bottom}) for the M5.0\,V star J04772+206 = \object{RX~J0447.2+2038} (compare with Fig.~\ref{fig.K2}).}
  \label{fig.Naves}
\end{figure}

\begin{figure}
  \centering
    \includegraphics[width=0.49\textwidth]{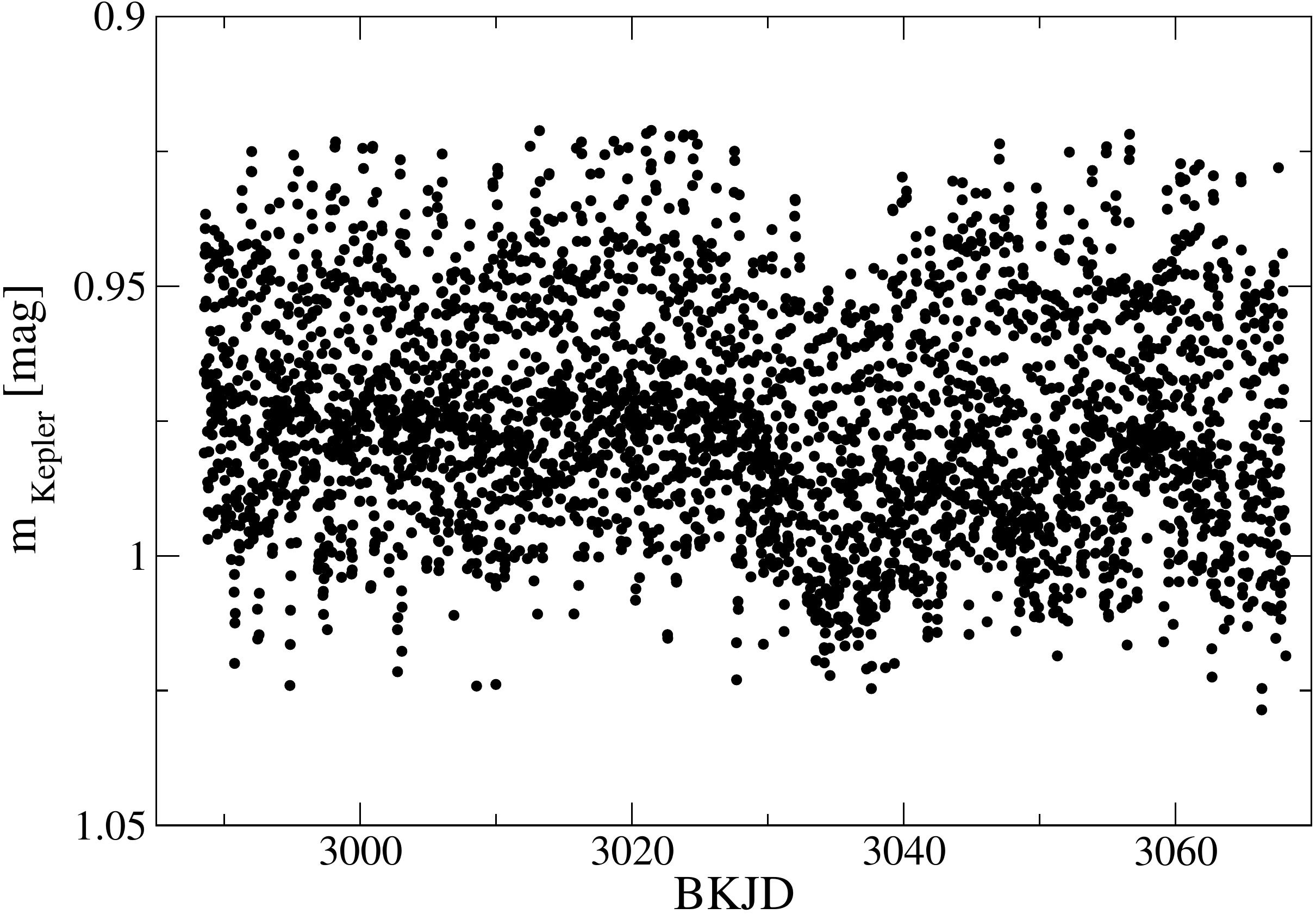}
    \includegraphics[width=0.49\textwidth]{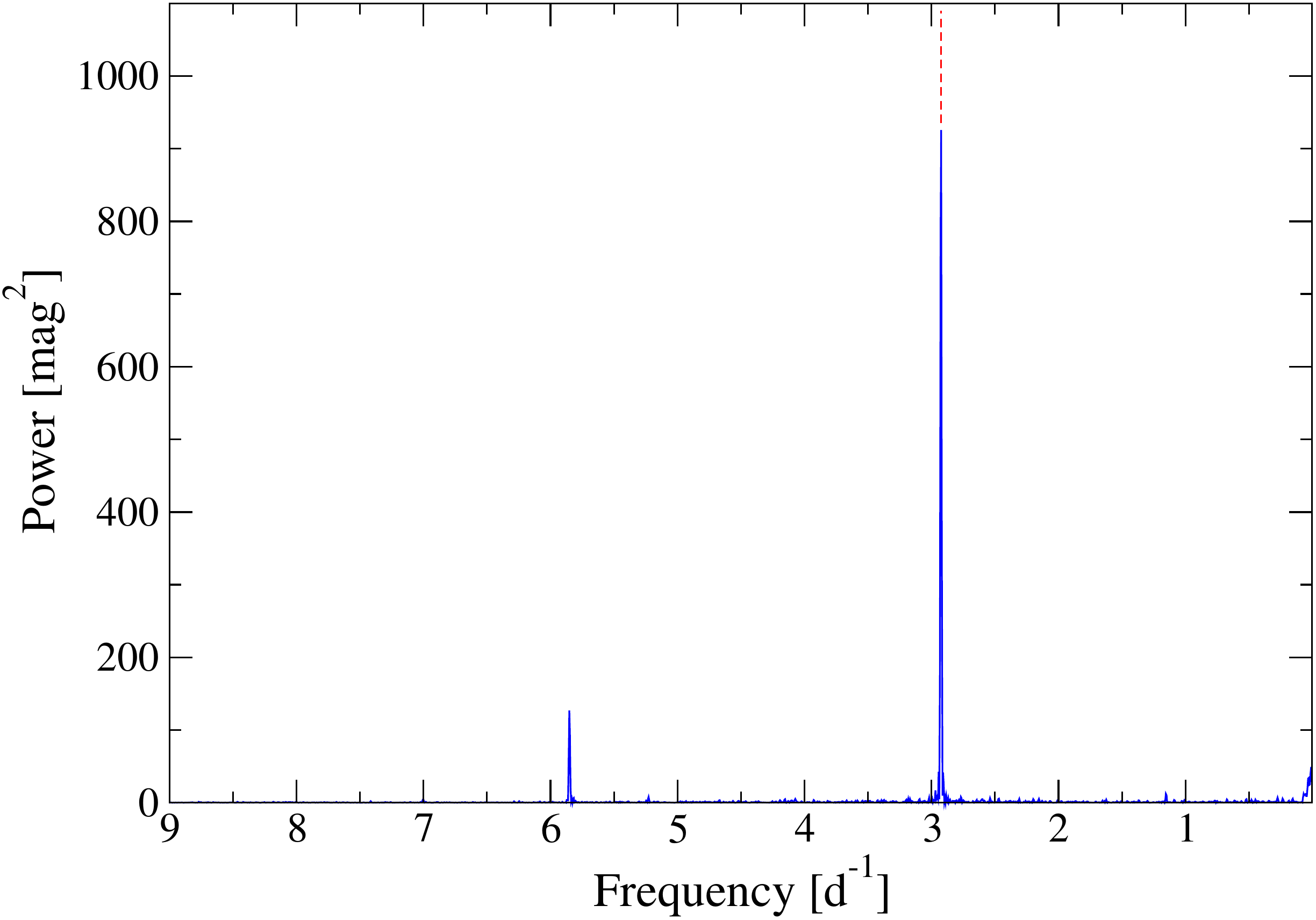}
    \includegraphics[width=0.49\textwidth]{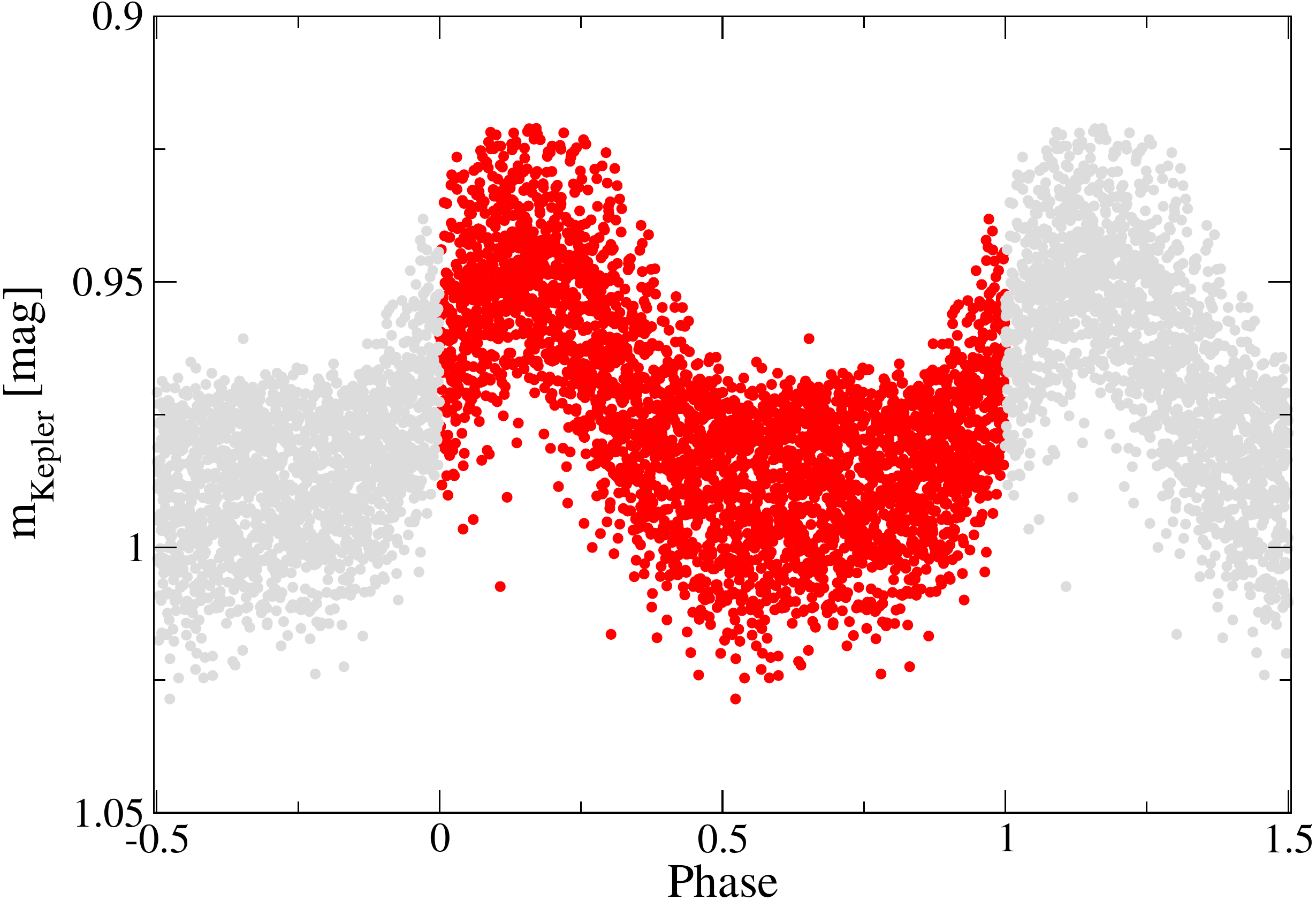}
  \caption{K2 photometric data ({\em top}), Lomb--Scargle periodogram ({\em middle}), and phase-folded rotation curve for $P$=0.342\,d ({\em bottom}) for the M5.0\,V star J04772+206 = \object{RX~J0447.2+2038} (compare with Fig.~\ref{fig.Naves}).}
  \label{fig.K2}
\end{figure}

\begin{figure}
  \centering
    \includegraphics[width=0.49\textwidth]{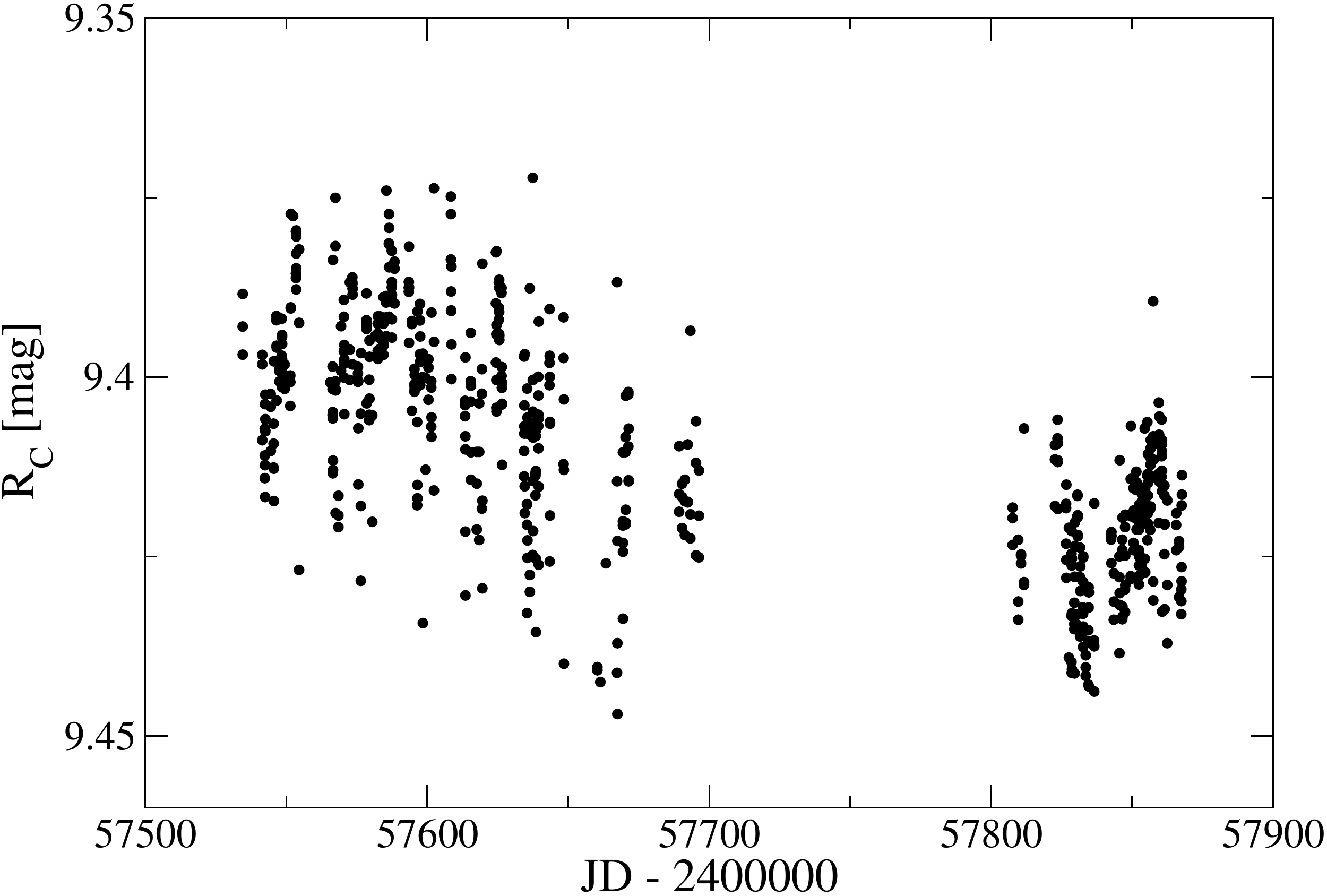}
    \includegraphics[width=0.49\textwidth]{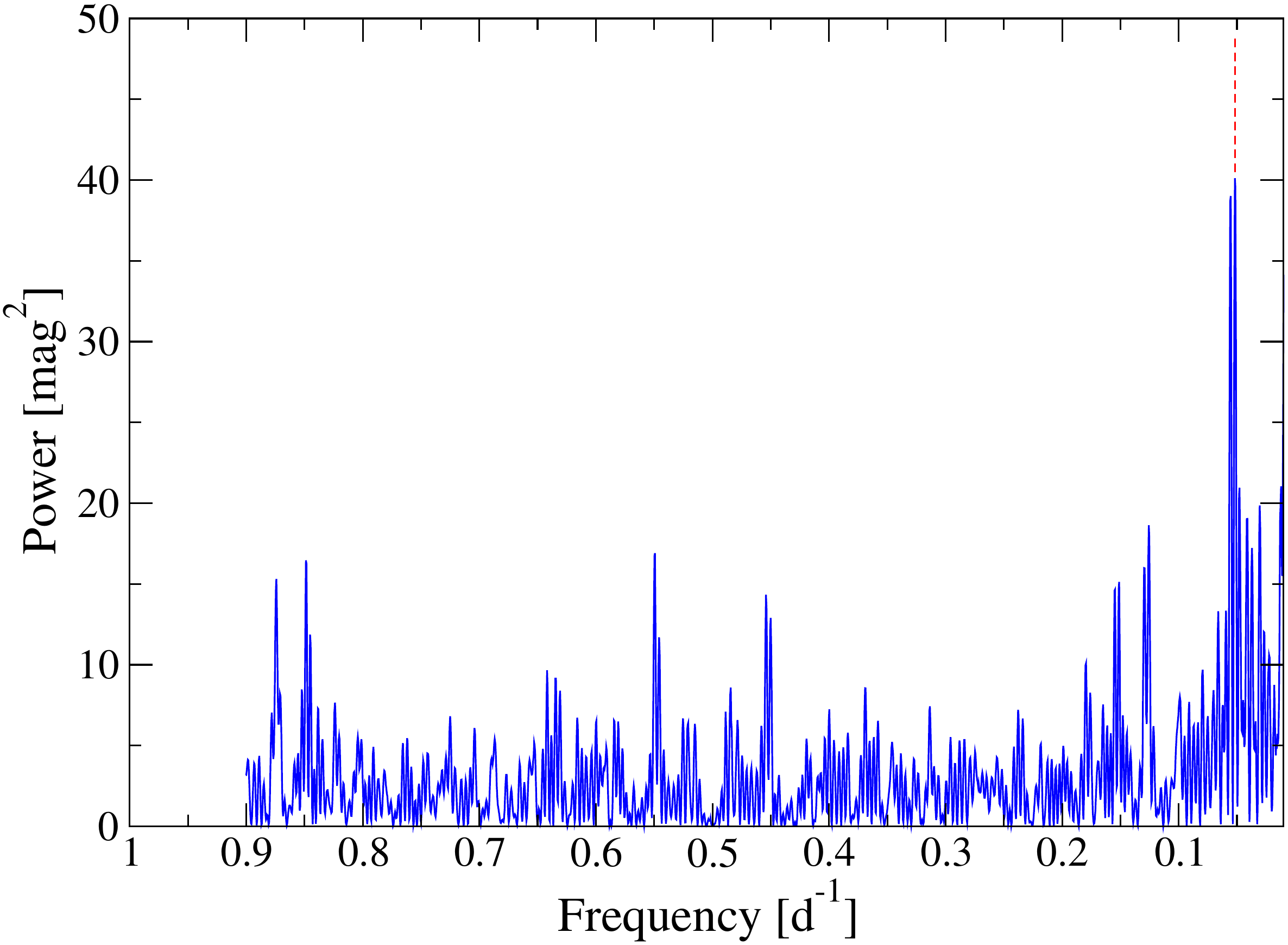}
    \includegraphics[width=0.49\textwidth]{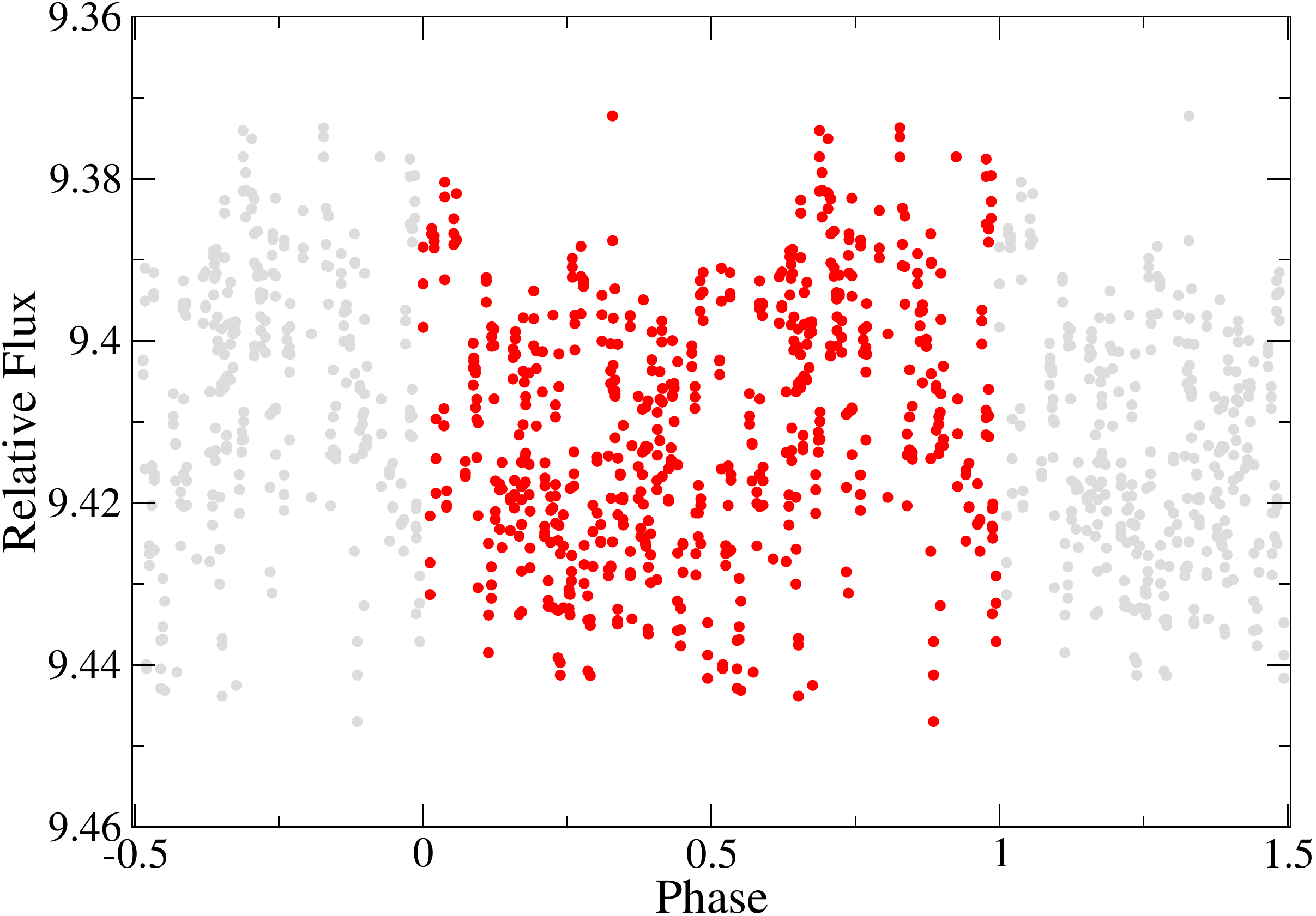}
  \caption{Montsec $R_C$-band photometric data ({\em top}), Lomb--Scargle periodogram ({\em middle}), and phase-folded rotation curve for $P$=19.3\,d ({\em bottom}) for the M0.5\,V star J17355+616 = \object{BD+61~1678C}.}
  \label{fig.montsec}
\end{figure}

\begin{figure}
  \centering
    \includegraphics[width=0.49\textwidth]{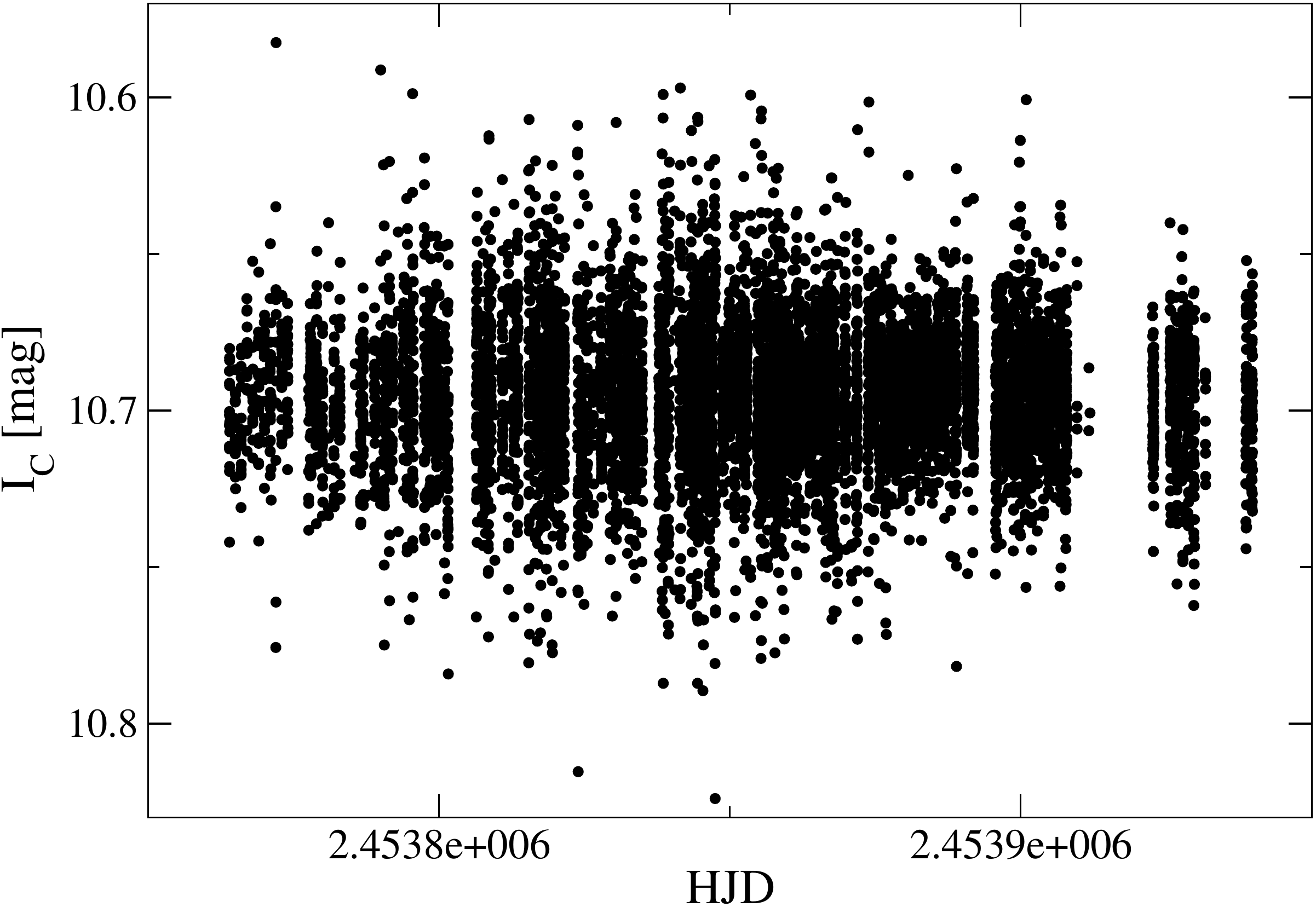}
    \includegraphics[width=0.49\textwidth]{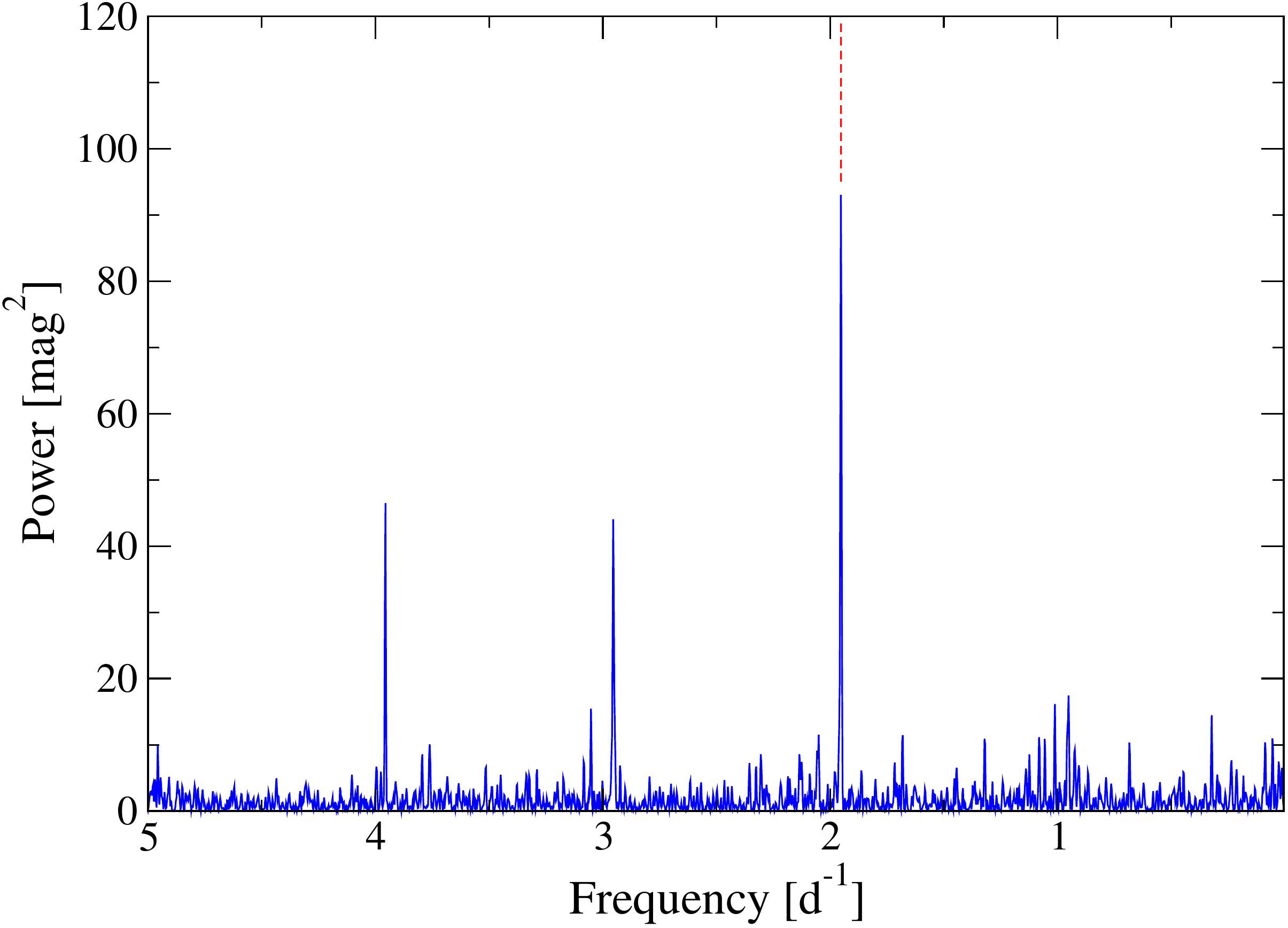}
    \includegraphics[width=0.49\textwidth]{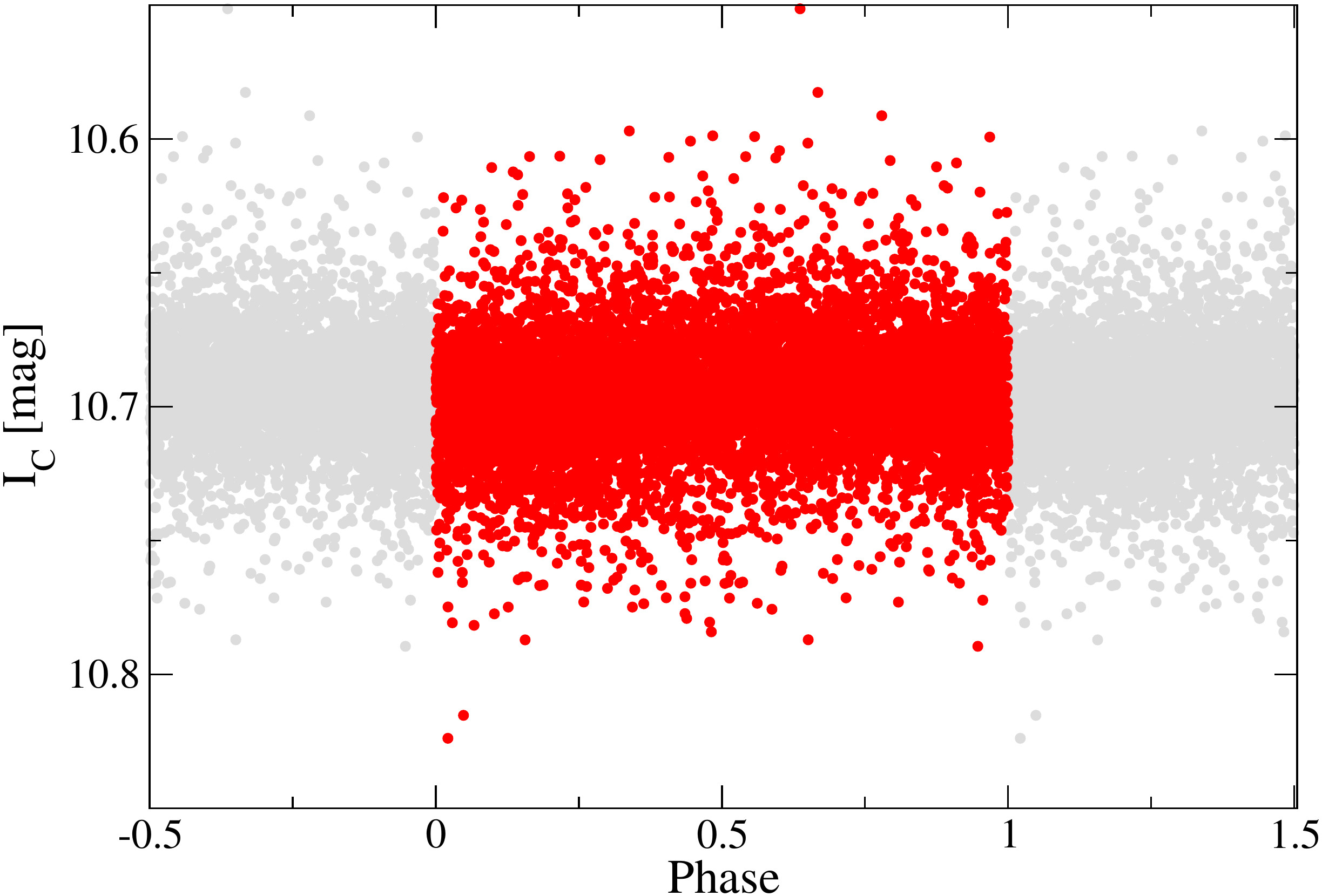}
  \caption{HATNet $I_C$-band photometric data ({\em top}), Lomb--Scargle periodogram ({\em middle}), and phase-folded rotation curve for $P$=0.512\,d ({\em bottom}) for the M5.0\,V star J16313+408 = \object{G~180--060}.}
  \label{fig.hatnet}
\end{figure}


\end{document}

%% file: pkT30.tex
\begin{small}


{\tiny
\begin{longtable}{ll cc lcc ccc}

\label{table.allstars}\\ %
\caption[]{M dwarfs searched for photometric periods.}\\ %
   \hline
   \hline
   \noalign{\smallskip}
Karmn  		& Name				& $\alpha$		& $\delta$		& Survey			& $N_{\rm obs}/N_{\rm used}$	& $\Delta t$	& $\overline{m}$	& $\sigma_m$	& $\overline{\delta m}$	\\
			&					& (J2000.0)		& (J2000.0)	&				&						& [a]			& [mag]			& [mag]		& [mag]				\\ 
\noalign{\smallskip}
    \hline
    \noalign{\smallskip}		
 \endfirsthead
\caption[]{M dwarfs searched for photometric periods (cont.).}\\ %
  \hline
  \hline
  \noalign{\smallskip}		
Karmn  		& Name				& $\alpha$		& $\delta$		& Survey			& $N_{\rm obs}/N_{\rm used}$	& $\Delta t$	& $\overline{m}$	& $\sigma_m$	& $\overline{\delta m}$	\\
			&					& (J2000.0)		& (J2000.0)	&				&						& [a]			& [mag]			& [mag]		& [mag]				\\ 
  \noalign{\smallskip}
  \hline
  \noalign{\smallskip}
  \endhead
  \noalign{\smallskip}
  \hline
  \endfoot
J00051+457	& GJ 2				&	00:05:10.78			&	+45:47:11.6		& SuperWASP		& 5126/4400				& 1.645			&	10.185			&0.052			&0.027					\\
			&					&				&			& NSVS		& 69/69				& 0.397			&	9.300			&	0.022		&	0.010			\\

J00067-075	& GJ 1002 & 00:06:43.26	& --07:32:14.7	&ASAS&181/172&7.863&13.899&0.282&0.122\\
&					&				&			
& Catalina&160/140&4.285&11.897&0.096&0.050\\
&					&				&			
& NSVS&106/105&0.397&11.889&0.032&0.020\\

J00162+198E	& LP 404--062	& 00:16:16.08	& +19:51:51.5	
&ASAS&317/282&6.855&11.690&0.139&0.114\\
&					&				&			
&MEarth-2010&935/910&2.562&0.000&0.007&0.004					\\

J00162+198W	& EZ Psc AB	& 00:16:14.63	& +19:51:37.6 
&MEarth-2010&935/933&0.720&0.000&0.009&0.003\\
\end{longtable}

}
\end{small}


\begin{longtable}{l c c c l c c c l}
\label{table.ourProt}\\ %
\caption[]{Rotation periods obtained for stars in our sample.} \\ 
   \hline
   \hline
   \noalign{\smallskip}
Karmn	& $P_{\rm rot}$ 	& $A$& FAP 	& Survey 		& $P_{\rm rot,lit}$ 	& $A_{\rm lit}$ 	& FAP$_{\rm lit}$ 	& Ref.$^{a}$ 	\\
 		& [d] 				& [mag] 		& (\%)	& 			& [d]				& [mag]				& (\%)			&	 		\\
    \hline
    \noalign{\smallskip}		
 \endfirsthead
\caption[]{Rotation periods obtained for stars in our sample (cont.).}\\ %
  \hline
  \hline
  \noalign{\smallskip}		
Karmn	& $P_{\rm rot}$ 	& $A$& FAP 	& Survey 		& $P_{\rm rot,lit}$ 	& $A_{\rm lit}$ 	& FAP$_{\rm lit}$ 	& Ref.$^{a}$ 	\\
 		& [d] 				& [mag] 		& [\%]	& 			& [d]				& [mag]				& [\%]			&	 		\\
  \noalign{\smallskip}
  \hline
  \noalign{\smallskip}
  \endhead
  \noalign{\smallskip}
  \hline
  \endfoot
J00051+457$^b$ &{15.37$\pm$0.09} & {0.003} &$<$ 10$^{-4}$&SuperWASP&21.2$\pm$0.5 & ... & $<$0.1 &SM17 \\ 
J00162+198E&105$\pm$44&0.006&$<$ 10$^{-4}$&MEarth& 104 & 0.0049 & ... & Irw11 \\
J00162+198W &4.83$\pm$0.06&0.007&$<$ 10$^{-4}$&MEarth& 4.79 & ... & ... & Nor07 \\
J00389+306&50.2$\pm$1.3&0.009&$<$ 10$^{-4}$&SuperWASP& ... & ... & ... & ... \\
J01019+541& {0.2779$\pm$0.0006} & {0.007} &$<$ 10$^{-4}$&MEarth& 0.278 & ... & ... & New16 \\
J01025+716 & 51.5$\pm$2.6 & 0.025 & $<$ 10$^{-4}$ & ASAS-SN & ... & ... & ...&... \\
J01026+623&19.9$\pm$0.4&0.006&$<$ 10$^{-4}$&MEarth&18.4$\pm$0.7& ... & $<$1 &SM17\\
J01033+623&1.02$\pm$0.01&0.051&$<$ 10$^{-4}$&MEarth&1.06&0.046& ... &Irw11\\
\end{longtable}
\begin{list}{}{}
\item[$^{a}$] {\em References}.
GR98: \cite{1998IBVS.4652....1G};
Fek00: \cite{2000AJ....120.3265F};
Tes04: \cite{2004ApJ...617..508T};
Riv05: \cite{2005ApJ...634..625R};
KS07: \cite{2007AcA....57..149K};
Nor07: \cite{2007A&A...467..785N};
Hal08: \cite{2008ApJ...684..644H};
Har11: \cite{2011AJ....141..166H};
Irw11: \cite{2011ApJ...727...56I};
Kir12: \cite{2012AcA....62...67K};
KS13: \cite{2013AcA....63...53K};
SM15: \cite{2015MNRAS.452.2745S};
Wes15: \cite{2015ApJ...812....3W};
Dav16: \cite{2016Natur.534..658D};
New16: \cite{2016ApJ...821...93N};
SM16: \cite{2016A&A...595A..12S};
SM17:\cite{2018A&A...612A..89M};
Clo17: \cite{2017A&A...608A..35C};
Vid17: \cite{2017ApJ...841..124V};
Lot18:\cite{2018AJ....155...66L}.
\item[$^b$] J00051+457 = \object{GJ~2}: the periodogram of the ASAS light curve displays several secondary peaks around 15.37\,d, including at 21.2\,d as found by  \cite{2015MNRAS.452.2745S}.
\item[$^c$] J04472+206 = \object{RX~J0447.2+2038}: we also measured $P_{\rm rot}$ = 0.342$\pm$0.002\,d in the Montcabrer dataset (see Fig~\ref{fig.Naves}).
\item[$^d$] J06318+414 = \object{LP~205--044}: the periodogram of the SuperWASP light curve displays a second peak at 0.482\,d of approximately the same power as at 0.299\,d.
\item[$^e$] J10508+068 = \object{EE~Leo}: we increased the frequency search to less than half of the inverse of the time baseline of about 80\,d, so the tabulated period is uncertain. 
\end{list}


%% file: pk30.bbl
\begin{thebibliography}{}

\bibitem[Alonso et al.(2007)]{2007ASPC..366...13A} Alonso, R., Brown, T.~M., Charbonneau, D., et al.\ 2007, Transiting Extrapolar Planets Workshop, 366, 13 

\bibitem[Alonso-Floriano et al.(2015a)]{2015A&A...577A.128A} Alonso-Floriano, F.~J., Morales, J.~C., Caballero, J.~A., et al.\ 2015a, \aap, 577, A128

\bibitem[Alonso-Floriano et al.(2015b)]{2015A&A...583A..85A} Alonso-Floriano, F.~J., Caballero, J.~A., Cort{\'e}s-Contreras, M., Solano, E., \& Montes, D.\ 2015b, \aap, 583, A85

\bibitem[Anglada-Escud{\'e} et~al.(2016)]{2016Natur.536..437A} Anglada-Escud{\'e}, G., Amado, P.~J., Barnes, J., et~al.\ 2016, \nat, 536, 437

\bibitem[Angus et al.(2018)]{2018MNRAS.474.2094A} Angus, R., Morton, T., Aigrain, S., Foreman-Mackey, D., \& Rajpaul, V.\ 2018, \mnras, 474, 2094 

\bibitem[Artigau et~al.(2014)]{2014SPIE.9147E..15A} Artigau, {\'E}., Kouach, D., Donati, J.-F., et~al.\ 2014, \procspie, 9147, 914715 

\bibitem[Auvergne et al.(2009)]{2009A&A...506..411A} Auvergne, M., Bodin, P., Boisnard, L., et al.\ 2009, \aap, 506, 411 

\bibitem[Baliunas \& Vaughan(1985)]{1985ARA&A..23..379B} Baliunas, S.~L., \& Vaughan, A.~H.\ 1985, \araa, 23, 379

\bibitem[Baliunas et~al.(1995)]{1995ApJ...438..269B} Baliunas, S.~L., Donahue, R.~A., Soon, W.~H., et~al.\ 1995, \apj, 438, 269

\bibitem[Baliunas et al.(1996)]{1996ApJ...460..848B} Baliunas, S.~L., Nesme-Ribes, E., Sokoloff, D., \& Soon, W.~H.\ 1996, \apj, 460, 848

\bibitem[Baluev et al.(2015)]{2015MNRAS.450.3101B} Baluev, R.~V., Sokov, E.~N., Shaidulin, V.~S., et al.\ 2015, \mnras, 450, 3101

\bibitem[Bakos et al.(2004)]{2004PASP..116..266B} Bakos, G., Noyes, R.~W., Kov{\'a}cs, G., et al.\ 2004, \pasp, 116, 266

\bibitem[Bakos et al.(2013)]{2013PASP..125..154B} Bakos, G.~{\'A}., Csubry, Z., Penev, K., et al.\ 2013, \pasp, 125, 154 

\bibitem[Bakos(2018)]{2018arXiv180100849B} Bakos, G.~{\'A}.\ 2018, Handbook of Exoplanets, eds. H. Deeg \& J. A. Belmonte, in press (eprint arXiv:1801.00849) 

\bibitem[Barnes et al.(2011)]{2011MNRAS.412.1599B} Barnes, J.~R., Jeffers, S.~V., \& Jones, H.~R.~A.\ 2011, \mnras, 412, 1599

\bibitem[Baroch et al.(2018)]{2018arXiv180806895B} Baroch, D., Morales, J.~C., Ribas, I., et al.\ 2018, arXiv:1808.06895 

\bibitem[Barros et al.(2011)]{2011A&A...525A..54B} Barros, S.~C.~C., Faedi, F., Collier Cameron, A., et al.\ 2011, \aap, 525, A54 

\bibitem[Berdyugina \& J{\"a}rvinen(2005)]{2005AN....326..283B} Berdyugina, S.~V., \& J{\"a}rvinen, S.~P.\ 2005, AN, 326, 283

\bibitem[Berta et~al.(2013)]{2013ApJ...775...91B} Berta, Z.~K., Irwin, J., \& Charbonneau, D.\ 2013, \apj, 775, 91 

\bibitem[Boisse et~al.(2011)]{2011IAUS..273..281B} Boisse, I., Bouchy, F., H{\'e}brard, G., et~al.\ 2011, Physics of Sun and Star Spots, 273, 281 

\bibitem[Bonfils et~al.(2005)]{2005A&A...443L..15B} Bonfils, X., Forveille, T., Delfosse, X., et~al.\ 2005, \aap, 443, L15 

\bibitem[Borucki et al.(2010)]{2010Sci...327..977B} Borucki, W.~J., Koch, D., Basri, G., et al.\ 2010, Science, 327, 977 

\bibitem[Bouvier et~al.(1993)]{1993A&A...272..176B} Bouvier, J., Cabrit, S., Fernandez, M., Martin, E.~L., \& Matthews, J.~M.\ 1993, \aap, 272, 176

\bibitem[Brown et al.(2013)]{2013PASP..125.1031B} Brown, T.~M., Baliber, N., Bianco, F.~B., et al.\ 2013, \pasp, 125, 1031


\bibitem[Browning(2008)]{2008ApJ...676.1262B} Browning, M.~K.\ 2008, \apj, 676, 1262

\bibitem[Browning et al.(2010)]{2010AJ....139..504B} Browning, M.~K., Basri, G., Marcy, G.~W., West, A.~A., \& Zhang, J.\ 2010, \aj, 139, 504 

\bibitem[Butler et~al.(2004)]{2004ApJ...617..580B} Butler, R.~P., Vogt, S.~S., Marcy, G.~W., et~al.\ 2004, \apj, 617, 580

\bibitem[Butters et al.(2010)]{2010A&A...520L..10B} Butters, O.~W., West, R.~G., Anderson, D.~R., et al.\ 2010, \aap, 520, L10

\bibitem[Caballero et al.(2010)]{2010AN....331..257C} Caballero, J.~A., Cornide, M., \& de Castro, E.\ 2010, AN, 331, 257

\bibitem[Caballero et al.(2016)]{2016csss.confE.148C} Caballero, J.~A., Cort{\'e}s-Contreras, M., Alonso-Floriano, F.~J., et al.\ 2016, 19th Cambridge Workshop on Cool Stars, Stellar Systems, and the Sun (CS19), 148

\bibitem[Chabrier et al.(2007)]{2007A&A...472L..17C} Chabrier, G., Gallardo, J., \& Baraffe, I.\ 2007, \aap, 472, L17

\bibitem[Charbonneau et al.(2009)]{2009Natur.462..891C} Charbonneau, D., Berta, Z.~K., Irwin, J., et al.\ 2009, \nat, 462, 891 

\bibitem[Claudi et~al.(2016)]{2016SPIE.9908E..1AC} Claudi, R., Benatti, S., Carleo, I., et~al.\ 2016, \procspie, 9908, 99081A

\bibitem[Cloutier et al.(2017)]{2017A&A...608A..35C} Cloutier, R., Astudillo-Defru, N., Doyon, R., et al.\ 2017, \aap, 608, A35

\bibitem[Correia et al.(2010)]{2010A&A...511A..21C} Correia, A.~C.~M., Couetdic, J., Laskar, J., et al.\ 2010, \aap, 511, A21

\bibitem[Cort{\'e}s-Contreras et al.(2017)]{2017A&A...597A..47C} Cort{\'e}s-Contreras, M., B{\'e}jar, V.~J.~S., Caballero, J.~A., et al.\ 2017, \aap, 597, A47

\bibitem[Collier Cameron et al.(2007)]{2007MNRAS.375..951C} Collier Cameron, A., Bouchy, F., H{\'e}brard, G., et al.\ 2007, \mnras, 375, 951 

\bibitem[Collins et al.(2017)]{2017AJ....153...77C} Collins, K.~A., Kielkopf, J.~F., Stassun, K.~G., \& Hessman, F.~V.\ 2017, \aj, 153, 77

\bibitem[Christensen et al.(2015)]{2015DPS....4730819C} Christensen, E.~J., Carson Fuls, D., Gibbs, A.~R., et al.\ 2015, AAS/Division for Planetary Sciences Meeting Abstracts, 47, 308.19 

\bibitem[Crossfield et al.(2015)]{2015ApJ...804...10C} Crossfield, I.~J.~M., Petigura, E., Schlieder, J.~E., et al.\ 2015, \apj, 804, 10 

\bibitem[Cumming(2004)]{2004MNRAS.354.1165C} Cumming, A.\ 2004, \mnras, 354, 1165

\bibitem[Dahn et al.(1985)]{1985IBVS.2796....1D} Dahn, C., Green, R., Keel, W., et al.\ 1985, Information Bulletin on Variable Stars, 2796, 1 

\bibitem[David et al.(2016)]{2016Natur.534..658D} David, T.~J., Hillenbrand, L.~A., Petigura, E.~A., et al.\ 2016, \nat, 534, 658 

\bibitem[Delfosse et~al.(1998)]{1998A&A...331..581D} Delfosse, X., Forveille, T., Perrier, C., \& Mayor, M.\ 1998, \aap, 331, 581

\bibitem[Distefano et al.(2016)]{2016A&A...591A..43D} Distefano, E., Lanzafame, A.~C., Lanza, A.~F., Messina, S., \& Spada, F.\ 2016, \aap, 591, A43

\bibitem[Distefano et al.(2017)]{2017A&A...606A..58D} Distefano, E., Lanzafame, A.~C., Lanza, A.~F., Messina, S., \& Spada, F.\ 2017, \aap, 606, A58


\bibitem[Dittmann et al.(2017)]{2017Natur.544..333D} Dittmann, J.~A., Irwin, J.~M., Charbonneau, D., et al.\ 2017, \nat, 544, 333

\bibitem[Donati et al.(2006)]{2006Sci...311..633D} Donati, J.-F., Forveille, T., Collier Cameron, A., et al.\ 2006, Science, 311, 633

\bibitem[Drake et~al.(2009)]{2009ApJ...696..870D} Drake, A.~J., Djorgovski, S.~G., Mahabal, A., et~al.\ 2009, \apj, 696, 870

\bibitem[Drake et al.(2014)]{2014ApJS..213....9D} Drake, A.~J., Graham, M.~J., Djorgovski, S.~G., et al.\ 2014, \apjs, 213, 9

\bibitem[Dressing \& Charbonneau(2015)]{2015ApJ...807...45D} Dressing, C.~D., \& Charbonneau, D.\ 2015, \apj, 807, 45 

\bibitem[Dumusque et~al.(2012)]{2012Natur.491..207D} Dumusque, X., Pepe, F., Lovis, C., et~al.\ 2012, \nat, 491, 207

\bibitem[Dupuy \& Liu(2012)]{2012ApJS..201...19D} Dupuy, T.~J., \& Liu, M.~C.\ 2012, \apjs, 201, 19

\bibitem[Favata et al.(2000)]{2000A&A...353..987F} Favata, F., Reale, F., Micela, G., et al.\ 2000, \aap, 353, 987 

\bibitem[Fekel \& Henry(2000)]{2000AJ....120.3265F} Fekel, F.~C., \& Henry, G.~W.\ 2000, \aj, 120, 3265 

\bibitem[Foreman-Mackey et al.(2017)]{2017AJ....154..220F} Foreman-Mackey, D., Agol, E., Ambikasaran, S., \& Angus, R.\ 2017, \aj, 154, 220 

\bibitem[France et~al.(2013)]{2013ApJ...763..149F} France, K., Froning, C.~S., Linsky, J.~L., et~al.\ 2013, \apj, 763, 149 

\bibitem[Freedman \& Diaconis(1981)]{FD81} Freedman, D. \& Diaconis, P. 1981, Probability Theory and Related Field, 57, 4, 453


\bibitem[Fuhrmeister et al.(2018)]{2018A&A...615A..14F} Fuhrmeister, B., Czesla, S., Schmitt, J.~H.~M.~M., et al.\ 2018, \aap, 615, A14 

\bibitem[Gillon et al.(2007)]{2007A&A...472L..13G} Gillon, M., Pont, F., Demory, B.-O., et al.\ 2007, \aap, 472, L13 

\bibitem[Gillon et al.(2012)]{2012A&A...542A...4G} Gillon, M., Triaud, A.~H.~M.~J., Fortney, J.~J., et al.\ 2012, \aap, 542, A4 

\bibitem[Gillon et al.(2016)]{2016Natur.533..221G} Gillon, M., Jehin, E., Lederer, S.~M., et al.\ 2016, \nat, 533, 221

\bibitem[Gillon et al.(2017)]{2017Natur.542..456G} Gillon, M., Triaud, A.~H.~M.~J., Demory, B.-O., et al.\ 2017, \nat, 542, 456

\bibitem[Gomes da Silva et~al.(2012)]{2012A&A...541A...9G} Gomes da Silva, J., Santos, N.~C., Bonfils, X., et~al.\ 2012, \aap, 541, A9

\bibitem[Greimel \& Robb(1998)]{1998IBVS.4652....1G} Greimel, R., \& Robb, R.~M.\ 1998, Information Bulletin on Variable Stars, 4652, 1 

\bibitem[Hallinan et al.(2008)]{2008ApJ...684..644H} Hallinan, G., Antonova, A., Doyle, J.~G., et al.\ 2008, \apj, 684, 644

\bibitem[Haro \& Chavira(1966)]{1966VA......8...89H} Haro, G., \& Chavira, E.\ 1966, Vistas in Astronomy, 8, 89 
 
\bibitem[Hartman et al.(2011)]{2011AJ....141..166H} Hartman, J.~D., Bakos, G.~{\'A}., Noyes, R.~W., et al.\ 2011, \aj, 141, 166

\bibitem[Hawley et al.(1996)]{1996AJ....112.2799H} Hawley, S.~L., Gizis, J.~E., \& Reid, I.~N.\ 1996, \aj, 112, 2799 

\bibitem[Heinze et al.(2018)]{2018arXiv180402132H} Heinze, A.~N., Tonry, J.~L., Denneau, L., et al.\ 2018, ApJ, subm., eprint arXiv:1804.02132

\bibitem[Herbig(1956)]{1956PASP...68..531H} Herbig, G.~H.\ 1956, \pasp, 68, 531 

\bibitem[Herrero et~al.(2011)]{2011A&A...526L..10H} Herrero, E., Morales, J.~C., Ribas, I., \& Naves, R.\ 2011, \aap, 526, L10 

\bibitem[Horne \& Baliunas(1986)]{1986ApJ...302..757H} Horne, J.~H., \& Baliunas, S.~L.\ 1986, \apj, 302, 757

\bibitem[Howell et al.(2014)]{2014PASP..126..398H} Howell, S.~B., Sobeck, C., Haas, M., et al.\ 2014, \pasp, 126, 398

\bibitem[Irwin et al.(2007)]{2007MNRAS.375.1449I} Irwin, J., Irwin, M., Aigrain, S., et al.\ 2007, \mnras, 375, 1449

\bibitem[Irwin et al.(2009)]{2009IAUS..253...37I} Irwin, J., Charbonneau, D., Nutzman, P., \& Falco, E.\ 2009, Transiting Planets, IAU symposium, 253, 37

\bibitem[Irwin et~al.(2011)]{2011ApJ...727...56I} Irwin, J., Berta, Z.~K., Burke, C.~J., et~al.\ 2011, \apj, 727, 56

\bibitem[Jankovics et al.(1978)]{1978IBVS.1454....1J} Jankovics, I., Tsvetkova, K.~P., \& Tsvetkov, M.~K.\ 1978, Information Bulletin on Variable Stars, 1454, 1 

\bibitem[Jeffers et al.(2018)]{2018A&A...614A..76J} Jeffers, S.~V., Sch{\"o}fer, P., Lamert, A., et al.\ 2018, \aap, 614, A76

\bibitem[Johnson et~al.(2007)]{2007ApJ...670..833J} Johnson, J.~A., Butler, R.~P., Marcy, G.~W., et~al.\ 2007, \apj, 670, 833 

\bibitem[Joshi et~al.(1997)]{1997Icar..129..450J} Joshi, M.~M., Haberle, R.~M., \& Reynolds, R.~T.\ 1997, \icarus, 129, 450

\bibitem[Kaminski et al.(2018)]{2018arXiv180801183K} Kaminski, A., Trifonov, T., Caballero, J.~A., et al.\ 2018, \aap, in press, eprint arXiv:1808.01183

\bibitem[Kiraga \& Stepien(2007)]{2007AcA....57..149K} Kiraga, M., \& Stepien, K.\ 2007, \actaa, 57, 149

\bibitem[Kiraga(2012)]{2012AcA....62...67K} Kiraga, M.\ 2012, \actaa, 62, 67

\bibitem[Kiraga \& St{\c e}pie{\'n}(2013)]{2013AcA....63...53K} Kiraga, M., \& St{\c e}pie{\'n}, K.\ 2013, \actaa, 63, 53 

\bibitem[Kochanek et al.(2017)]{2017PASP..129j4502K} Kochanek, C.~S., Shappee, B.~J., Stanek, K.~Z., et al.\ 2017, \pasp, 129, 104502 

\bibitem[Kopparapu et~al.(2013)]{2013ApJ...765..131K} Kopparapu, R.~K., Ramirez, R., Kasting, J.~F., et~al.\ 2013, \apj, 765, 131

\bibitem[Korhonen et al.(2010)]{2010AN....331..772K} Korhonen, H., Vida, K., Husarik, M., et al.\ 2010, AN, 331, 772 

\bibitem[Kron(1947)]{1947PASP...59..261K} Kron, G.~E.\ 1947, \pasp, 59, 261

\bibitem[K{\"u}ker et al.(2018)]{2018arXiv180402925K} K{\"u}ker, M., R{\"u}diger, G., Ol{\'a}h, K., \& Strassmeier, K.~G.\ 2018, eprint arXiv:1804.02925

\bibitem[L{\'e}ger et~al.(2009)]{2009A&A...506..287L} L{\'e}ger, A., Rouan, D., Schneider, J., et~al.\ 2009, \aap, 506, 287

\bibitem[Liebert et al.(1978)]{1978PASP...90..718L} Liebert, J., Kron, R.~G., \& Spinrad, H.\ 1978, \pasp, 90, 718

\bibitem[Lothringer et al.(2018)]{2018arXiv180100412L} Lothringer, J.~D., Benneke, B., Crossfield, I.~J.~M., et al.\ 2018, ApJ, subm., eprint arXiv:1801.00412

\bibitem[Luger et al.(2017)]{2017NatAs...1E.129L} Luger, R., Sestovic, M., Kruse, E., et al.\ 2017, Nature Astronomy, 1, 0129 

\bibitem[Mahadevan et~al.(2014)]{2014SPIE.9147E..1GM} Mahadevan, S., Ramsey, L.~W., Terrien, R., et~al.\ 2014, \procspie, 9147, 91471G 

\bibitem[Maldonado et al.(2017)]{2017A&A...598A..27M} Maldonado, J., Scandariato, G., Stelzer, B., et al.\ 2017, \aap, 598, A27 

\bibitem[McCullough et al.(2005)]{2005PASP..117..783M} McCullough, P.~R., Stys, J.~E., Valenti, J.~A., et al.\ 2005, \pasp, 117, 783 

\bibitem[McQuillan et al.(2013)]{2013MNRAS.432.1203M} McQuillan, A., Aigrain, S., \& Mazeh, T.\ 2013, \mnras, 432, 1203 

\bibitem[Messina \& Guinan(2002)]{2002A&A...393..225M} Messina, S., \& Guinan, E.~F.\ 2002, \aap, 393, 225

\bibitem[Messina et~al.(2010)]{2010A&A...520A..15M} Messina, S., Desidera, S., Turatto, M., Lanzafame, A.~C., \& Guinan, E.~F.\ 2010, \aap, 520, A15

\bibitem[Mochnacki et al.(2002)]{2002AJ....124.2868M} Mochnacki, S.~W., Gladders, M.~D., Thomson, J.~R., et al.\ 2002, \aj, 124, 2868

\bibitem[Mohanty \& Basri(2003)]{2003ApJ...583..451M} Mohanty, S., \& Basri, G.\ 2003, \apj, 583, 451

\bibitem[Morin et~al.(2008)]{2008MNRAS.390..567M} Morin, J., Donati, J.-F., Petit, P., et~al.\ 2008, \mnras, 390, 567

\bibitem[Mullan \& MacDonald(2001)]{2001ApJ...559..353M} Mullan, D.~J., \& MacDonald, J.\ 2001, \apj, 559, 353

\bibitem[Newton et~al.(2016)]{2016ApJ...821...93N} Newton, E.~R., Irwin, J., Charbonneau, D., et~al.\ 2016, \apj, 821, 93 

\bibitem[Newton et al.(2017)]{2017ApJ...834...85N} Newton, E.~R., Irwin, J., Charbonneau, D., et al.\ 2017, \apj, 834, 85 

\bibitem[Norton et al.(2007)]{2007A&A...467..785N} Norton, A.~J., Wheatley, P.~J., West, R.~G., et al.\ 2007, \aap, 467, 785

\bibitem[Nutzman \& Charbonneau(2008)]{2008PASP..120..317N} Nutzman, P., \& Charbonneau, D.\ 2008, \pasp, 120, 317

\bibitem[Ochsenbein et al.(2000)]{2000A&AS..143...23O} Ochsenbein, F., Bauer, P., \& Marcout, J.\ 2000, \aaps, 143, 23 

\bibitem[Ol{\'a}h et al.(2009)]{2009A&A...501..703O} Ol{\'a}h, K., Koll{\'a}th, Z., Granzer, T., et al.\ 2009, \aap, 501, 703

\bibitem[Osten et al.(2005)]{2005ApJ...621..398O} Osten, R.~A., Hawley, S.~L., Allred, J.~C., Johns-Krull, C.~M., \& Roark, C.\ 2005, \apj, 621, 398 

\bibitem[Osten et al.(2010)]{2010ApJ...721..785O} Osten, R.~A., Godet, O., Drake, S., et al.\ 2010, \apj, 721, 785 

\bibitem[Paunzen \& Vanmunster(2016)]{2016AN....337..239P} Paunzen, E., \& Vanmunster, T.\ 2016, AN, 337, 239

\bibitem[Pepper et al.(2007)]{2007PASP..119..923P} Pepper, J., Pogge, R.~W., DePoy, D.~L., et al.\ 2007, \pasp, 119, 923 

\bibitem[Pettersen et al.(1984)]{1984ApJ...282..214P} Pettersen, B.~R., Coleman, L.~A., \& Evans, D.~S.\ 1984, \apj, 282, 214 

\bibitem[Pojma\'nski(1997)]{1997AcA....47..467P} Pojma\'nski, G.\ 1997, \actaa, 47, 467 

\bibitem[Pojma\'nski(2002)]{2002AcA....52..397P} Pojma\'nski, G.\ 2002, \actaa, 52, 397
    
\bibitem[Pollacco et~al.(2006)]{2006PASP..118.1407P} Pollacco, D.~L., Skillen, I., Collier Cameron, A., et~al.\ 2006, \pasp, 118, 1407

\bibitem[Queloz et~al.(2001)]{2001A&A...379..279Q} Queloz, D., Henry, G.~W., Sivan, J.~P., et~al.\ 2001, \aap, 379, 279

\bibitem[Quirrenbach et~al.(2012)]{2012SPIE.8446E..0RQ} Quirrenbach, A., Amado, P.~J., Seifert, W., et~al.\ 2012, \procspie, 8446, E0R

\bibitem[Quirrenbach et~al.(2014)]{2014SPIE.9147E..1FQ} Quirrenbach, A., Amado, P.~J., Caballero, J.~A., et~al.\ 2014, \procspie, 9147, E1F 

\bibitem[Quirrenbach et~al.(2015)]{2015csss...18..897Q} Quirrenbach, A., Caballero, J.~A., Amado, P.~J. et~al.\ 2015, 18th Cambridge Workshop on Cool Stars, Stellar Systems, and the Sun, Proceedings of the conference held at Lowell Observatory, 8-.14 June 2014. Edited by G. van Belle and H.C. Harris., pp. 897--906 

\bibitem[Quirrenbach et~al.(2016)]{2016SPIE.9908E..12Q} Quirrenbach, A., Amado, P.~J., Caballero, J.~A., et~al.\ 2016, \procspie, 9908, E12 

\bibitem[Rasmussen \& Williams(2005)]{RW05} Rasmussen, C. E. \& Williams, C. K. I., Gaussian Processes for Machine Learning, The MIT Press, November 2006. ISBN: 9780262182539

\bibitem[Reiners \& Basri(2007)]{2007ApJ...656.1121R} Reiners, A., \& Basri, G.\ 2007, \apj, 656, 1121

\bibitem[Reiners \& Basri(2008)]{2008ApJ...684.1390R} Reiners, A., \& Basri, G.\ 2008, \apj, 684, 1390

\bibitem[Reiners \& Basri(2010)]{2010ApJ...710..924R} Reiners, A., \& Basri, G.\ 2010, \apj, 710, 924 

\bibitem[Reiners et al.(2009)]{2009ApJ...692..538R} Reiners, A., Basri, G., \& Browning, M.\ 2009, \apj, 692, 538 

\bibitem[Reiners et~al.(2010)]{2010ApJ...710..432R} Reiners, A., Bean, J.~L., Huber, K.~F., et~al.\ 2010, \apj, 710, 432 

\bibitem[Reiners et al.(2018a)]{2018A&A...609L...5R} Reiners, A., Ribas, I., Zechmeister, M., et al.\ 2018a, \aap, 609, L5

\bibitem[Reiners et al.(2018b)]{2018A&A...612A..49R} Reiners, A., Zechmeister, M., Caballero, J.~A., et al.\ 2018b, \aap, 612, A49

\bibitem[Ricker et~al.(2015)]{2015JATIS...1a4003R} Ricker, G.~R., Winn, J.~N., Vanderspek, R., et~al.\ 2015, JATIS, 1, 014003

\bibitem[Rivera et al.(2005)]{2005ApJ...634..625R} Rivera, E.~J., Lissauer, J.~J., Butler, R.~P., et al.\ 2005, \apj, 634, 625 

\bibitem[Robertson et al.(2013)]{2013ApJ...764....3R} Robertson, P., Endl, M., Cochran, W.~D., \& Dodson-Robinson, S.~E.\ 2013, \apj, 764, 3

\bibitem[Saar \& Donahue(1997)]{1997ApJ...485..319S} Saar, S.~H., \& Donahue, R.~A.\ 1997, \apj, 485, 319
      
\bibitem[Savanov(2012)]{2012ARep...56..716S} Savanov, I.~S.\ 2012, Astronomy Reports, 56, 716

\bibitem[Scalo et~al.(2007)]{2007AsBio...7...85S} Scalo, J., Kaltenegger, L., Segura, A.~G., et~al.\ 2007, Astrobiology, 7, 85

\bibitem[Segura et~al.(2005)]{2005AsBio...5..706S} Segura, A., Kasting, J.~F., Meadows, V., et~al.\ 2005, Astrobiology, 5, 706

\bibitem[Scargle(1982)]{1982ApJ...263..835S} Scargle, J.~D.\ 1982, \apj, 263, 835 

\bibitem[Schwarzenberg-Czerny(1991)]{1991MNRAS.253..198S} Schwarzenberg-Czerny, A.\ 1991, \mnras, 253, 198 

\bibitem[Shkolnik et al.(2009)]{2009ApJ...699..649S} Shkolnik, E., Liu, M.~C., \& Reid, I.~N.\ 2009, \apj, 699, 649 

\bibitem[Skrutskie et al.(2006)]{2006AJ....131.1163S} Skrutskie, M.~F., Cutri, R.~M., Stiening, R., et al.\ 2006, \aj, 131, 1163

\bibitem[Stauffer et al.(1984)]{1984ApJ...280..202S} Stauffer, J.~R., Hartmann, L., Soderblom, D.~R., \& Burnham, N.\ 1984, \apj, 280, 202 

\bibitem[Strassmeier(2009)]{2009A&ARv..17..251S} Strassmeier, K.~G.\ 2009, \aapr, 17, 251

\bibitem[Stokes et al.(2000)]{2000Icar..148...21S} Stokes, G.~H., Evans, J.~B., Viggh, H.~E.~M., Shelly, F.~C., \& Pearce, E.~C.\ 2000, \icarus, 148, 21 

\bibitem[Su{\'a}rez Mascare{\~n}o et~al.(2015)]{2015MNRAS.452.2745S} Su{\'a}rez Mascare{\~n}o, A., Rebolo, R., Gonz{\'a}lez Hern{\'a}ndez, J.~I., \& Esposito, M.\ 2015, \mnras, 452, 2745  

\bibitem[Su{\'a}rez Mascare{\~n}o et~al.(2016)]{2016A&A...595A..12S} Su{\'a}rez Mascare{\~n}o, A., Rebolo, R., \& Gonz{\'a}lez Hern{\'a}ndez, J.~I.\ 2016, \aap, 595, A12

\bibitem[Mascare{\~n}o et al.(2018)]{2018A&A...612A..89M} Su{\'a}rez Mascare{\~n}o, A., Rebolo, R., Gonz{\'a}lez Hern{\'a}ndez, J.~I., et al.\ 2018, \aap, 612, A89

\bibitem[Tal-Or et al.(2018)]{2018A&A...614A.122T} Tal-Or, L., Zechmeister, M., Reiners, A., et al.\ 2018, \aap, 614, A122

\bibitem[Tamura et al.(2012)]{2012SPIE.8446E..1TT} Tamura, M., Suto, H., Nishikawa, J., et al.\ 2012, \procspie, 8446, 84461T

\bibitem[Tarter et~al.(2007)]{2007AsBio...7...30T} Tarter, J.~C., Backus, P.~R., Mancinelli, R.~L., et~al.\ 2007, Astrobiology, 7, 30

\bibitem[Testa et al.(2004)]{2004ApJ...617..508T} Testa, P., Drake, J.~J., \& Peres, G.\ 2004, \apj, 617, 508

\bibitem[Vida et al.(2013)]{2013AN....334..972V} Vida, K., Kriskovics, L., \& Ol{\'a}h, K.\ 2013, AN, 334, 972

\bibitem[Vida et al.(2014)]{2014MNRAS.441.2744V} Vida, K., Ol{\'a}h, K., \& Szab{\'o}, R.\ 2014, \mnras, 441, 2744

\bibitem[Vida et al.(2016)]{2016A&A...590A..11V} Vida, K., Kriskovics, L., Ol{\'a}h, K., et al.\ 2016, \aap, 590, A11 

\bibitem[Vida et al.(2017)]{2017ApJ...841..124V} Vida, K., K{\H o}v{\'a}ri, Z., P{\'a}l, A., Ol{\'a}h, K., \& Kriskovics, L.\ 2017, \apj, 841, 124 

\bibitem[Wargelin et al.(2017)]{2017MNRAS.464.3281W} Wargelin, B.~J., Saar, S.~H., Pojma{\'n}ski, G., Drake, J.~J., \& Kashyap, V.~L.\ 2017, \mnras, 464, 3281

\bibitem[West et al.(2015)]{2015ApJ...812....3W} West, A.~A., Weisenburger, K.~L., Irwin, J., et al.\ 2015, \apj, 812, 3

\bibitem[Wo{\'z}niak et~al.(2004)]{2004AJ....127.2436W} Wo{\'z}niak, P.~R., Vestrand, W.~T., Akerlof, C.~W., et~al.\ 2004, \aj, 127, 2436

\bibitem[Zechmeister \& K{\"u}rster(2009)]{2009A&A...496..577Z} Zechmeister, M., \& K{\"u}rster, M.\ 2009, \aap, 496, 577 

\bibitem[Zhilyaev et al.(2000)]{2000A&A...364..641Z} Zhilyaev, B.~E., Romanyuk, Y.~O., Verlyuk, I.~A., et al.\ 2000, \aap, 364, 641 

\end{thebibliography}
